\pdfoutput = 1

\documentclass[12pt,a4paper]{article}

\usepackage{amsmath}
\usepackage{amssymb}
\usepackage{cite}
\usepackage{hyperref}
\usepackage{graphicx}
\usepackage{booktabs}
\usepackage[caption=false]{subfig}

\newcommand{\lsim}{\mathrel{\hbox{\rlap{\hbox{\lower4pt\hbox{$\sim$}}}\hbox{$<$}}}}

\newcommand{\euler}{e}
\newcommand{\imaginary}{i}
\newcommand{\myexp}[1]{\ensuremath{\stackrel{\text{#1}}{=}}}

\usepackage{xspace}

\newcommand{\B}{\ensuremath{B}\xspace}
\newcommand{\Bbar}{\kern 0.18em\overline{\kern -0.18em B}{}\xspace}

\newcommand{\Bd }{\ensuremath{\B^0_d}\xspace}

\newcommand{\uBd}{\ensuremath{\B_d^{\vphantom{+}}}\xspace}
\newcommand{\Bs }{\ensuremath{\B^0_s}\xspace}
\newcommand{\Bsb}{\ensuremath{\Bbar^0_s}\xspace}
\newcommand{\uBs}{\ensuremath{\B_s^{\vphantom{+}}}\xspace}

\newcommand{\Bq }{\ensuremath{\B^0_q}\xspace}
\newcommand{\Bqb}{\ensuremath{\Bbar^0_q}\xspace}
\newcommand{\uBq}{\ensuremath{\B_q^{\vphantom{+}}}\xspace}
\newcommand{\Kbar}{\kern 0.18em\overline{\kern -0.18em K}{}\xspace}

\newcommand{\Bdtopipi}{\mbox{\ensuremath{\Bd\to \pi^-\pi^+}}\xspace}
\newcommand{\BdtoKpi}{\mbox{\ensuremath{\Bd\to \pi^-K^+}}\xspace}
\newcommand{\BdtoKK}{\mbox{\ensuremath{\Bd\to K^-K^+}}\xspace}
\newcommand{\uBdtopipi}{\mbox{\ensuremath{\uBd\to \pi^-\pi^+}}\xspace}
\newcommand{\uBdtoKpi}{\mbox{\ensuremath{\uBd\to \pi^-K^+}}\xspace}
\newcommand{\uBdtoKK}{\mbox{\ensuremath{\uBd\to K^-K^+}}\xspace}
\newcommand{\BstoKK}{\mbox{\ensuremath{\Bs\to K^-K^+}}\xspace}
\newcommand{\BstoKpi}{\mbox{\ensuremath{\Bs\to K^-\pi^+}}\xspace}
\newcommand{\Bstopipi}{\mbox{\ensuremath{\Bs\to \pi^-\pi^+}}\xspace}
\newcommand{\uBstoKK}{\mbox{\ensuremath{\uBs\to K^-K^+}}\xspace}
\newcommand{\uBstoKpi}{\mbox{\ensuremath{\uBs \to K^-\pi^+}}\xspace}
\newcommand{\uBstopipi}{\mbox{\ensuremath{\uBs\to \pi^-\pi^+}}\xspace}
\newcommand{\BdtoKantiK}{\mbox{\ensuremath{\Bd\to K^0\Kbar^0}}\xspace}
\newcommand{\BstoKantiK}{\mbox{\ensuremath{\Bs\to K^0\Kbar^0}}\xspace}
\newcommand{\uBdtoKantiK}{\mbox{\ensuremath{\uBd\to K^0\Kbar^0}}\xspace}
\newcommand{\uBstoKantiK}{\mbox{\ensuremath{\uBs\to K^0\Kbar^0}}\xspace}

\begin{document}


\begin{titlepage}

\vspace*{-0.0truecm}

\begin{flushright}
Nikhef-2016-039\\
SI-HEP-2016-28\\
QFET-2016-19
\end{flushright}

\vspace*{0.3truecm}

\begin{center}
{\Large \bf \boldmath Towards New Frontiers in the Exploration of 

\vspace*{0.3truecm}

Charmless Non-Leptonic $B$ Decays}
\end{center}

\vspace{0.9truecm}

\begin{center}
{\bf Robert Fleischer,\,${}^{a,b}$  Ruben Jaarsma,\,${}^{a}$ and  K. Keri Vos\,${}^{a,c,d}$}

\vspace{0.5truecm}

${}^a${\sl Nikhef, Science Park 105, NL-1098 XG Amsterdam, Netherlands}

${}^b${\sl  Department of Physics and Astronomy, Vrije Universiteit Amsterdam,\\
NL-1081 HV Amsterdam, Netherlands}

${}^c${\sl Van Swinderen Institute for Particle Physics and Gravity, University of Groningen,\\
NL-9747 AG Groningen, Netherlands}

${}^d${\sl Theoretische Physik 1, Naturwissenschaftlich-Technische Fakult\"at, \\
Universit\"at Siegen, D-57068 Siegen, Germany}

\end{center}

\vspace*{1.7cm}


\begin{abstract}
\noindent
Non-leptonic $B$ decays into charmless final states offer an important laboratory to study CP 
violation and the dynamics of strong interactions. Particularly interesting are $B^0_s\to K^-K^+$ 
and $B^0_d\to\pi^-\pi^+$ decays, which are related by the $U$-spin symmetry of strong interactions,
and allow for the extraction of CP-violating phases and tests of the Standard Model. The theoretical precision 
is limited by $U$-spin-breaking corrections and innovative methods are needed in view of the impressive 
future experimental precision expected in the era of Belle II and the LHCb upgrade. We have 
recently proposed a novel method to determine the $B_s^0$--$\bar{B}_s^0$ mixing phase $\phi_s$ from 
the $B_s^0\to K^-K^+$, $B_d^0\to \pi^-\pi^+$ system, where semileptonic $B^0_s\to K^-\ell^+\nu_\ell$,
$B^0_d\to \pi^-\ell^+\nu_\ell$ decays are a new ingredient and the theoretical situation is very favourable. 
We discuss this strategy in detail, with a focus on penguin contributions as well as exchange and penguin-annihilation topologies which can be probed by a variety of non-leptonic $B$ decays into charmless 
final states. We show that a theoretical precision as high as ${\cal O}(0.5^\circ)$ for $\phi_s$ 
can be attained in the future, thereby offering unprecedented prospects for the search for new sources 
of CP violation.
\end{abstract}


\vspace*{2.1truecm}

\vfill

\noindent
December 2016

\end{titlepage}


\thispagestyle{empty}

\vbox{}

\tableofcontents

\newpage

\setcounter{page}{1}


\section{Introduction}
CP-violating asymmetries of $B$ mesons are powerful probes in the search for physics beyond the 
Standard Model (SM) of particle physics. New sources of CP violation might be revealed when comparing 
the experimental observables determined from different decays with the corresponding SM expectations.
Since CP asymmetries are generated through interference effects, non-leptonic decays govern this territory 
of the $B$ physics landscape. As new heavy particles may well enter the loop contributions (see, for
instance, Ref.~\cite{Bur14}), decays with penguin topologies are particularly interesting. In order to fully exploit the physics potential of these channels in the era of 
Belle II \cite{Belle-II} and the LHCb upgrade \cite{LHCbup}, an unprecedented precision of the corresponding
SM predictions is essential to match experiment.

The decay \BstoKK is dominated by QCD penguin topologies and is hence a particularly promising probe
to search for footprints of New Physics (NP) through studies of CP violation. However, the corresponding
hadronic parameters suffer from significant theoretical uncertainties through non-perturbative effects. 
Fortunately, this  decay is related to \Bdtopipi through the $U$-spin flavour symmetry of the strong interaction, 
which relates -- in analogy to the well-known isospin symmetry -- the $d$ and $s$ quarks to each other. 
Applying the $U$-spin symmetry, the hadronic parameters characterizing the 
\Bdtopipi and \BstoKK modes can be related to each other, allowing the extraction of the angle $\gamma$
of the unitarity triangle (UT) of the Cabibbo--Kobayashi--Maskawa (CKM) matrix and the 
$B_s^0$--$\bar{B}_s^0$
mixing phase $\phi_s$ \cite{Fle99, Fle07, Fle10}. First measurements of this $U$-spin method have been
performed by the LHCb Collaboration, yielding results for $\gamma$ and $\phi_s$ in agreement with the 
SM and uncertainties at the $7^\circ$ level  \cite{Aaij:2013tna,Aaij:2014xba}.  

The theoretical precision of this strategy, which is limited by non-factorizable $U$-spin-breaking corrections,
is unfortunately not sufficient to fully exploit the future measurements of CP violation in the 
\Bdtopipi, \BstoKK system at Belle II and the LHCb upgrade. In view of this situation, we have proposed 
a new method which is very robust with respect to theoretical uncertainties. It uses $\gamma$, which 
can eventually be determined with ${\cal O}(1^\circ)$ precision through pure tree decays, as input and
allows the determination of $\phi_s$ with a theoretical precision of up to $0.5^\circ$ at Belle II and the 
LHCb upgrade \cite{SHORT}. As the main new ingredient, it uses the \Bdtopipi, \BstoKK system in 
combination with the semileptonic $B_d^0\rightarrow \pi^- \ell^+ \nu_\ell$, $B_s^0\rightarrow K^- \ell^+ \nu_\ell$ decays.
Following these lines, the application of the $U$-spin symmetry 
can be limited to theoretically well behaved quantities and valuable tests of the $U$-spin symmetry 
can be obtained.  As we pointed out in Ref.~\cite{SHORT}, the current experimental picture is very promising. 

In the present paper, we explore the technical details of this new strategy and the attainable precision 
of $\phi_s$ in a more comprehensive way. The leading $U$-spin-breaking corrections enter through a
ratio of colour-allowed tree amplitudes, which are well-behaved with respect to factorization and can
be analysed within QCD factorization. The major limiting uncertainties enter through certain penguin 
topologies as well as exchange and penguin-annihilation topologies. The latter are expected to play a 
minor role in the \Bdtopipi and \BstoKK system on the basis of dynamical arguments 
\cite{Gro94, Gro95, Bob14}. Here we present a detailed analysis to constrain these contributions through
experimental data, where $B^0_s\to\pi^-\pi^+$, $B^0_d\to K^-K^+$ modes play the key role as they
emerge exclusively from exchange and penguin-annihilation topologies. In order to determine the relevant
penguin contributions, the $B^0_{s,d}\to K^0\bar K^0$ system will be in the spotlight. We will give a 
roadmap for exploiting the physics information offered by these $U$-spin-related systems at Belle II and
the LHCb upgrade, allowing valuable new insights into hadron dynamics and $U$-spin-breaking effects.

The outline of this paper is as follows: in Section~\ref{sec:topologiesAndAsymmetries},
we introduce the \Bdtopipi and \BstoKK decays and the relevant observables. In Section~\ref{sec:ori}, we discuss the
original $U$-spin strategy and its prospects for the LHCb upgrade. The new strategy is presented in 
Section~\ref{sec:newstrat}, exploring also the picture arising from the current data. In 
Sections~\ref{sec:penguindy} and \ref{sec:InformationFromBtohhDecays}, we explore the dynamics of
penguin topologies and exchange, penguin-annihilation topologies, respectively. In the latter section, 
we discuss also the expected pattern of the CP asymmetries in the \BdtoKK, \Bstopipi decays and various
future scenarios. The prospects of our new strategy are discussed in Section~\ref{sec:Discussion}, and
our main conclusions are summarized in Section~\ref{sec:Conclusion}. Throughout this paper we
shall assume that all decay amplitudes are described by their SM expressions.


\section{Decay Amplitudes and CP Asymmetries} \label{sec:topologiesAndAsymmetries}


\subsection{Topologies}

\begin{figure}[tp]
	\centering
	\subfloat[Tree ($T$)]{\label{fig:treeDiagram} \includegraphics[width=0.49\textwidth]{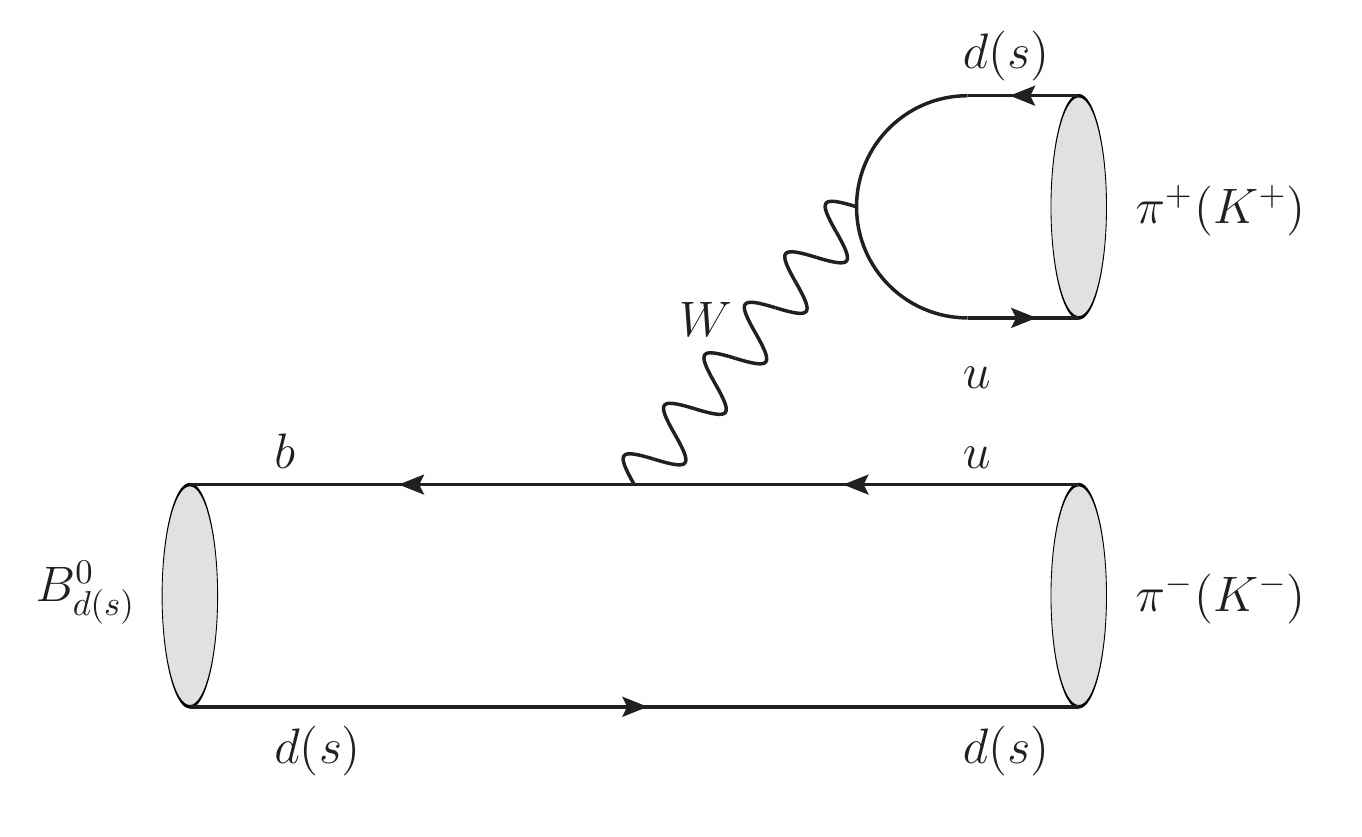}}
	\subfloat[Penguin ($P$)]{\label{fig:penguinDiagram} \includegraphics[width=0.49\textwidth]{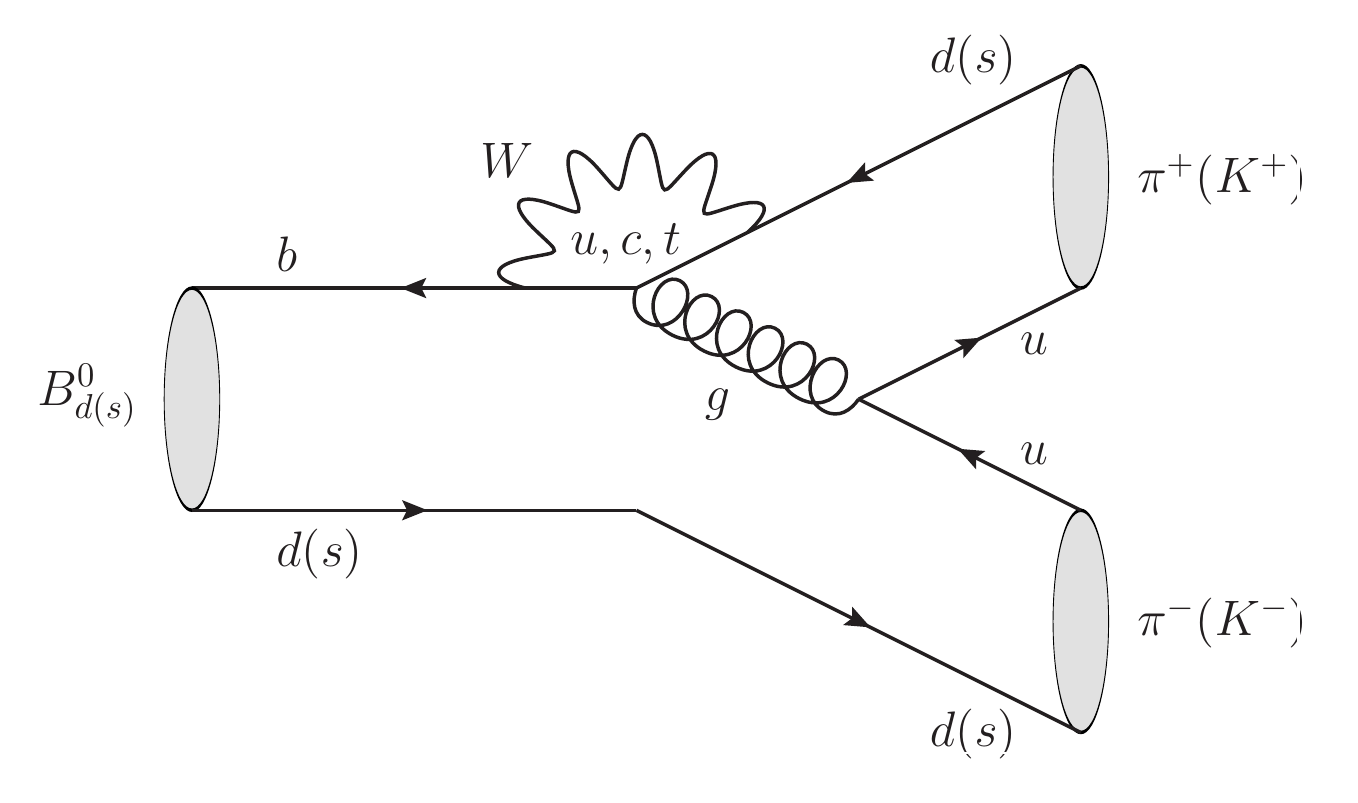}}\\
	\subfloat[Exchange ($E$)]{\label{fig:exchangeDiagram} \includegraphics[width=0.49\textwidth]{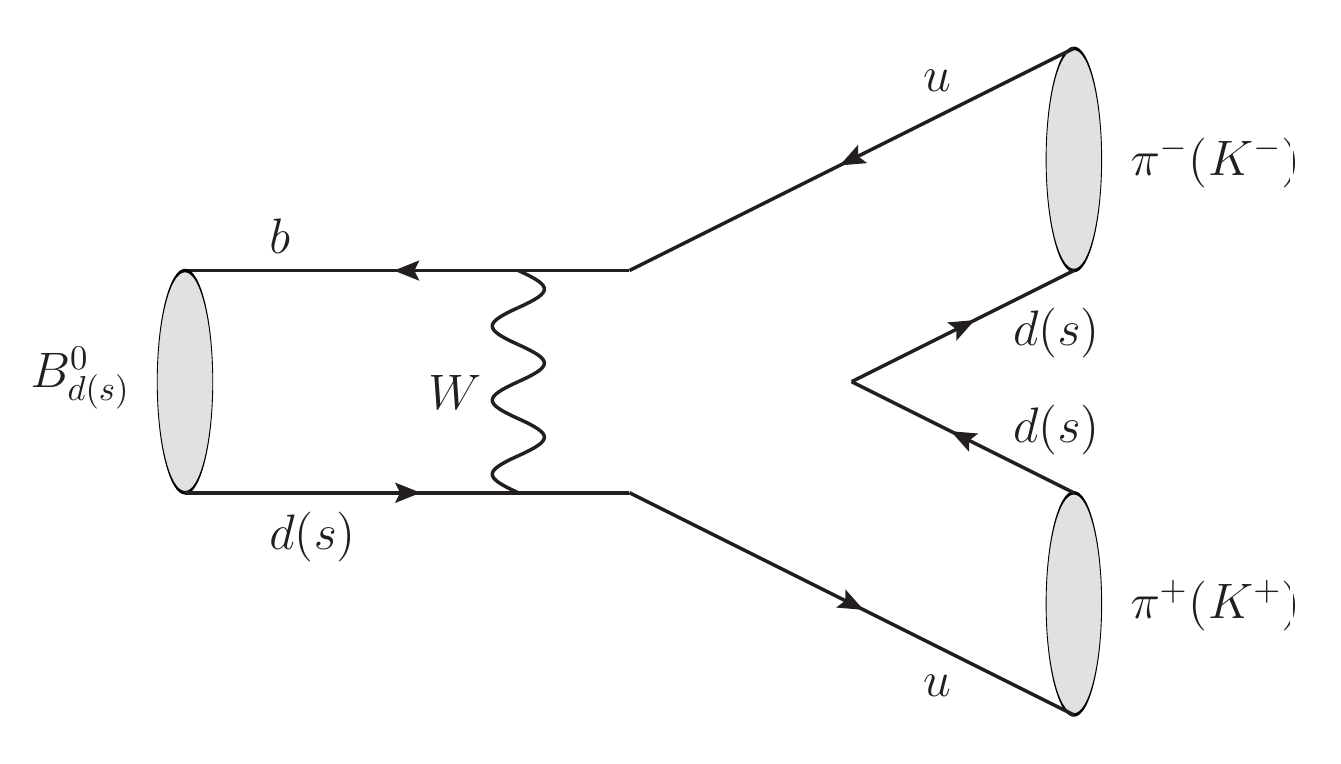}}
	\subfloat[Penguin-annihilation ($PA$)]{\label{fig:penguinannihilationDiagram} \includegraphics[width=0.49\textwidth]{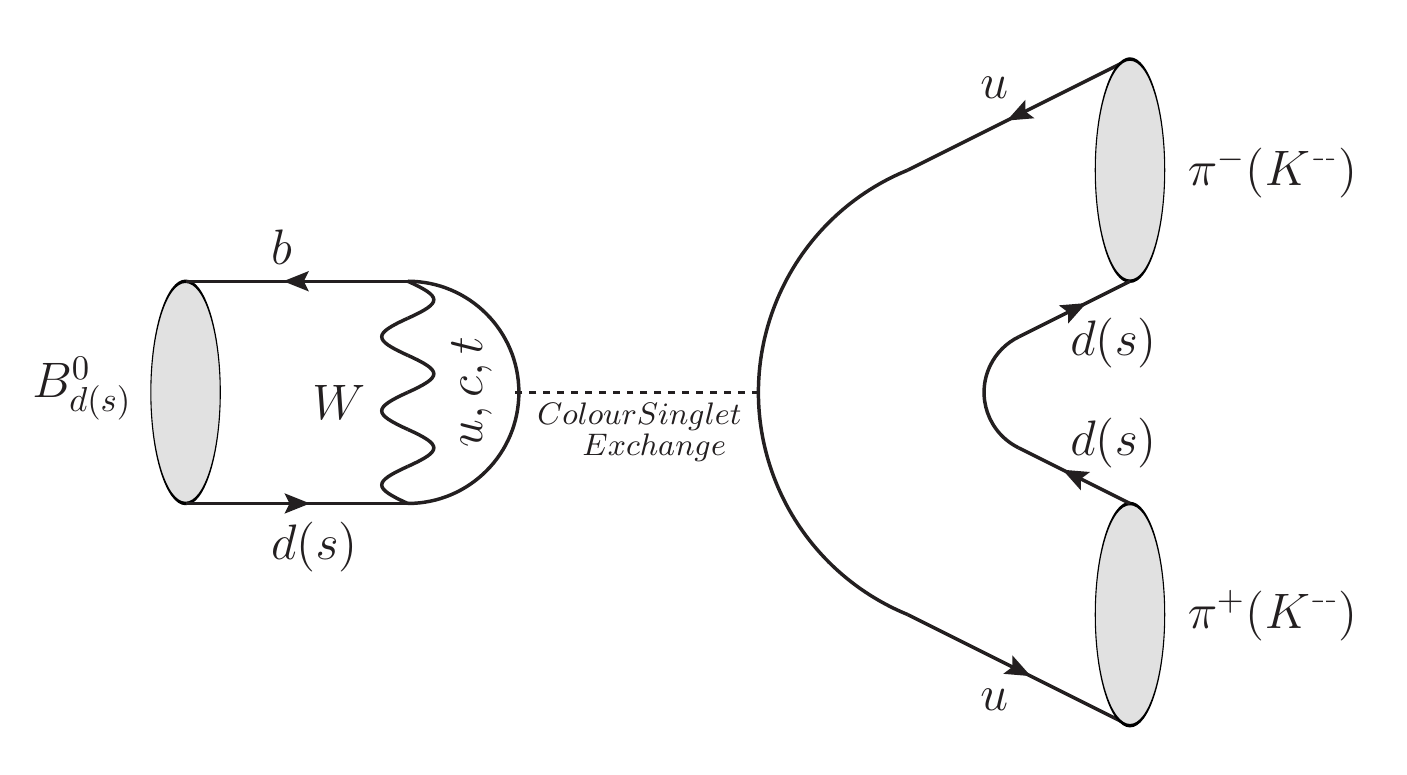}}
	\caption{Topologies of the $\Bdtopipi$ and $\BstoKK$ decays.}
	\label{fig:diagrams}
\end{figure}

The non-leptonic decay \Bdtopipi, characterized by a $\bar{b}\rightarrow \bar{u}u\bar{d}$ transition, is governed by the decay topologies depicted in Fig.~\ref{fig:diagrams}. The decay amplitude is dominated by contributions from the tree ($T$) and penguin ($P$) topologies, but also receives contributions from exchange ($E$) and penguin-annihilation ($PA$) topologies. In the SM, we have \cite{Fle99}
\begin{equation}\label{Eq:Bdtopipi-Amp}
A\left(\Bdtopipi\right) = e^{i\gamma}\mathcal{C}\left[1- d e^{i\theta}e^{-i\gamma}\right]
\end{equation}
and
\begin{equation}\label{eq:Cpipi}
\mathcal{C} \equiv \lambda^3 A R_b\left[T+E+P^{(ut)}+PA^{(ut)}\right]
\end{equation}
\begin{equation}\label{eq:dpipi}
de^{i\theta} \equiv \frac{1}{R_b}\left[\frac{P^{(ct)}+PA^{(ct)}}
{T+E+P^{(ut)}+PA^{(ut)}}\right] 
\end{equation}
with
\begin{equation}
P^{(qt)} \equiv P^{(q)}-P^{(t)}, \quad\quad PA^{(qt)}\equiv PA^{(q)}-PA^{(t)} \ .
\end{equation} 
Both $\mathcal{C}$ and $de^{i\theta}$ are CP-conserving hadronic parameters, while $\gamma$
provides a CP-violating phase. On the other hand,
\begin{equation}
R_b \equiv \left(1-\frac{\lambda^2}{2}\right)\frac{1}{\lambda}\left|\frac{V_{ub}}{V_{cb}}\right| = 0.390 \pm 0.030
\end{equation}
measures one side of the UT, with $\lambda$ and $A$ denoting the Wolfenstein parameters
of the CKM matrix \cite{Wol83,BLO}. For the numerical value, we have used the following results 
\cite{Cha15}:
\begin{equation}
\lambda \equiv |V_{us}|=0.22543 \pm 0.00042, \quad
A\equiv |V_{cb}|/\lambda^2= 0.8227^{+ 0.0066}_{- 0.0136}.
\end{equation}

The decay \BstoKK originates from a $\bar{b}\rightarrow \bar{u}u \bar{s}$ transition and is related 
to the \Bdtopipi channel through the $U$-spin symmetry  \cite{Fle99, Fle07, Fle10}. In the SM, the
$\BstoKK$ transition amplitude can be written in the following form:
\begin{equation}\label{eq:bstoKK}
A\left(\BstoKK\right)= \sqrt{\epsilon}e^{i\gamma}\mathcal{C}'\left[1+ \frac{1}{\epsilon}d' e^{i\theta'}e^{-i\gamma}\right]\:,
\end{equation}
where $\mathcal{C'}$ and $d'e^{i\theta'}$are the primed equivalents of Eqs.~\eqref{eq:Cpipi} and 
\eqref{eq:dpipi}, respectively. The decay topologies are given in Fig.~\ref{fig:diagrams}. The suppression of the overall amplitude and the enhancement of the penguin parameters $d'e^{i\theta'}$ is given by 
\begin{equation}
\epsilon \equiv \frac{\lambda^2}{1-\lambda^2} = 0.0535 \pm 0.0002\:.
\end{equation}

The $U$-spin symmetry  \cite{Fle99} implies
\begin{equation}\label{eq:Uspinrel}
d e^{i \theta} = d' e^{i\theta'} \ , 
\end{equation}
which is only sensitive to non-factorizable $U$-spin-breaking corrections because the factorizable contributions cancel in these ratios of amplitudes. Contrary, the $U$-spin relation 
\begin{equation} \label{eq:UspinrelC}
\mathcal{C} = \mathcal{C}'
\end{equation}
is affected by both factorizable and non-factorizable $U$-spin-breaking effects.


\subsection{CP Asymmetries}
Thanks to quantum-mechanical oscillations between \Bq and \Bqb mesons, an initially present \Bq meson evolves in time into a linear combination of \Bq and \Bqb states. CP violation is probed by the following 
time-dependent decay rate asymmetry \cite{Fle02}:
\begin{equation}
\begin{split}
\mathcal{A}_\text{CP}(t) &= \frac{|A(B^0_q(t)\rightarrow f)|^2-|A(\bar{B}^0_q(t)\rightarrow f)|^2}{|A(B^0_q(t)\rightarrow f)|^2+|A(\bar{B}^0_q(t)\rightarrow f)|^2} \\
&=\frac{\mathcal{A}_{\rm CP}^{\rm dir}(B_q\rightarrow f )\cos(\Delta M_qt)+\mathcal{A}_{\rm CP}^{\rm mix}(B_q\rightarrow f)\sin(\Delta M_qt)}{\cosh(\Delta\Gamma_qt/2)+\mathcal{A}_{\Delta\Gamma}(B_q\rightarrow f)\sinh(\Delta\Gamma_qt/2)}\:,
\end{split}
\end{equation}
where \mbox{$\Delta M_q\equiv M^{(q)}_{\rm H}-M^{(q)}_{\rm L}$} and \mbox{$\Delta\Gamma_q\equiv \Gamma_{\rm L}^{(q)}-\Gamma_{\rm H}^{(q)}$} denote the mass and decay width differences between the 
``heavy" and ``light" $\uBq$ mass eigenstates, respectively.

For the $B_s^0\rightarrow K^-K^+$ channel, we obtain the following expressions \cite{Fle99}:
\begin{align}
\label{eq:cpasym}
\mathcal{A}_{\rm CP}^{\rm dir}(\uBstoKK)&=\frac{2 \epsilon d' \sin\theta'\sin\gamma}
{d^{\prime 2}+2\epsilon d'\cos\theta'\cos\gamma+\epsilon^2}\:, \\
\label{eq:cpasymmix}
\mathcal{A}_{\rm CP}^{\rm mix}(\uBstoKK)&=\phantom{-}\left[
\frac{d^{\prime 2}\sin\phi_s +2 \epsilon d' \cos\theta'\sin(\phi_s+\gamma)+\epsilon^2\sin(\phi_s+2\gamma)}
{d^{\prime 2}+2\epsilon d'\cos\theta'\cos\gamma+\epsilon^2}\right]\:, \\
\label{eq:cpasymDG}
\mathcal{A}_{\Delta\Gamma}(\uBstoKK)&=-\left[
\frac{d^{\prime 2}\cos\phi_s +2 \epsilon d' \cos\theta'\cos(\phi_s+\gamma)+\epsilon^2\cos(\phi_s+2\gamma)}
{d^{\prime 2}+2\epsilon d'\cos\theta'\cos\gamma+\epsilon^2}\right]\:.
\end{align}
These observables are not independent from one another, satisfying the general relation
\begin{equation}\label{eq:sumrule}
\left[\mathcal{A}_{\rm CP}^{\rm dir}(\uBstoKK)\right]^2+
\left[\mathcal{A}_{\rm CP}^{\rm mix}(\uBstoKK)\right]^2 +
\left[\mathcal{A}_{\Delta\Gamma}(\uBstoKK)\right]^2 =1 \ .
\end{equation}
The CP-violating asymmetries for the \Bdtopipi channel can be straightforwardly obtained 
through the following replacements:
\begin{equation}
d' \to d \ ,\qquad \theta' \to \theta \ ,\qquad \phi_s \to \phi_d \ ,\qquad \epsilon \to -1 \ .
\end{equation}

While the direct CP asymmetries $\mathcal{A}_{\rm CP}^{\rm dir}$ of \Bdtopipi and \BstoKK originate from
interference between tree and penguin topologies, the mixing-induced CP asymmetries 
$\mathcal{A}_{\rm CP}^{\rm mix}$ are induced by interference between \Bq--\Bqb mixing and decay processes. The latter observables involve the $B^0_q$--$\bar B^0_q$ mixing phases
\begin{equation}
\phi_d=2\beta + \phi_d^{\rm NP}\:, \qquad \phi_s= -2\beta_s + \phi_s^{\rm NP}\:,
\end{equation} 
where $\beta$ is the usual angle of the UT and $\phi_s$ is a doubly Cabibbo-suppressed phase in
the SM. The fits of the UT allow us to calculate the SM value of $\phi_s$ with high precision
\cite{Cha15, Art15}:
\begin{equation}\label{eq:Phis_SM}
\phi_s^{\rm SM} = -2\beta_s = -(2.092^{+0.075}_{-0.069})^{\circ}\:.
\end{equation}
The phases $\phi_{d}^{\rm NP}$ and $\phi_{s}^{\rm NP}$ describe possible CP-violating NP contributions
to $B^0_d$--$\bar B^0_d$ and $B^0_s$--$\bar B^0_s$ mixing, respectively.


\subsection{Untagged Decay Rates}\label{ssec:untagged}
Branching ratios contain information from the untagged decay rates \cite{DFN}. In experiments, the branching ratio is typically defined by using the time-integrated untagged rate, while theoretical expressions require the untagged decay rate at time $t = 0$ \cite{DeBruyn:2012wj}. For the $B_s$ meson system there is -- in 
contrast to the $B_d$-meson system -- a sizeable difference between the decay widths of the 
mass eigenstates  \cite{Amh14}:
\begin{equation}
y_s \equiv \frac{\Delta\Gamma_s}{2\Gamma_s} \equiv \frac{\Gamma_\text{L}^{(s)}-\Gamma_\text{H}^{(s)}}{2\Gamma_s} = 0.0625 \pm 0.0045.
\end{equation}
Consequently, the experimental branching ratio needs to be converted into the theoretical branching ratio
by means of the following expression \cite{DeBruyn:2012wj}:
\begin{equation}\label{eq:theoexp}
\mathcal{B}(B_s \to f)_\text{theo} = \left[\frac{1-y_s^2}{1+\mathcal{A}_{\Delta\Gamma}^fy_s}\right]\mathcal{B}(B_s \to f)_\text{exp}.
\end{equation}
For decays into a flavour-specific final state, such as $\BstoKpi$, only the $[1-y_s^2]$ factor contributes
in (\ref{eq:theoexp}). Using the effective lifetime
\begin{equation}
\tau_f\equiv \frac{\int^\infty_0 t \left\langle \Gamma(B_s(t)\rightarrow f)\right\rangle  dt}{\int^\infty_0  \left\langle \Gamma(B_s(t)\rightarrow f)\right\rangle  dt} 
\end{equation}
of the $B_s$ decay at hand, the conversion between the experimental and theoretical branching ratios 
can be obtained with the help of the relation
\begin{equation}\label{eq:BRtheo}
\mathcal{B}(B_s \to f)_\text{theo} = \left[2-(1-y_s^2)\frac{\tau_f}{\tau_{B_s}}\right]\mathcal{B}(B_s \to f)_\text{exp},
\end{equation}
which does not explicitly depend on the $\mathcal{A}_{\Delta\Gamma}^f$ observable \cite{DeBruyn:2012wj}. 

For the conversion of the experimental $\BstoKK$ branching ratio into its theoretical counterpart, 
we use the measurement of the LHCb Collaboration \cite{Aaij:2014fia}
\begin{equation}
\begin{split}
\tau_{K^+K^-} & = \left[1.407 \pm 0.016 \text{ (stat) } \pm 0.007  \text{ (syst)}\right] \text{ps} , 
\end{split}
\end{equation}
which leads to a difference between the experimental and theoretical branching ratios of about $7\%$. 

It is useful to introduce the following quantity \cite{Fle07,Fle10}:
\begin{align} \label{eq:kObservable}
K &\equiv \frac{1}{\epsilon}\left|\frac{\mathcal{C}}{\mathcal{C}^\prime}\right|^2\left[\frac{m_{B_s}}{m_{B_d}}\frac{\Phi(m_\pi/m_{B_d},m_\pi/m_{B_d})}{\Phi(m_K/m_{B_s},m_K/m_{B_s})}\frac{\tau_{B_d}}{\tau_{B_s}}\right]\frac{\mathcal{B}(B_s \to K^-K^+)_\text{theo}}{\mathcal{B}(B_d \to \pi^-\pi^+)} \nonumber \\
&= \frac{1+2(d^\prime/\epsilon)\cos\theta^\prime\cos\gamma+(d^\prime/\epsilon)^2}{1-2d\cos\theta\cos\gamma+d^2} ,
\end{align}
where
\begin{equation} \label{eq:phaseSpaceFunction}
\Phi(X,Y) = \sqrt{[1-(X+Y)^2][1-(X-Y)^2]}
\end{equation}
is the usual phase-space function. The factorizable $U$-spin-breaking contributions
to the ratio $|\mathcal{C}/\mathcal{C}'|$ are given as follows:
\begin{equation}\label{eq:factcc}
\left|\frac{\mathcal{C}}{\mathcal{C}^\prime}\right|_\textrm{fact} = \frac{f_\pi}{f_K}\left[\frac{m_{B_d}^2-m_\pi^2}{m_{B_s}^2-m_K^2}\right]\left[\frac{F_0^{B_d\pi}(m_\pi^2)}{F_0^{B_sK}(m_K^2)}\right] = 0.71_{-0.11}^{+0.06} , 
\end{equation}
where we have used the QCD light-cone sum rule (LCSR) calculation $F_0^{B_sK}(0)/F_0^{B_d\pi}(0)=1.15^{+0.17}_{-0.09}$ \cite{Dub08}, which is in agreement with previous results in \cite{Kho04}, and $f_K/f_\pi = 1.1928\pm 0.0026$ \cite{Ros15}. The form factors for the $\bar{B}^0_d \to \pi^+$ transition
are defined through
\begin{align} \label{eq:ffDef}
\left\langle \pi^+(k)| \bar{u}\gamma_\mu b |\bar{B}^0_d(p)\right\rangle = F_0^{B_d\pi}(q^2)\left(\frac{m_{B_d}^2-m_\pi^2}{q^2}\right)q_\mu \nonumber \\
+F_1^{B_d\pi}(q^2)\left[(p+k)_\mu - \left(\frac{m_{B_d}^2 - m_\pi^2}{q^2}\right) q_\mu\right]
\end{align}
with $q\equiv p-k$; the $\bar{B}^0_s \to K^+$  form factors $F_{0,1}^{B_sK}(q^2)$ are defined 
in an analogous way. Finally, we obtain
\begin{equation}\label{eq:Kval}
K \myexp{exp} 51.4_{-15.7}^{+9.0} \ , 
\end{equation}
where we have neglected non-factorizable $U$-spin-breaking corrections to the ratio $|\mathcal{C}/\mathcal{C}'|$. We shall return to this quantity in Section~\ref{sec:Discussion}. 
Using the $U$-spin relations in Eq.~\eqref{eq:Uspinrel}, we may also write 
\begin{equation}
K = -\frac{1}{\epsilon}\left[\frac{\mathcal{A}_\text{CP}^\text{dir}(\uBdtopipi)}{\mathcal{A}_\text{CP}^\text{dir}(\uBstoKK)}\right]  \myexp{exp} 41.4 \pm 33.2 \ ,
\end{equation}
which is in agreement with Eq.~\eqref{eq:Kval}, but has a much larger error due to the currently 
large uncertainties of the $\BstoKK$ CP asymmetries.


\section{The Original Strategy}\label{sec:ori}
Before discussing the new method, it is instructive to have a closer look at the original strategy 
\cite{Fle99, Fle07, Fle10}, where $\gamma$ and $\phi_s$ can be extracted from the \Bdtopipi, \BstoKK 
system with the help of the $U$-spin symmetry. Using information on the corresponding branching
ratios, CP violation in the \Bdtopipi mode and the first measurement of CP violation in the 
\BstoKK channel \cite{Aaij:2013tna}, the LHCb collaboration has reported the following results \cite{Aaij:2014xba}:
\begin{equation}\label{eq:Uspinnow}
\gamma = (63.5^{+7.2}_{-6.7})^\circ, \qquad \phi_s = -(6.9^{+9.2}_{-8.0})^\circ \ ,
\end{equation}
which are in agreement with the picture of the previous analyses in Refs.~\cite{Fle99, Fle07, Fle10}.

\boldmath
\subsection{The UT Angle $\gamma$}
\unboldmath
The UT angle $\gamma$ can be determined in a theoretically clean way from pure tree decays of
the kind $B\rightarrow D^{(*)}K^{(*)}$ \cite{gw,ADS} (for an overview, see \cite{FR-gam}). The averages
of the corresponding experimental results performed by the CKMfitter \cite{Charles:2004jd} and UTfit \cite{UTfit} collaborations yield
\begin{equation}\label{eq:gammasfit}
\gamma = (73.2_{-7.0}^{+6.3})^\circ \qquad\mbox{and}\qquad \gamma = (68.3\pm7.5)^{\circ},
\end{equation}
respectively. The results in (\ref{eq:gammasfit}) are in remarkable agreement with the $\gamma$ 
measurement in (\ref{eq:Uspinnow}), and it is interesting to note that the current uncertainties of 
both determinations are at the same level. In the future era of Belle II and the LHCb upgrade, the 
uncertainty of the $\gamma$ determination from pure $B\rightarrow D^{(*)}K^{(*)}$ tree decays 
can be reduced to the $1^\circ$ level, which is very impressive \cite{Belle-II,LHCbup}.

\begin{table}[t]
	\centering
	\begin{tabular}{l | r | r }
		& Current \cite{Amh14, PDG} & Upgrade  \cite{LHCbup} \\
		\hline\hline
		$\mathcal{A}_{\textrm{CP}}^{\textrm{dir}}(\uBdtopipi)$ & $ -0.31 \pm 0.05$ & $-0.31 \pm 0.008$\\
		$\mathcal{A}_{\textrm{CP}}^{\textrm{mix}}(\uBdtopipi)$ & $0.66 \pm 0.06 $ & $0.66 \pm 0.008$ \\
		$\mathcal{A}_{\textrm{CP}}^{\textrm{dir}}(\uBstoKK)$ & $0.14 \pm 0.11$ & $0.087 \pm 0.008$\\
		$\mathcal{A}_{\textrm{CP}}^{\textrm{mix}}(\uBstoKK)$ & $-0.30\pm 0.13$ & $-0.19 \pm 0.008$ \\
	\end{tabular}
	\caption{Overview of the current measurements and the expected accuracy at the LHCb upgrade. The upgrade central values for \BstoKK are calculated by applying the $U$-spin symmetry to $(d,\theta)$ obtained from the \Bdtopipi CP asymmetries. }
	\label{tab:CPs}
\end{table}

The current values of the CP asymmetries \cite{Amh14,PDG}
are listed in Table~\ref{tab:CPs}. Let us now explore the prospects of the $U$-spin strategy. Contrary to the pure tree determination of 
$\gamma$, the \Bdtopipi, \BstoKK system obtains significant contributions from penguin loop topologies, 
which may receive NP contributions. Within the current precision at the level of $7^\circ$, there is not
any sign of CP-violating NP effects of this kind in the data and an effort has to be made to achieve a
much higher precision.

\begin{figure}[t]
	\centering 
 	\includegraphics[width=0.49\textwidth]{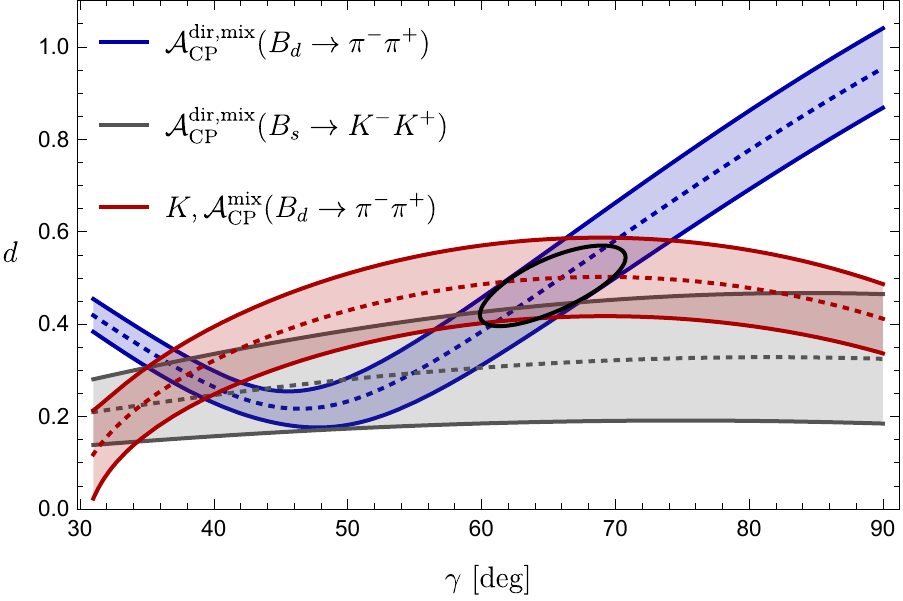}
	\caption{Illustration of the determination of $\gamma$ from the CP asymmetries of \Bdtopipi, \BstoKK and the observable $K$ for the current data. }
	\label{fig:dgammaK}
\end{figure}

Let us use the mixing phases $\phi_d= 43.2\pm 1.8$ \cite{DeBruyn:2014oga}, as determined from $B_{d,s}^0 \rightarrow J/\psi K_s^0$ decays by taking penguin effects into account, and the PDG average $\phi_s = -(0.68 \pm 2.2)^\circ$ \cite{PDG} (see Subsection~\ref{ssec:phis}). Moreover, we assume the $U$-spin relations in Eq.~\eqref{eq:Uspinrel}. In Fig.~\ref{fig:dgammaK}, we show the contours in the $d$--$\gamma$ plane which can then be fixed -- in a theoretically clean way -- through the CP asymmetries of the \Bdtopipi and \BstoKK decays. We observe that currently only poor constraints on $\gamma$ can be obtained by using only the CP asymmetries, which is mainly due to the large uncertainty of the CP violation measurements of the \BstoKK channel.

Consequently, the current LHCb determination in Eq.~\eqref{eq:Uspinnow} is governed by CP 
violation in \Bdtopipi and the branching ratio information encoded in the $K$ observable given in 
 Eq.~\eqref{eq:kObservable}. We illustrate this feature in Fig.~\ref{fig:dgammaK}, where we have used the value of $K$ in Eq.~\eqref{eq:Kval}\footnote{To be conservative, we consider only the largest uncertainty 
 for $K$ in Eq.~\eqref{eq:Kval}.}, containing the factorizable form-factor contributions to the ratio $|\mathcal{C}/\mathcal{C}'|$ given in Eq.~\eqref{eq:factcc}. We have neglected any non-factorizable contributions to $|\mathcal{C}/\mathcal{C}'|$, and have assumed the $U$-spin relations in Eq.~\eqref{eq:Uspinrel}. In
 Fig.~\ref{fig:dgammaK}, we show also the $1\sigma$ contour from a $\chi^2$ fit to the current data. 
 We obtain the following results: 
\begin{equation}\label{eq:gammacur}
\gamma = (66^{+5}_{-6})^\circ, \qquad d=0.49^{+ 0.08} _{-0.09}, \qquad \theta = (147 ^{+ 7}_{-10})^\circ \ ,
\end{equation}
where $\gamma$ is in agreement with Eq.~\eqref{eq:Uspinnow}.

As we can see from the fit, the determination of $\gamma$ in Eq.~\eqref{eq:gammacur} is essentially fully 
driven by the CP asymmetries of $\Bdtopipi$ and $K$, while CP violation in $\BstoKK$ has a minor impact.
To quantify this, we perform a $\chi^2$ fit to only the CP asymmetries of \Bdtopipi and $K$. We then find
\begin{equation}
	\gamma = (66^{+5}_{-6})^\circ, \qquad d=0.50^{+0.09} _{-0.10}, \qquad \theta = (147^{+8}_{-10})^\circ \ ,
\end{equation}
which is in very good agreement with the results in Eq.~\eqref{eq:gammacur}. This now allows us to determine the CP asymmetries of \BstoKK. Employing the $U$-spin relations in Eq.~(\ref{eq:Uspinrel}) yields
\begin{align}\label{eq:precpval}
\mathcal{A}_{\rm CP}^{\rm dir}(\uBstoKK) {}& = 0.11^{+0.03}_{-0.02}|_{d}\;{}^{+0.03}_{-0.02}|_{\theta}\;{}^{+0.00}_{-0.00}|_{\gamma}= 0.11^{+0.04}_{-0.03} \nonumber \\
\mathcal{A}_{\rm CP}^{\rm mix}(\uBstoKK) {}&= -0.18^{+0.03}_{-0.04}|_{d}\;{}^{+0.02}_{-0.02}|_{\theta}\;{}^{+0.01}_{-0.01}|_{\gamma} = -0.18^{+0.04}_{-0.04} \ .
\end{align}

\begin{figure}[t]
	\centering
	\subfloat[Current situation]{\label{fig:now} \includegraphics[width=0.49\textwidth]{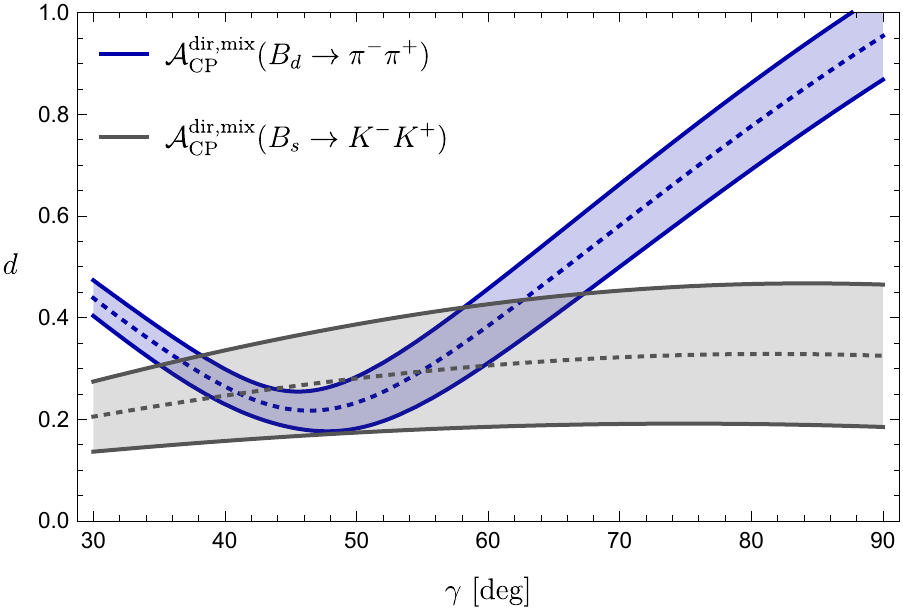}}
	\subfloat[Upgrade situation]{\label{fig:dversusgamma_up} \includegraphics[width=0.49\textwidth]{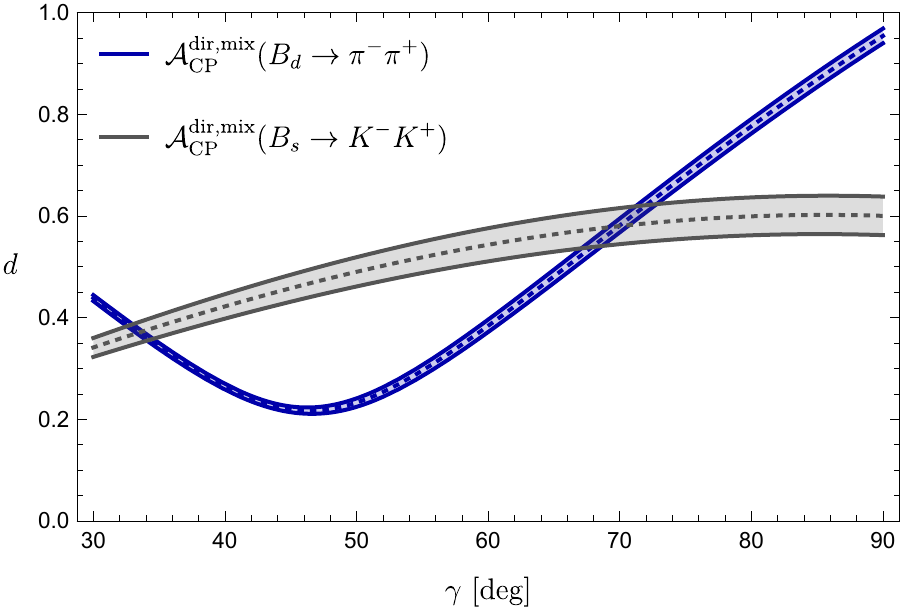}}
	\caption{Illustration of the determination of $\gamma$ from the CP asymmetries of \Bdtopipi and \BstoKK as given in Table~\ref{tab:CPs}. }
	\label{fig:dversusgamma}
\end{figure}

In view of the expected much more precise measurements of the CP asymmetries of $\BstoKK$ at the 
LHCb upgrade there is great potential in this strategy. In fact, the $K$ observable can then be avoided
and $\gamma$ can be extracted using only the CP asymmetries of \Bdtopipi and \BstoKK, thereby 
resulting in a much more favourable situation \cite{Fle99, Fle07, Fle10}. In Fig.~\ref{fig:dversusgamma}, we compare the contours from the \Bdtopipi and \BstoKK CP asymmetries for (a) the current situation, and (b) the LHCb upgrade scenario with $\phi_s = -(2.1 \pm 0.5)^\circ$, as given in Table~\ref{tab:CPs}. In this scenario, we use the expected uncertainties given in \cite{LHCbup}, and we use the $U$-spin relations in Eq.~\eqref{eq:Uspinrel} combined with Eqs.~\eqref{eq:cpasym}~and~\eqref{eq:cpasymmix} to calculate the central values for the \BstoKK CP asymmetries, because of the large current uncertainties. Assuming the $U$-spin relation $d=d'$, the upgrade scenario leads to $\gamma = (69.9_{-2.1}^{+2.4})^\circ$ \footnote{A somewhat better precision is reached if the value of $\gamma$ is lower.}. However, $U$-spin-breaking corrections limit the precision of $\gamma$. 
In order to illustrate these effects, we parametrize them as
\begin{equation}\label{eq:uspinbreaking}
\xi \equiv \frac{d'}{d},  \qquad \Delta \equiv \theta' - \theta,
\end{equation}
and consider $U$-spin breaking effects of $20 \%$, i.e. $\xi=1.0\pm 0.2$ and $\Delta=(0 \pm 20)^\circ$. This leads to $\gamma = (70^{+8}_{-6})^\circ$, which is comparable to the current situation described above.

The impact of $U$-spin-breaking contributions was also studied in Ref.~\cite{Ciuchini:2012gd}, where 
the $U$-spin method was combined with the $B \to \pi\pi$ isospin analysis \cite{Gro90} 
to reduce $U$-spin breaking effects. In Ref.~\cite{Aaij:2014xba}, it was found that the corresponding results agree with Ref.~\cite{Fle99} for corrections of up to $50\%$, while the  $B \to \pi\pi$ system stabilizes the situation for even larger
corrections. We shall discuss $U$-spin-breaking effects in more detail below, showing that such 
anomalously large effects are not supported by the experimental data.

%
\boldmath
\subsection{The $\Bs$--$\Bsb$ Mixing Phase $\phi_s$ }\label{ssec:phis}
\unboldmath
The phase $\phi_s$ can be determined from $B_s^0\rightarrow J/\psi \phi$ and decays with similar
dynamics, which are dominated by tree topologies \cite{DDLR,DDF}. The theoretical precision is limited by penguin contributions (see Ref.~\cite{DeBruyn:2014oga} and references therein). The current average from the Particle Data Group (PDG) \cite{PDG} reads
\begin{equation} \label{eq:phisPDG}
\phi_s = -(0.68 \pm 2.2)^\circ,
\end{equation}
which is in agreement with the LHCb result in Eq.~\eqref{eq:Uspinnow}. In the future, we may
extract $\phi_s$ from CP-violating effects in $B_s^0\rightarrow J/\psi \phi$ and penguin control channels
with a precision as high as $\mathcal{O}(0.5^\circ)$ \cite{DeBruyn:2014oga}.

The $B^0_s$--$\bar B^0_s$ mixing phase can also be extracted from \BstoKK decays. The corresponding
CP asymmetries allow us to determine the ``effective mixing phase" 
\begin{equation}
\phi_s^{\rm eff} \equiv \phi_s + \Delta \phi_{KK}
\end{equation}
through
\begin{equation}
\sin{\phi_s^{\textrm{eff}}}= \frac{\mathcal{A}_{\rm CP}^{\rm mix}(\uBstoKK)}{\sqrt{1-\mathcal{A}_{\rm CP}^{\rm dir}(\uBstoKK)^2}} \ ,
\end{equation}
where the hadronic phase shift $\Delta\phi_{KK}$ takes the following form 
\cite{Fleischer:2011cw,DeBruyn:2014oga,Bel15}:
\begin{equation}\label{eq:tandelphikk}
\tan{\Delta\phi_{KK}} = 2\epsilon \sin{\gamma} \left[
\frac{d' \cos{\theta'}+\epsilon \cos\gamma}{d'^2+2\epsilon d' \cos\theta' \cos\gamma+\epsilon^2 \cos{2\gamma}}\right] \ .
\end{equation} 
Let us now use $\gamma= (70\pm1)^\circ$ and $\phi_d= (43.2\pm0.6)^\circ$ \cite{DeBruyn:2014oga} as an input. Using also Table~\ref{tab:CPs}, we then
find for the LHCb upgrade scenario
\begin{equation}\label{eq:phiseff}
\phi_s^{\rm eff} = -(11.0\pm0.5)^\circ \ ,
\end{equation}
which would match the expected precision for $\phi_s$ from $B^0_s\to J/\psi \phi$ and related decays. 
However, in order to extract $\phi_s$ from this phase, we need the hadronic phase shift $\Delta\phi_{KK}$. It 
can be calculated by applying the $U$-spin symmetry to $d$ and $\theta$ extracted from the \Bdtopipi CP asymmetries.
Assuming $U$-spin-breaking corrections of $20\%$ as before, i.e.\ $\xi=1.0\pm 0.2$ and 
$\Delta =0\pm 20^\circ$, yields 
\begin{equation} \label{eq:DeltaPhiKKoriginalUpgrade}
\Delta\phi_{KK} = -(8.9 \pm 2.6)^\circ \ ,
\end{equation}
leading to $\phi_s= -(2.1 \pm 2.6)^\circ$. Consequently, we cannot match the precision of $\phi_s$ 
from $B^0_s\to J/\psi \phi$ and related decays due to the $U$-spin-breaking corrections and cannot fully exploit the experimental precision at the LHCb upgrade. To this end, an innovative method is needed, which we describe in the next section.


\section{The New Strategy} \label{sec:newstrat}

\begin{figure}[h]
	\centering
	\includegraphics[width=0.69\textwidth]{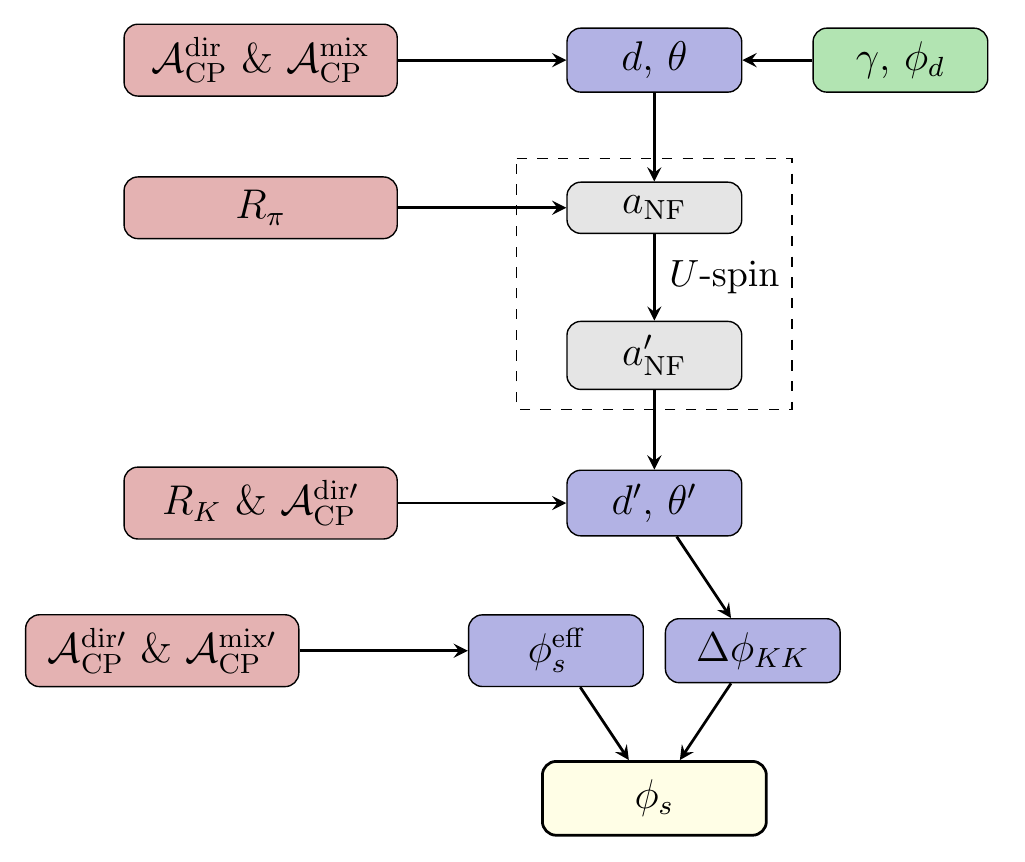}
	\caption{Flowchart of the new strategy as discussed in detail in Section~\ref{sec:newstrat}. 
	The $\mathcal{A}_\text{CP}^\text{dir}$, $\mathcal{A}_\text{CP}^\text{mix}$ and ${\mathcal{A}_\text{CP}^\text{dir}}'$, ${\mathcal{A}_\text{CP}^\text{mix}}'$ denote the direct, mixing-induced CP asymmetries of the decays \Bdtopipi and \BstoKK, respectively.}
	\label{fig:flowchart}
\end{figure}

In order to take full advantage of the huge amount of data to be collected at Belle II and the LHCb 
upgrade, we proposed a new strategy for the \Bdtopipi, \BstoKK system. It uses $\gamma$ as an 
input and makes minimal use of the $U$-spin symmetry, allowing the extraction of the 
$B^0_s$--$\bar B^0_s$ mixing phase $\phi_s$ with a future theoretical precision as high as 
$\mathcal{O}(0.5^\circ)$ \cite{SHORT}.  Moreover, valuable insights into 
$U$-spin-breaking effects can be obtained. The new key elements are the differential rates of the semileptonic decays $B^0_s\to K^-\ell^+\nu_\ell$ and $B^0_d\to \pi^-\ell^+\nu_\ell$, which we 
combine with the $B^0_s\to K^-K^+$ and $B^0_d\to\pi^-\pi^+$ decay rates; the corresponding 
information is encoded in observables $R_K$ and $R_\pi$, respectively. The flow chart of this 
strategy is shown in Fig.~\ref{fig:flowchart}.


\subsection{Semileptonic Decay Rates} \label{sec:SemileptonicDecayRates}

\begin{table}[t]
	\centering
	\begin{tabular}{l | r}
		Decay & \multicolumn{1}{l}{Branching ratio} \\
		\hline\hline
		\uBdtopipi & $(5.12\pm0.19 ) \times 10^{-6}$ \\
		\uBdtoKpi & $(1.96\pm0.05 ) \times 10^{-5}$ \\
		\uBdtoKK & $(8.03\pm1.49 ) \times 10^{-8}$ \\
		\hline
		\uBstoKK & $(2.49\pm0.17 ) \times 10^{-5}$ \\
		\uBstoKpi & $(5.5\pm0.6 ) \times 10^{-6}$ \\
		\uBstopipi & $(6.71\pm0.83 ) \times 10^{-7}$ \\
		\hline
		$B^\pm\rightarrow K^\pm K$ & $(1.32\pm0.14 ) \times 10^{-6}$ \\
		$B^\pm\rightarrow \pi^\pm K$ & $(23.79\pm0.75 ) \times 10^{-6}$ \\
	\end{tabular}
	\caption{Overview of the experimental branching ratios \cite{PDG,Amh14}. For the \BdtoKK and \Bstopipi modes recent LHCb results \cite{Aaij:2016elb} were used to calculate new averages according to the PDG method \cite{PDG}.}
	\label{tab:BtohhDecayModes}
\end{table}

For the upgrade scenario, we assume a determination of $\gamma = (70\pm 1)^\circ$ from the pure 
tree decays \cite{LHCbup}. In addition, we use $\phi_d = (43.2 \pm 0.6)^\circ$ \cite{DeBruyn:2014oga}, 
as well as the CP asymmetries given for the upgrade in Table~\ref{tab:CPs}. These inputs allow a determination of $d$ and $\theta$ from the \Bdtopipi CP asymmetries \cite{Fle99}. We find
\begin{equation}\label{eq:dval}
d = 0.58 \pm 0.02, \quad \quad \theta=(151.4 \pm 1.1)^\circ \, ,
\end{equation}
where the precision for these non-perturbative parameters is remarkable.

Additional information is encoded in the branching ratios, as we have seen in Eq.~\eqref{eq:kObservable}.
However, the observable $K$ is affected by the $U$-spin-breaking form-factor ratio, as well as
non-factorizable effects. It is more advantageous to consider ratios of non-leptonic decay rates with 
respect to differential rates of semileptonic modes, as was done for an extensive analysis of 
$B \to D\bar{D}$ decays in Ref.~\cite{Bel15}. These ratios also provide a well-known test for
the factorisation of hadronic matrix elements of non-leptonic 
decays~\cite{Bjo88, Bor90, Ros90, Neu97, Ben99, Ben01, Ben10, Fle10b}.

For our transitions at hand, we define 
\begin{equation}\label{eq:Rpi}
	R_{\pi}  \equiv  \frac{ \Gamma(\uBdtopipi)}{|d\Gamma(B^0_d\rightarrow \pi^- \ell^+ 
	\nu_\ell)/dq^2|_{q^2=m_\pi^2}}= 6\pi^2 |V_{ud}|^2 f_\pi^2 X_{\pi} r_\pi |a_{\rm NF}|^2 \ ,
\end{equation}
where
\begin{equation}\label{eq:rpi}
r_\pi \equiv 1+d^2-2d\cos\theta\cos\gamma		 \ ,
\end{equation}
$f_\pi$ denotes the charged pion decay constant, $V_{ud}$ is the corresponding CKM matrix element, and 
\begin{equation}
X_\pi \equiv \frac{ (m_{B_d}^2-m_\pi^2)^2}{m_{B_d}^2 (m_{B_d}^2-4m_\pi^2)}\left[ \frac{F_0^{B_d\pi}(m_\pi^2)}{F_1^{B_d\pi}(m_\pi^2)}\right]^2 \ .
\end{equation}
The form factors were defined in Eq.~\eqref{eq:ffDef}, and satisfy the relation
\begin{equation}\label{kin-rel}
\frac{F_0^{B_d\pi}(0)}{F_1^{B_d\pi}(0)} = 1
\end{equation}
due to kinematic constraints which are also implemented in lattice QCD calculations 
\cite{lattice-1,lattice-2}. We assume $F_0^{B_d\pi}(m_\pi^2)/F_1^{B_d\pi}(m_\pi^2) = 1$, i.e.\
a negligible deviation from this result for the small momentum transfer $q^2=m_\pi^2$. 
The non-factorizable contributions are parameterized by the following quantity:
\begin{equation} \label{eq:aNFd}
	a_{\rm NF} \equiv (1+r_P)(1+x) a^T_{\rm NF} \ ,
\end{equation}
where
 \begin{equation}\label{eq:x}
 r_P\equiv \frac{P^{(ut)}}{T}  \ , \qquad x\equiv |x|e^{i\sigma}\equiv \frac{E+PA^{(ut)}}{T+P^{(ut)}} \ .
 \end{equation}
 The non-factorizable contributions to the colour-allowed tree topology $T$ are characterized by 
 the deviation of $a_{\rm NF}^T$ from one. This parameter can be described within the QCD factorization framework \cite{Ben99, Ben01}. The current state-of-the-art calculation \cite{Ben10}, including 
 two-loop (NNLO) QCD effects,  yields 
\begin{equation}\label{QCDF-calc}
a_{\rm NF}^T=1.000^{+0.029}_{-0.069}+(0.011^{+0.023}_{-0.050})i \ .
\end{equation}
The colour-allowed tree amplitude is theoretically very favourable with respect to the factorization of
hadronic matrix elements, which is also reflected by the sophisticated analysis devoted to the 
parameter in (\ref{QCDF-calc}). On the other hand, penguin topologies are much more challenging
and are affected by non-factorizable effects and long-distance contributions, such as those attributed to 
``charming penguins"  \cite{charm-pen}. 

The branching ratio of the \Bdtopipi channel is given in Table~\ref{tab:BtohhDecayModes}. The differential decay rate at low $q^2$ unfortunately suffers from sizable experimental uncertainties. We may estimate the required partial branching fraction of the semileptonic rate by averaging the low $q^2$ measurements of the BaBar 
and Belle collaborations
\cite{Belle13, BaBar12, PDG}. We find $d\text{BR}/dq^2 \sim (6 \pm 1)\text{GeV}^{-2}$. A more sophisticated analysis of this quantity lies outside the scope of this paper. However, we note that our estimate is in agreement with the analyses in, e.g., Refs.~\cite{Ims14}~and~\cite{Bal06}, where this rate is used to extract the CKM matrix element $|V_{ub}|$. Finally, we obtain 
\begin{equation}
R_\pi = (0.85 \pm 0.15) \rm{GeV}^2 \ , 
\end{equation}
which corresponds to a relative error of $17\%$. We advocate to extract this ratio directly from the
experimental Belle (II) and LHCb data.

Using $(d,\theta)$ from the \Bdtopipi CP asymmetries in Eq.~\eqref{eq:dval}, we may extract 
$r_\pi$ in Eq.~\eqref{eq:rpi}. Combining this parameter with $R_\pi$ and the experimental value for \cite{Ros15}
\begin{equation} \label{eq:decayConstantspiK}
|V_{ud}|f_{\pi} = (127.13 \pm 0.02 \pm 0.13 ) \textrm{MeV} 
\end{equation}
gives
\begin{equation}\label{eq:anf}
|a_{\rm NF}|  = 0.73\pm 0.06 \ .
\end{equation}

Concerning $B^0_s\to K^-K^+$, we introduce in analogy to $R_\pi$ the following ratio:
\begin{equation}\label{eq:rk}
R_K \equiv \frac{\Gamma(B_s \to K^-K^+)}{|d\Gamma(B_s^0 \to K^-\ell^+\nu_\ell)/dq^2|_{q^2=m_K^2}} 
 =6\pi^2 |V_{us}|^2 f_K^2 X_K r_K |a_{\rm NF}'|^2 \ ,
\end{equation}
where
\begin{equation}
r_K\equiv  1+ \left(\frac{d'}{\epsilon}\right)^2+2\frac{d'}{\epsilon} \cos\theta'\cos\gamma \ ,
\end{equation}
\begin{equation}
X_K \equiv \frac{ (m_{B_s}^2-m_K^2)^2}{m_{B_s}^2 (m_{B_s}^2-4m_K^2)}\left[ \frac{{F^{B_sK}_0}(m_K^2)}{{F^{B_sK}_1}(m_K^2)}\right]^2 \ ,
\end{equation}
 $f_K$ denotes the charged kaon decay constant, and $V_{us}$ is the corresponding 
 CKM matrix element.


\boldmath
\subsection{Determination of $\Delta \phi_{KK}$} \label{sec:DeterminationOfDeltaPhiKK}
\unboldmath
In order to determine the hadronic parameters $d'$ and $\theta'$ of the $B^0_s\to K^-K^+$ decay, we use
the following expression:
\begin{equation}\label{eq:rkrel}
	r_K = r_\pi \frac{R_K}{R_\pi}\left[\frac{|V_{ud}|f_\pi}{|V_{us}|f_K}\right]^2\frac{X_{\pi}}{X_K}(\xi^a_{\rm NF})^2\ .
\end{equation}
As we have seen above, $r_\pi$ can be determined from the CP asymmetries in \Bdtopipi, and the 
only unknown quantity in the game is the following parameter  \cite{SHORT}:
\begin{equation}\label{eq:xianf}
\xi^a_{\rm NF} \equiv \left|\frac{a_{\rm NF}}{a_{\rm NF}'}\right|=
\left|\frac{1+r_P}{1+r^{'}_P}\right|\left|\frac{1+x}{1+x'}\right|\left|\frac{a_{\rm NF}^T}{a_{\rm NF}^{T'}}\right| \ .
\end{equation}
It can be determined with the help of the $U$-spin symmetry. We will show below that $\xi^a_{\rm NF}$ has actually
a structure which is very favourable with respect to $U$-spin-breaking corrections. We may then
determine $r_K$, which we may combine with the direct CP asymmetry of the $B^0_s\to K^-K^+$
decay to extract its hadronic parameters $d'$ and $\theta'$:
\begin{equation}\label{eq:dprimeana}
d' = \epsilon\left[r_K+\cos2\gamma \pm \sqrt{(r_K+\cos2\gamma)^2-(r_K-1)^2-(r_K\mathcal{A}_\text{CP}^{\textrm{dir} \prime}/\tan\gamma)^2}\right]^{1/2} \ ,
\end{equation}
\begin{equation}
	\cos\theta' = \frac{\epsilon^2 (r_K-1) -d^{\prime 2}}{2\epsilon d'\cos\gamma}, \qquad \sin\theta' = \frac{\epsilon r_K \mathcal{A}_\text{CP}^{\textrm{dir}\prime}}{2d'\sin \gamma} \,.
\end{equation}
Here we have defined $\mathcal{A}_{\rm CP}^{\rm dir \prime} \equiv \mathcal{A}_{\rm CP}^{\rm dir}(\uBstoKK)$. Finally, we may calculate the hadronic phase shift $\Delta\phi_{KK}$ using 
Eq.~\eqref{eq:tandelphikk}, which yields
\begin{equation}\label{eq:tan2}
\tan\Delta\phi_{KK} = \frac{2\sin\gamma}{\epsilon}\left[\frac{d' \cos\theta' + \epsilon \cos\gamma}{r_K -1 
+\cos 2\gamma} \right ]\ .
\end{equation}
The $B^0_s$--$\bar B^0_s$ mixing phase $\phi_s = \phi_s^{\rm eff} - \Delta\phi_{KK}$ 
can then be extracted from the
measured effective mixing phase $\phi_s^{\rm eff}$. 

Unfortunately, the semileptonic decay $B_s^0\rightarrow K^- \ell^+ \nu_\ell$ has not yet been measured. 
We advocate analyses of this channel at Belle (II) and LHCb, preferably by a direct measurement 
of the double ratio $R_\pi/R_K$. Here only the double ratio of the form factors enters through $X_\pi/X_K$, 
which strongly reduces the sensitivity to small deviations from (\ref{kin-rel}) for the momentum transfers
$q^2=m_\pi^2$ and $m_K^2$, thereby yielding a double ratio of form factors equal to one with 
excellent precision. In addition, the ratio $|V_{us}| f_K/|V_{ud}| f_\pi = 0.27599\pm0.00037$ can be 
determined with tiny uncertainties from experimental data \cite{Ros15}. It is interesting to note that $R_K$ does not depend on the ratio of the $B_{s,d}^0$ fragmentation functions $f_s/f_d$, 
which is the major limiting factor for measurements of \Bs branching ratios \cite{FST-fs}.

\begin{figure}[t]\centering
	\includegraphics[width = 0.55\linewidth]{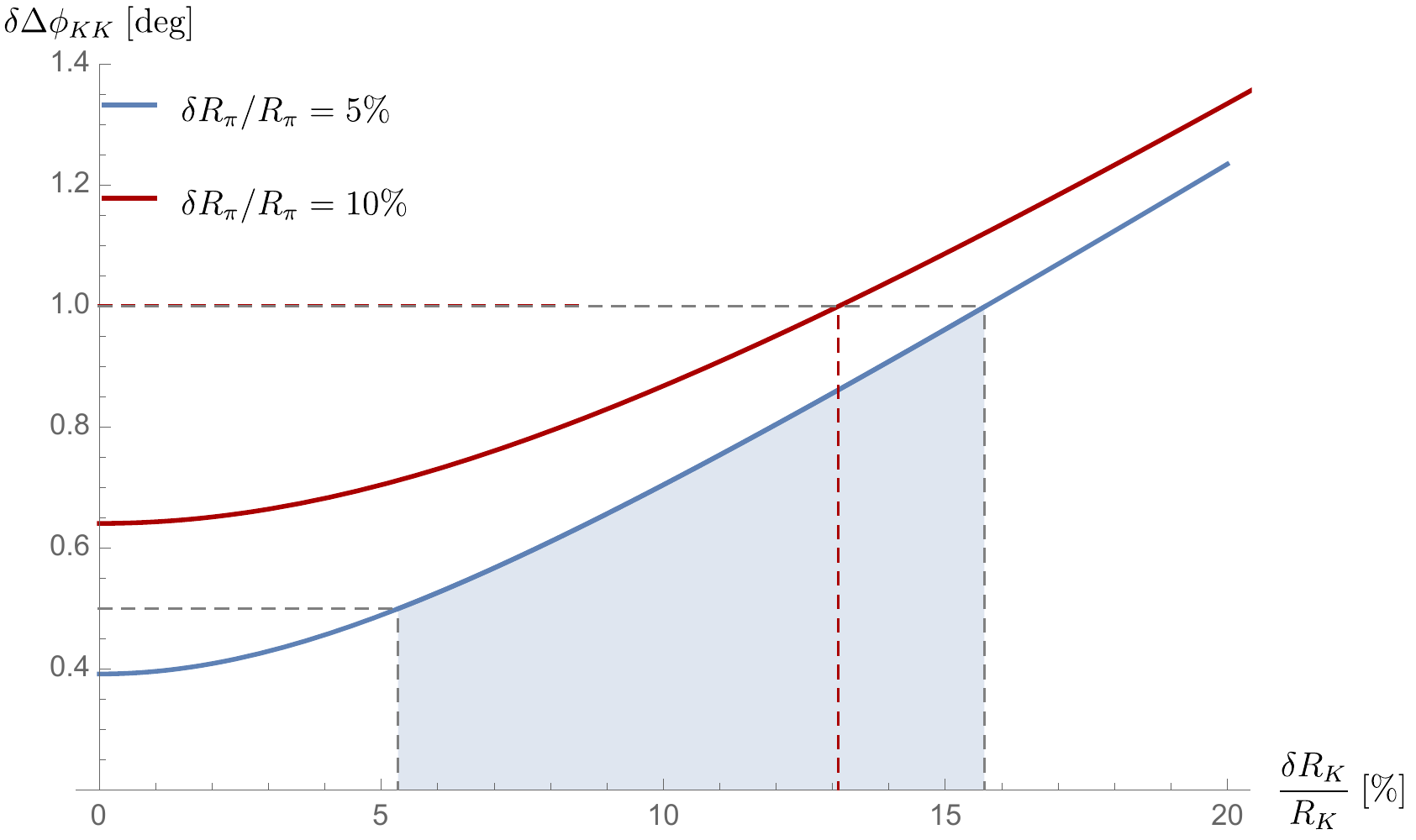}
	\caption{The experimental error for $\Delta\phi_{KK}$ as a function of the relative precision of $R_K$
	 for a relative precision of $R_\pi$ of $5\%$ and $10\%$, assuming a perfect theoretical situation. }
	\label{fig:errorphikk_Rk}
\end{figure}

\begin{figure}
	\centering
		\includegraphics[width=0.5
		\linewidth]{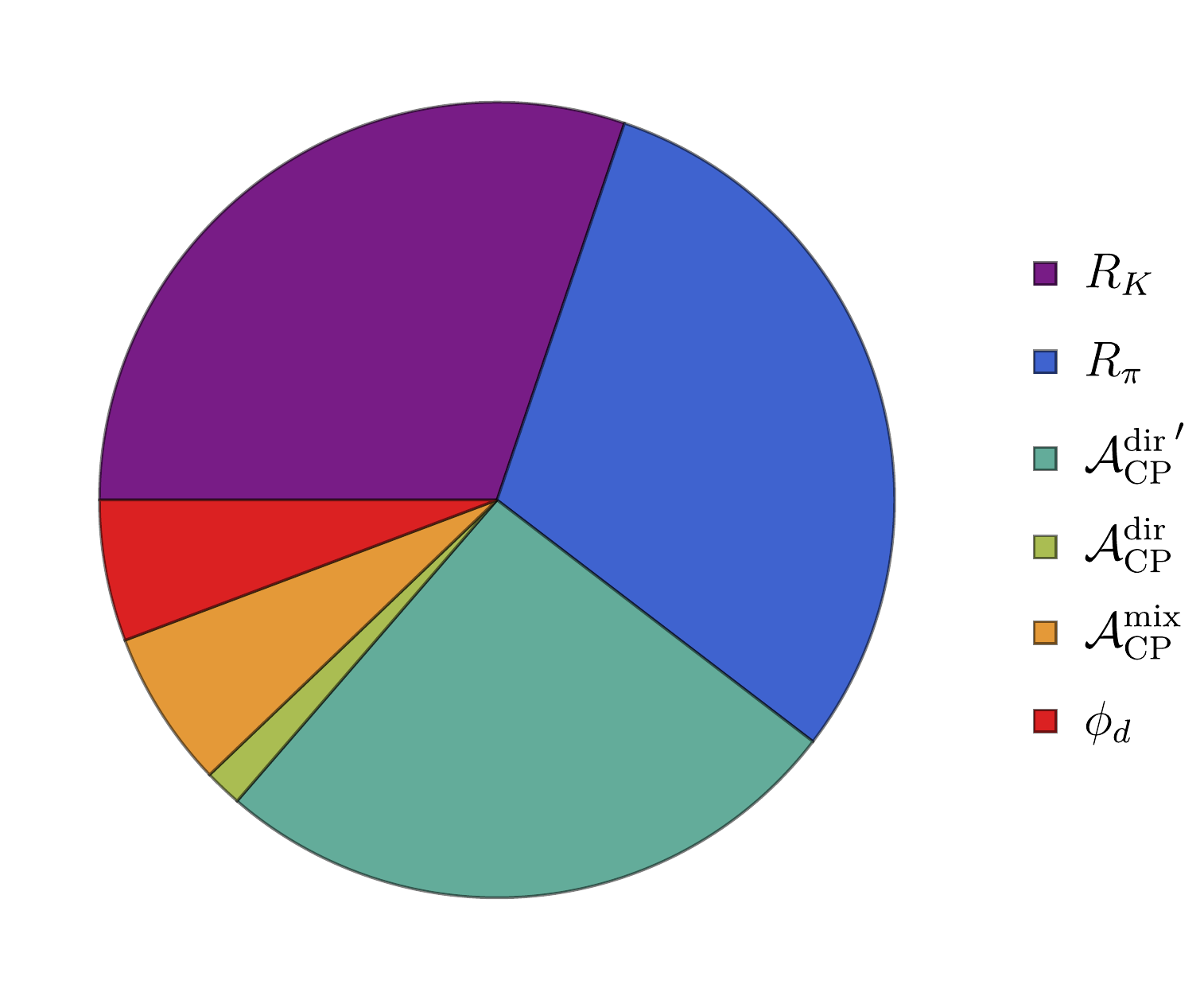}
		\caption{Experimental error budget for $\Delta\phi_{KK}$. Here we have assumed a relative precision of $5 \%$ for $R_K$ and $R_\pi$ and a perfect theoretical situation.}
	\label{fig:errorbudget}
\end{figure}

We illustrate the future experimental precision for $\Delta\phi_{KK}$ that can eventually be achieved
with our new strategy for a perfect theoretical situation in Fig.~\ref{fig:errorphikk_Rk}.
There we show the sensitivity as a function of the relative precision of $R_K$, while assuming 
measurements of $R_\pi$ in the upgrade era with relative precisions of $5\%$ and $10\%$. Getting to the precision of $0.5^\circ$ for $\Delta\phi_{KK}$ requires a determination of $R_K$ and $R_\pi$ with a relative error of $5\%$. In Fig.~\ref{fig:errorbudget}, we show the experimental error budget of $\Delta\phi_{KK}$, considering 
a relative error of $5\%$ for $R_K$ and $R_\pi$.

Interestingly, for values of $\gamma$ around $70^\circ$, the dependence of $\Delta\phi_{KK}$ on $\gamma$ is essentially negligible. This can be understood as $\tan \Delta\phi_{KK}$ in Eq.~\eqref{eq:tan2} is then
given by
\begin{equation}
\tan \Delta\phi_{KK} \sim \frac{2 \sin\gamma}{\sqrt{r_K}}, 
\end{equation}
while 
\begin{equation}
r_K \propto r_\pi  \propto \sin^2 \gamma.
\end{equation}
Consequently, if we used $\phi_s$ as an input for our strategy and were aiming to determine $\gamma$, 
we would have a small sensitivity for this angle. It is hence much more advantageous to use $\gamma$
as input and determine $\phi_s$.

The theoretical precision of the new strategy is limited by the $U$-spin-breaking corrections affecting
$\xi^a_{\rm NF}$ in Eq.~\eqref{eq:xianf}. The structure of $\xi_\text{NF}^a$, which depends on 
\begin{equation} \label{eq:defrat}
\Xi_P \equiv \left|\frac{1+r_P}{1+r^{'}_P}\right|\ , \qquad \Xi_x \equiv \left|\frac{1+x}{1+x'}\right| \ , 
\end{equation}
and the ratio of the non-factorizable, colour-allowed tree-level contributions, is very favourable in this respect. As 
both $r_P$ and $x$ are small parameters, the ratios entering Eq.~\eqref{eq:xianf} are very robust 
concerning $U$-spin-breaking corrections. We will come back to this feature in 
Subections~\ref{sec:impl} and \ref{sec:impl2} after we have explored the implications of 
the current data for $r_P$ and $x$.

Corrections to the $U$-spin relation $a_{\rm NF}^T = a_{\rm NF}^{T'}$ for the non-factorizable 
contributions to the colour-allowed tree amplitudes can be quantified within the framework of 
QCD factorization \cite{Ben10}. So far only the $\Bdtopipi$ decay has been analyzed, with the
result in Eq.~\eqref{QCDF-calc}. Following Ref.~\cite{SHORT}, we write
\begin{equation}
a_{\rm NF}^{T(')} = 1+\Delta_{\rm NF}^{T(')} 
\end{equation}
with $\Delta_{\rm NF}^{T'} = \Delta_{\rm NF}^T(1-\xi_{\rm NF}^T)$, such that we obtain
\begin{equation}\label{eq:anfrat}
\frac{a_{\rm NF}^T}{a_{\rm NF}^{T'}} = 1+ \Delta_{\rm NF}^T \xi_{\rm NF}^T + \mathcal{O}((\Delta_{\rm NF}^T)^2) \ .
\end{equation}
Using Eq.~\eqref{QCDF-calc}, we estimate $\Delta_{\rm NF}^T\sim 0.05$. Allowing for $U$-spin-breaking corrections of $20\%$ for the non-factorizable contributions gives a tiny correction of $\mathcal{O}(1\%)$ 
to the ratio in Eq.~\eqref{eq:anfrat}. Even larger $U$-spin-breaking corrections would not have a significant
impact on this picture. It would be interesting to extend the QCD factorization analysis of the colour-allowed
tree amplitude to the $\BstoKK$ decay.

The advantage of our new strategy concerning $U$-spin-breaking effects in comparison with 
the original method can be clearly seen by rewriting the parameter $\xi$ in Eq.~\eqref{eq:uspinbreaking} as
\begin{equation}
\xi=\xi_{\rm NF}^a\left|\frac{T_{\rm fact}}{T'_{\rm fact}}\right|\biggl|\frac{P^{(ct)'}+PA^{(ct)'}}{P^{(ct)}+PA^{(ct)}}\biggr|.
\end{equation}
Here the leading $U$-spin-breaking corrections are associated with penguin topologies, which 
are challenging, with issues such as ``charming'' penguins \cite{charm-pen}. Therefore, the uncertainty 
of $\xi_\text{NF}^a$ is significantly smaller than that of $\xi$.

\subsection{Picture from Current Data} \label{sec:proof}

Since the differential semileptonic $B_s^0\rightarrow K^0\ell^+\nu_\ell$ decay rate has not yet been measured, we cannot apply our new strategy to current data. However, as a demonstration, we can consider the \BdtoKpi decay. This channel only receives contributions from tree and penguin topologies. Neglecting exchange and penguin-annihilation contributions to $\BstoKK$, the $\BdtoKpi$ decay topologies only differ at 
the spectator-quark level. The transition amplitude can be written as follows \cite{Fle99, Fle07}:
\begin{equation} \label{eq:TPd}
A(B_d^0 \to \pi^-K^+) = \sqrt{\epsilon}\euler^{\imaginary\gamma}\tilde{\mathcal{C}}^\prime\left[1+\frac{1}{\epsilon}\tilde{d}^\prime\euler^{\imaginary\tilde{\theta}^\prime}\euler^{-\imaginary\gamma}\right] 
\end{equation}
with
\begin{equation} \label{eq:cdtilpdef}
\tilde{\mathcal{C}}^\prime \equiv \lambda^3AR_b\left[\tilde{T}^\prime+\tilde{P}^{(ut)\prime}\right] \ , \qquad \tilde{d}^\prime\euler^{\imaginary\tilde{\theta}^\prime} \equiv \frac{1}{R_b}\left[\frac{\tilde{P}^{(ct)\prime}}{\tilde{T}^\prime+\tilde{P}^{(ut)\prime}}\right] \ .
\end{equation}

Using Eq.~\eqref{eq:bstoKK} and the $SU(3)$ relation
\begin{equation}\label{eq:su3rel}
\frac{P^{(ct)^\prime}}{T^\prime+P^{(ut)\prime}} = \frac{\tilde{P}^{(ct)^\prime}}{\tilde{T}^\prime+\tilde{P}^{(ut)\prime}}
\end{equation}
gives
\begin{equation} \label{eq:penguinSU3relation}
 \tilde{d}^\prime\euler^{\imaginary\tilde{\theta}^\prime} =\zeta' d^\prime\euler^{\imaginary\theta^\prime} \ ,
\end{equation}
where 
\begin{equation}
\zeta' \equiv \frac{1+x'}{1+r_{PA}'} \quad\mbox{and}\quad 
r_{PA}  \equiv \frac{PA^{(ct)}}{P^{(ct)}} 
\end{equation} 
parametrize the exchange and penguin-annihilation topologies. Neglecting these topologies gives $\zeta'=1$, leading to a direct relation between the hadronic parameters of $\BdtoKpi$ and $\BstoKK$. We discuss
the parameter $\zeta'$ further in Section~\ref{sec:InformationFromBtohhDecays}. Non-factorizable contributions to the $SU(3)$ relation in Eq.~\eqref{eq:su3rel} are expected to be small as the tree and penguin topologies differ only at the spectator-quark level.

In analogy to $R_K$, we introduce 
\begin{equation}\label{eq:rktilde}
\tilde{R}_K  \equiv \frac{\Gamma(B_d \to \pi^-K^+)}{|d\Gamma(B_d^0 \to \pi^-\ell^+\nu_\ell)/dq^2|_{q^2=m_K^2}} =6\pi^2 |V_{us}|^2 f_K^2 \tilde{X}_K \tilde{r}_K |\tilde{a}_\text{NF}'|^2 \ ,
\end{equation}
where
\begin{equation}
\tilde{r}_K \equiv 1+2\frac{\tilde{d}'}{\epsilon}\cos\tilde{\theta}'\cos\gamma+\left(\frac{\tilde{d}'}{\epsilon}\right)^2
\end{equation}
and
\begin{equation}
\tilde{X}_K \equiv \frac{(m_{B_d}^2-m_\pi^2)^2}{[m_{B_d}^2-(m_\pi+m_K)^2][m_{B_d}^2-(m_\pi-m_K)^2]}\left[\frac{F_0^{B_d\pi}(m_K^2)}{F_1^{B_d\pi}(m_K^2)}\right]^2.
\end{equation}
The non-factorizable contributions are parametrized by
\begin{equation}
\tilde{a}_\text{NF}' \equiv (1+\tilde{r}_P')\tilde{a}_\text{NF}^{T \prime} \ .
\end{equation}

In analogy to Eq.~\eqref{eq:rkrel}, we can now write
\begin{equation}
\tilde{r}_K=\frac{\tilde{R}_K}{R_\pi} \left(\frac{|V_{ud}|f_\pi}{|V_{us}|f_K}\right)^2\frac{X_{\pi}}{\tilde{X}_K}(\tilde{\xi}_\text{NF}^a)^2(1-2d\cos\theta\cos\gamma+d^2) 
\end{equation}
with
\begin{equation}
\tilde{\xi}_\text{NF}^a \equiv \left|\frac{1+r_P}{1+\tilde{r}'_P}\right| \left|1+x\right| \left|\frac{a_{\rm NF}^T}{\tilde{a}_{\rm NF}^{T'}}\right|,
\end{equation}
where now only a single $|1+x|$ term occurs, which vanishes if the $E$ and $PA$ topologies are neglected.
Interestingly, the semileptonic decay rates cancel in the ratio $\tilde{R}_K/R_\pi$
up to small corrections due to the difference in the corresponding kinematical points.

The direct CP asymmetry of \BdtoKpi has been measured as follows \cite{Amh14}:
\begin{equation}\label{eq:BdtoKpi}
\mathcal{A}_{\textrm {CP}}^{\rm dir}(\uBdtoKpi) \equiv \frac{|A(\B_d^0 \to \pi^-K^+)|^2 - |A(\bar{B}_d^0 \to \pi^+K^-)|^2}{|A(\B_d^0 \to \pi^-K^+)|^2 + |A(\bar{B}_d^0 \to \pi^+K^-)|^2} = 0.082 \pm 0.006\:.
\end{equation}
From the current data for the $\Bdtopipi$ CP asymmetries, using also $\gamma=(70\pm 7)^\circ$ and $\phi_d=(43.2\pm1.8)^\circ$ as input, we find
\begin{equation}\label{eq:dvalnow}
d = 0.58 \pm 0.16 \ , \qquad \theta = (151.4 \pm 7.6)^\circ \ .  
\end{equation}
Neglecting the $E$ and $PA$ topologies and applying the $U$-spin symmetry for $\tilde{\xi}^a_{\rm NF}$, 
we obtain $\tilde{r}_K$. Combined with the direct CP asymmetry for $\BdtoKpi$ this gives
\begin{equation}\label{eq:primed}
\tilde{d}'  = 0.50 \pm 0.03 \ , \qquad \tilde{\theta}' = (157.2 \pm 2.2)^\circ \ .
\end{equation}
Moreover, we can also determine the results
\begin{equation}
\tilde{\xi} \equiv \tilde{d}'/d = 0.87 \pm 0.20 , \qquad \tilde{\Delta} \equiv \tilde{\theta}'-\theta = (5.8 \pm 8.3)^\circ \ ,
\end{equation}
which are fully consistent with the $U$-spin symmetry. In particular, the 
anomalously large $U$-spin-breaking corrections of (50--100)\% considered in Ref.~\cite{Aaij:2014xba} are strongly disfavoured.

Finally, we determine the hadronic phase shift  as follows:
\begin{equation}
\Delta\phi_{KK} = -(10.8\pm 0.6)^\circ .
\end{equation} 
Already this precision for the current data is impressive and shows the exciting prospects for the method. 
Using the current data for the \BstoKK CP asymmetries, which yield $\phi_s^\text{eff} = -(17.6 \pm 7.9)^\circ$, we obtain
\begin{equation}
\phi_s = -(6.8 \pm 7.9)^\circ,
\end{equation}
where the uncertainty is dominated by the experimental data. This value is in excellent 
agreement with the result in Eq.~\eqref{eq:Uspinnow}, although obtained with a completely 
different method. As we have neglected the exchange and penguin-annihilation contributions, 
this agreement indicates that these topologies are actually playing a minor role. 


\begin{table}[t]
	\centering
	\begin{tabular}{l | r }
		Observable & Measurement \\
		\hline\hline
		$\mathcal{A}_{\textrm{CP}}^{\textrm{dir}}(\uBdtopipi)$ & $ -0.24 \pm 0.07$ \\
		$\mathcal{A}_{\textrm{CP}}^{\textrm{mix}}(\uBdtopipi)$ & $0.68 \pm 0.06 $ \\
		$\mathcal{A}_{\textrm{CP}}^{\textrm{dir}}(\uBstoKK)$ & $0.24 \pm 0.06$ \\
		$\mathcal{A}_{\textrm{CP}}^{\textrm{mix}}(\uBstoKK)$ & $-0.22 \pm 0.06$ \\
		$\mathcal{A}_{\Delta\Gamma}(\uBstoKK)$ & $-0.75 \pm 0.13$
	\end{tabular}
	\caption{Overview of the preliminary new LHCb measurements \cite{LHCbpre}.  }
	\label{tab:preval}
\end{table}

\boldmath
\subsection{News from LHCb}
\unboldmath
The LHCb collaboration has recently reported new preliminary measurements of the CP-violating observables
of the \BstoKK and \Bdtopipi  decays \cite{LHCbpre}. We have summarized these results 
in Table~\ref{tab:preval}. Comparing to the experimental data for the CP asymmetries in 
Table~\ref{tab:CPs}, which includes also our scenario for the LHCb upgrade, we find good agreement 
for the \Bdtopipi channel. 

However, while the mixing-induced CP asymmetry of \BstoKK is also in good agreement with the 
numbers in this table, the new measurement of the direct CP asymmetry is surprising. In particular, 
there is a large difference between the direct CP asymmetry of \BstoKK in Table~\ref{tab:preval} and 
the direct CP asymmetry of \BdtoKpi in  Eq.~\eqref{eq:BdtoKpi}. As we discussed in the previous section, 
these decays differ only through their spectator quarks. Since the underlying quark-level transitions are the
same, NP effects cannot be responsible for this difference. As exchange and penguin-annihilation topologies
contribute to the \BstoKK decay but have no counterparts in the \BdtoKpi mode, they could -- in principle -- 
be the origin of this surprising measurement. However, as we will show in detail in 
Sections~\ref{sec:penguindy} and \ref{sec:InformationFromBtohhDecays}, such a picture is not supported
by experimental data. Moreover, a similar relation arises between the direct CP asymmetries of the 
\Bdtopipi and \BstoKpi decays, which is perfectly satisfied by the data, thereby also not indicating any
anomalous behaviour.

\begin{figure}[t]
	\centering
	\includegraphics[scale=0.65]{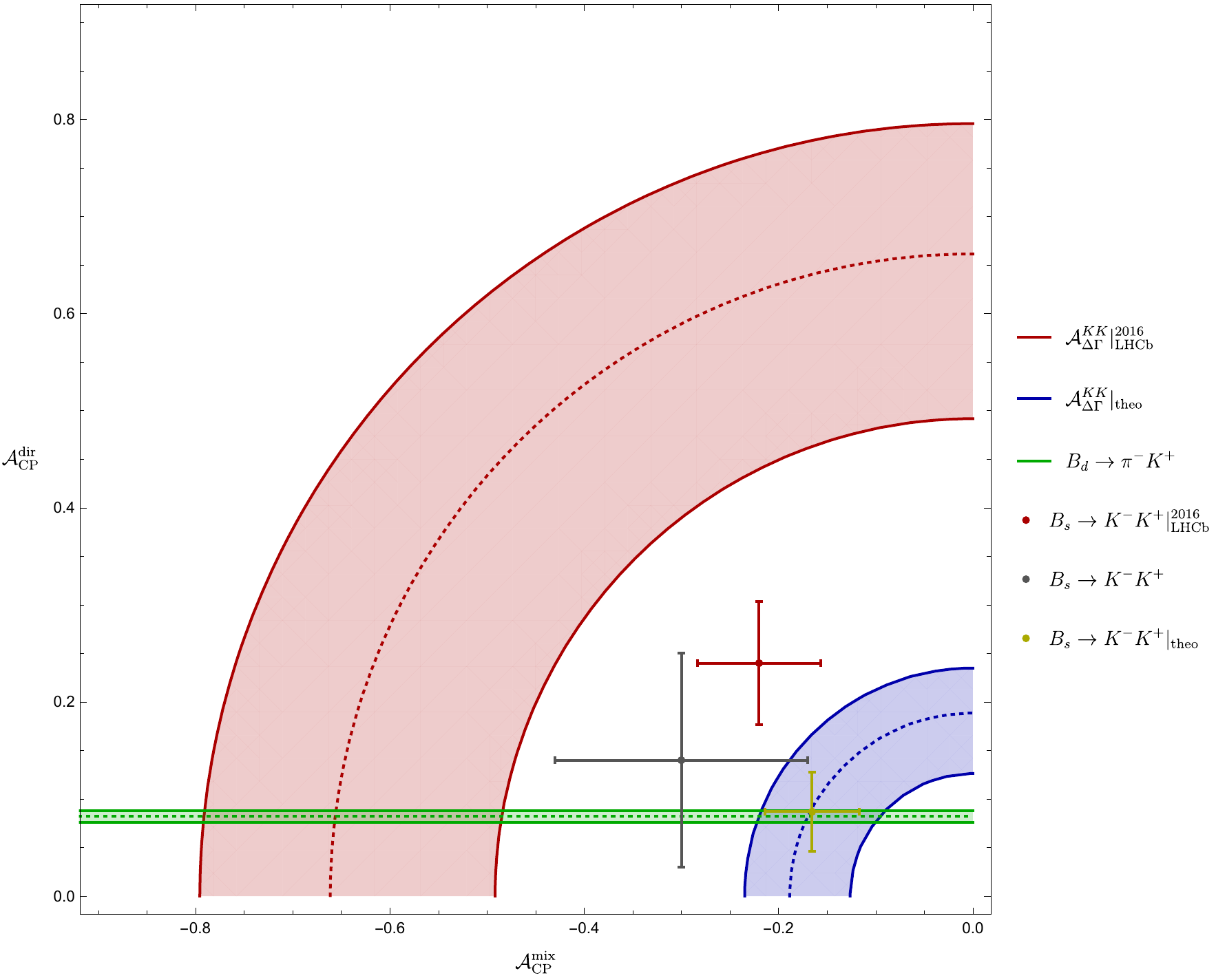}
	\caption{Comparison of the new LCHb data with the previous results and theoretically predicted values for the \BstoKK CP asymmetries, as discussed in the text. Moreover, we show contours corresponding to the new and predicted values of $\mathcal{A}_{\Delta\Gamma}(\BstoKK)$ as well as the current direct CP asymmetry of \BdtoKpi.}
	\label{fig:prelim}
\end{figure}

In combination with the CP asymmetries, the LHCb collaboration has also reported a new preliminary measurement of the observable $\mathcal{A}_{\Delta\Gamma}^{KK}\equiv \mathcal{A}_{\Delta\Gamma}(\uBstoKK)$ \cite{LHCbpre}, which we give in Table~\ref{tab:preval}. An important check for the internal
consistency of the data is provided by the sum rule in Eq.~\eqref{eq:sumrule}, which is a general feature
of the different observables and cannot be violated through NP effects. For the preliminary LHCb data, 
we find the following result:
\begin{equation}\label{eq:sumrule2}
\Delta_\text{SR}\equiv1-\left(\mathcal{A}_{\rm CP}^{\rm dir}{}'\right)^2-
\left(\mathcal{A}_{\rm CP}^{\rm mix}{}'\right)^2 -
\left(\mathcal{A}_{\Delta\Gamma}^{KK}\right)^2 =0.33 \pm 0.20 \,,
\end{equation}
which differs from zero at the $1.7\,\sigma$ level. We have illustrated this situation in Fig.~\ref{fig:prelim},
where we indicate the CP asymmetries of \BstoKK from Table~\ref{tab:CPs} and the preliminary new
results listed in Table~\ref{tab:preval} through grey and red data points, respectively. Moreover, we add
a red circular band corresponding to Eq.~\eqref{eq:sumrule2}, which clearly shows the inconsistency of the data. 
In Fig.~\ref{fig:prelim}, we have furthermore considered predictions of the \BstoKK CP asymmetries and $\mathcal{A}_{\Delta\Gamma}^{KK}$, calculated from Eqs.~\eqref{eq:cpasym},~\eqref{eq:cpasymmix}~and~\eqref{eq:cpasymDG} by applying the $U$-spin symmetry to $d$ and $\theta$ from Eq.~\eqref{eq:dvalnow}, which lead to the yellow data point and the blue circular band, respectively. They are in perfect agreement with the direct CP asymmetry 
of \BdtoKpi represented by the green horizontal band. We expect that the
central value of the new LHCb result for the direct CP asymmetry of the \BstoKK decay will move correspondingly in the future.


\section{Insights into Penguin Dynamics}\label{sec:penguindy}
The size of the parameters $r_P$ and $x$ introduced in Eq.~\eqref{eq:x} has to be quantified in order
to analyze the theoretical precision of our strategy in more detail. In this section, we discuss the penguin topologies contributing to $r_P$. Specifically, we write 
\begin{equation} \label{eq:onePlusrPzetarho}
1+r_P = \frac{1}{1-\zeta d e^{i\theta} \rho_P} \ ,
\end{equation}
where the penguin ratio $\rho_P$ is defined as
\begin{equation}\label{eq:rhop}
\rho_P\equiv |\rho_P| e^{i \theta_P} = R_b \frac{P^{(ut)}}{P^{(ct)}} \ ,
\end{equation}
and
\begin{equation}\label{eq:Repa}
\zeta\equiv |\zeta| e^{i \omega}= \frac{1+x}{1+r_{PA}} \ ,\quad \quad 
r_{PA}  \equiv \frac{PA^{(ct)}}{P^{(ct)}} \ .
\end{equation}
Completely analogous expressions hold for $1+r_P'$.

The parameter $\zeta'$ was already introduced in Eq.~\eqref{eq:penguinSU3relation}, and $\zeta^{(\prime)}$ is expected to be close to one as the exchange and penguin-annihilation topologies are expected to be small. 
We shall return to this quantity in Section~\ref{sec:InformationFromBtohhDecays}. Let 
us first focus on the parameter $\rho_P$, which is governed by the interplay of the QCD penguin topologies
with internal up, charm and top quarks \cite{BF-95}. This quantity can be studied with the pure penguin decays 
$\BdtoKantiK$,  $\BdtoKantiK$ and $B^+ \rightarrow K^+ \Kbar^0$, $B^+\rightarrow \pi^+ K^0$. The 
various decay topologies and their specific use in our new strategy are summarized in 
Table~\ref{tab:decaymodes}. In Subsection~\ref{sec:tilde}, we shall also discuss the 
$\BdtoKpi$, $\BstoKpi$ system \cite{GroRo}, 
which has only tree and penguin contributions and can hence also 
be used to study $U$-spin-breaking effects in the corresponding decay topologies. 


\boldmath
\subsection{$\BdtoKantiK$ and $\BstoKantiK$} \label{sec:rhokk}
\unboldmath
The decays \BdtoKantiK and \BstoKantiK are related by the $U$-spin symmetry and receive 
only contributions from penguin and penguin annihilation topologies \cite{RF-ang,DMV}. Consequently, 
they offer an excellent laboratory to study penguin contributions.

\begin{figure}[tp]
	\centering
	\subfloat[QCD penguin ($P_{KK}$)]{\label{fig:penguinDiagramK0K0} \includegraphics[width=0.49\textwidth]{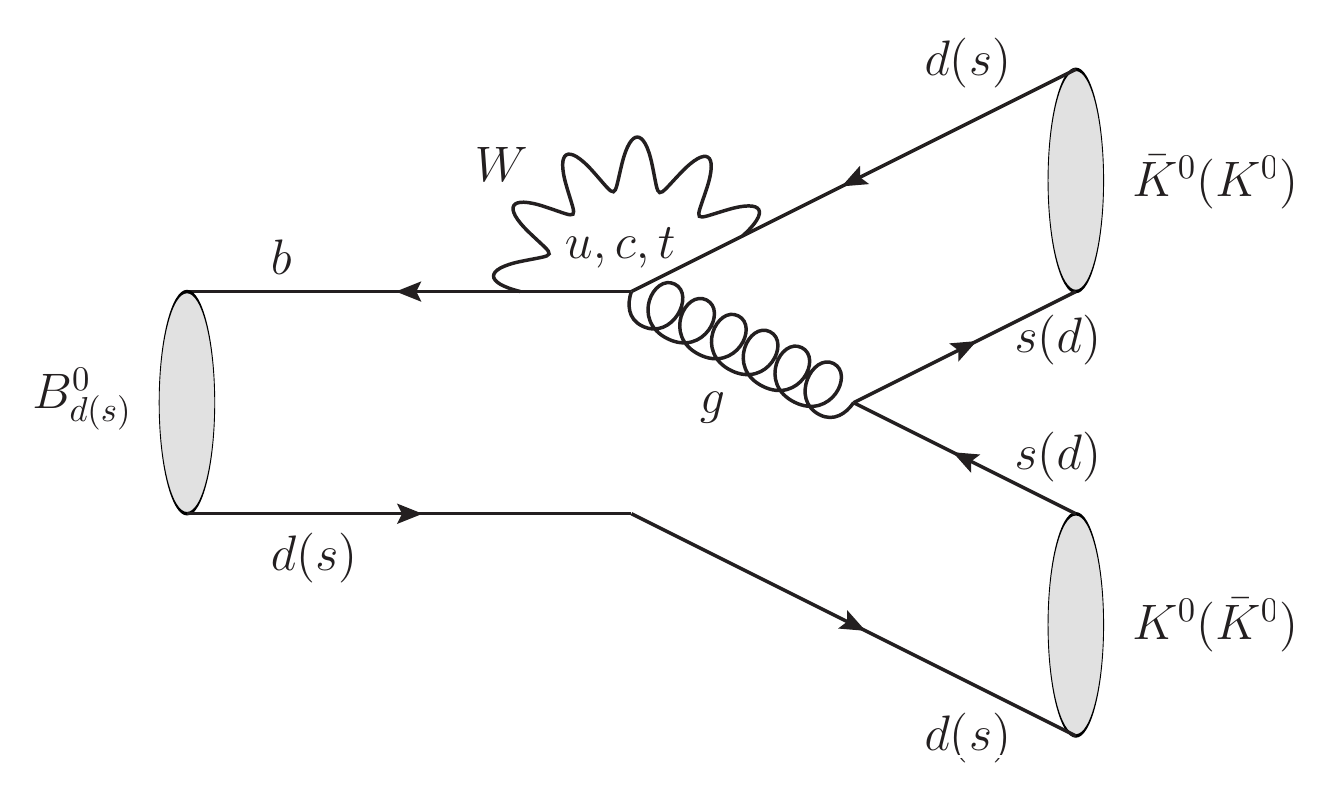}}
	\subfloat[Penguin annihilation ($PA_{KK}$)]{\label{fig:penguinannihilationDiagramK0K0} \includegraphics[width=0.49\textwidth]{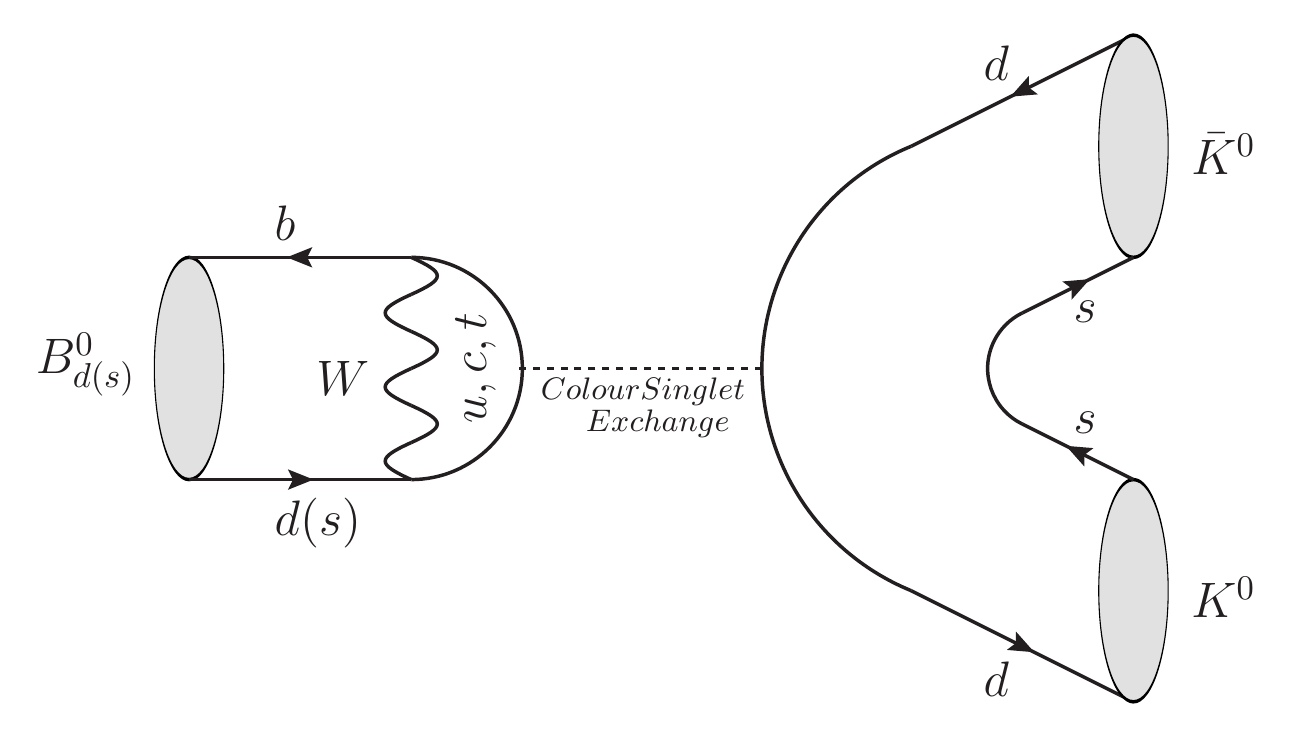}}
\caption{Topologies of the $\BdtoKantiK$ and $\BstoKantiK$ decays.}
\label{fig:diagramsK0K0}
\end{figure}

As can be seen in Fig.~\ref{fig:diagramsK0K0}, the penguin topologies of $\BdtoKantiK$ and 
$\BstoKantiK$ differ from those of \Bdtopipi and \BstoKK only through the quark pair that 
is generated by the gluon. Consequently, the $B^0_{d,s}\to K^0 \Kbar^0$ system offers the 
most suitable probes for $\rho_P$ and subsequently $r_P$. We shall neglect tiny contributions from colour-suppressed electroweak penguins.

The corresponding decay amplitudes can be written as \cite{Fle04}
\begin{align}
A(\BdtoKantiK) = -\mathcal{C}_{KK}\left[1-d_{KK}\euler^{\imaginary\theta_{KK}}\euler^{\imaginary\gamma}\right] , \nonumber \\
A(\BstoKantiK) = \frac{1}{\sqrt{\epsilon}}\mathcal{C}_{KK}^\prime\left[1+\epsilon d_{KK}^\prime\euler^{\imaginary\theta_{KK}^\prime}\euler^{\imaginary\gamma}\right] ,
\end{align}
with
\begin{equation} \label{eq:CKantiK}
\mathcal{C}_{KK} \equiv A\lambda^3\left[P_{KK}^{(ct)}+PA_{KK}^{(ct)}\right] \ , \qquad d_{KK}\euler^{\imaginary\theta_{KK}} \equiv R_b\left[\frac{P_{KK}^{(ut)}+PA_{KK}^{(ut)}}{P_{KK}^{(ct)}+PA_{KK}^{(ct)}}\right] \ ,
\end{equation}
and analogous expressions for $\mathcal{C}_{KK}'$ and $d_{KK}'\euler^{\imaginary\theta_{KK}'}$. In contrast to $\rho_P$, the parameter $d_{KK}\euler^{\imaginary\theta_{KK}}$ also receives contributions from $PA$ topologies. However, these topologies are suppressed in comparison with the leading penguin contributions and can therefore be neglected.
Since the decays \BdtoKantiK, \BstoKantiK and \Bdtopipi, \BstoKK are related to one another by 
the $SU(3)$ flavour symmetry, the extraction of $d_{KK}^{(\prime)} \euler^{\imaginary\theta_{KK}^{(\prime)}}$ allows a determination of $ \rho_P^{(\prime)}$.

The CP asymmetries are given as follows:
\begin{align} \label{eq:ACPdirBdtoKantiK} 
\mathcal{A}_{\rm CP}^\text{dir}(\uBdtoKantiK)& = \frac{2d_{KK}\sin\theta_{KK}\sin\gamma}{1-2d_{KK}\cos\theta_{KK}\cos\gamma+d_{KK}^{2}} \ , \nonumber \\ 
\mathcal{A}_{\rm CP}^\text{mix}(\uBdtoKantiK)& = \frac{\sin\phi_d -2d_{KK}\cos\theta_{KK}\sin(\phi_d+\gamma) + d_{KK}^{2}\sin(\phi_d +2\gamma)}{1-2d_{KK}\cos\theta_{KK}\cos\gamma+d_{KK}^{2}} \ ,
\end{align}
with analogous ``primed'' expressions for the CP asymmetries of \BstoKantiK. The CP asymmetries of \BdtoKantiK have been measured by the BaBar \cite{Aub06} and Belle collaborations \cite{Nak07}. We list them in Table~\ref{tab:CPsBdtoKantiK}, together with their PDG average \cite{PDG}. The experimental
situation is not conclusive and will hopefully be settled with future data.  

If we use the mixing phases $\phi_{d,s}$ as input, the experimental results for these CP asymmetries 
can be converted into theoretically clean values of the parameters $d_{KK}\euler^{\imaginary\theta_{KK}}$ 
and $d_{KK}^{\prime}\euler^{\imaginary\theta_{KK}^{\prime}}$, which will allow valuable insights 
into the dynamics of penguin topologies, shedding  light on the issue of the ``charming penguins"
and into $U$-spin-breaking effects in these penguin parameters. As form factors cancel, those effects
are genuinely related to non-factorizable effects. Using the $SU(3)$ flavour symmetry to relate 
the hadronic parameters of the \BstoKantiK, \BdtoKantiK system to those of the  \Bdtopipi, \BstoKK modes
allows the determination of both $\rho_P$ and $\rho_P^{\prime}$.

\begin{table}
	\centering
	\begin{tabular}{l|r|r|r}
		CP asymmetry & \multicolumn{1}{l|}{BaBar \cite{Aub06}} & \multicolumn{1}{l|}{Belle \cite{Nak07}} & \multicolumn{1}{l}{PDG \cite{PDG}} \\
		\hline\hline
		$\mathcal{A}_\text{CP}^\text{dir}(\uBdtoKantiK)$ & $-0.40 \pm 0.41 \pm 0.06$ & $0.38 \pm 0.38 \pm 0.05$ & $0.0 \pm 0.4$ \\
		$\mathcal{A}_\text{CP}^\text{mix}(\uBdtoKantiK)$ & $1.28\pm0 .80\pm 0.16$ & $0.38 \pm 0.77 \pm 0.09$ & $0.8 \pm 0.5$
	\end{tabular}
	\caption{Overview of the \BdtoKantiK CP asymmetries, where we have conservatively taken the largest uncertainty if the error was asymmetric. }
	\label{tab:CPsBdtoKantiK}
\end{table}

Since there is currently no measurement of the CP asymmetries in \BstoKantiK, we consider the 
following ratio of branching ratios: 
\begin{align}\label{eq:Hkk}
H_{KK} &\equiv \frac{1}{\epsilon}\left|\frac{\mathcal{C}_{KK}'}{\mathcal{C}_{KK}}\right|^2\left[\frac{m_{B_d}}{m_{B_s}}\frac{\Phi(m_K/m_{B_s},m_K/m_{B_s})}{\Phi(m_K/m_{B_d},m_K/m_{B_d})}\frac{\tau_{B_s}}{\tau_{B_d}}\right]\frac{\mathcal{B}(\uBdtoKantiK)}{\mathcal{B}(\uBstoKantiK)} \\
& = \frac{ 1-2d_{KK}\cos\theta_{KK}\cos\gamma+d_{KK}^2}{1+2\epsilon d_{KK}^\prime\cos\theta_{KK}^\prime\cos\gamma+\epsilon^2d_{KK}^{\prime2}} \nonumber \ ,
\end{align}
where the phase-space function $\Phi$ was introduced in Eq.~\eqref{eq:phaseSpaceFunction}. The various measurements of the \BdtoKantiK branching ratio are consistent with one another, and the PDG 
average \cite{PDG} reads
\begin{equation}
\mathcal{B}(\BdtoKantiK)  = (1.21 \pm 0.16) \times 10^{-6}.
\end{equation}
The Belle collaboration has recently announced the observation of the $\BstoKantiK$ channel
\cite{Pal15}, resulting in the branching ratio
\begin{equation}\label{eq:BRBstoKantiK}
\mathcal{B}(\BstoKantiK) = (19.6^{+6.2}_{-5.6}) \times 10^{-6} \ .
\end{equation} 

Using the factorization approximation, we obtain
\begin{equation}\label{eq:ccprime}
\left|\frac{ \mathcal{C}_{KK}^\prime}{\mathcal{C}_{KK}}\right|_{\textrm{fact}} = \left(\frac{m_{B_s}^2- m_K^2}{m_{B_d}^2-m_K^2}\right)\left[\frac{F_0^{B_sK}(m_K^2)}{F_0^{B_dK}(m_K^2)}\right] = 0.92\pm 0.13 ,
\end{equation}
where we have used LCSR results for the corresponding form factors \cite{Kho04}. Using
the information for the branching ratios then gives
\begin{equation}\label{eq:Hkkval}
H_{KK}  = 0.94 \; \pm 0.13|_{B_d}\; \pm 0.29|_{B_s} \; \pm 0.27|_{\mathcal{C}}= 0.94 \pm 0.42,
\end{equation}
where we show the individual contributions of the various quantities to the error budget. 

If we apply the $U$-spin relation
\begin{equation}
d_{KK}\euler^{\imaginary\theta_{KK}} = d_{KK}^\prime\euler^{\imaginary\theta_{KK}^\prime} ,
\end{equation}
the observable $H_{KK}$ and the CP asymmetries of the $\uBdtoKantiK$ channel allow the 
extraction of $\gamma$ and the hadronic parameters \cite{RF-ang}; further information can be
obtained through the measurement of CP violation in \BstoKantiK. However, due to the large current 
uncertainties for both the CP asymmetries of  \BdtoKantiK and the observable $H_{KK}$ only very 
weak constraints can be obtained.


\boldmath
\subsection{$B^+ \rightarrow K^+ \Kbar^0$ and $B^+\rightarrow \pi^+ K^0$} 
\unboldmath
Given the current experimental results for the \BdtoKantiK, \BstoKantiK system discussed in the
previous subsection, the charged $B^+ \rightarrow K^+ \Kbar^0$, $B^+\rightarrow \pi^+ K^0$ decays 
offer an interesting alternative. These modes were previously studied in Ref.~\cite{Fle07}. Let us
update this analysis using the current data. The decays $B^+ \rightarrow K^+ \Kbar^0$ and $B^+\rightarrow \pi^+ K^0$ are characterized by $\bar{b}\rightarrow \bar{s}s\bar{d}$ and $\bar{b}\rightarrow \bar{d}d\bar{s}$ transitions, respectively, and related to each other by the $U$-spin symmetry. The $B^+ \rightarrow K^+ \Kbar^0, B^+\rightarrow \pi^+ K^0$ modes can be related to the \BdtoKantiK, \BstoKantiK decays by applying the $SU(3)$ flavour symmetry at the spectator-quark level, thereby allowing us to determine $\rho_P$. 

The corresponding decay amplitudes can be written in the following form \cite{Fle07}:
\begin{align}
A(B^+\rightarrow \pi^+ K^0) &= \mathcal{P}_{\pi K} \left[ 1+ \epsilon \rho_{\pi K} e^{i \sigma_{\pi K}}e^{i\gamma}\right] \\
A(B^+\rightarrow K^+ \Kbar^0) &= \sqrt{\epsilon} \mathcal{P}_{KK} \left[ 1-\rho_{KK} e^{i\sigma_{KK}}e^{i\gamma}\right] \ ,
\end{align}
where 
\begin{equation}
\rho_{\pi K}\euler^{\imaginary\sigma_{\pi K}} \equiv R_b\frac{\mathcal{P}^{(ut) \prime}}{\mathcal{P}^{(ct) \prime}} \ ,
\end{equation}
 and in analogy
 \begin{equation}
 \rho_{KK}\euler^{\imaginary\sigma_{KK}} \equiv R_b\frac{\mathcal{P}^{(ut)}}{\mathcal{P}^{(ct)}}.
 \end{equation}
The CP asymmetry is defined by 
\begin{equation}
\begin{split}
\mathcal{A}^{\textrm{dir}}_{\rm CP}  (B^\pm\rightarrow \pi^\pm K) &\equiv \frac{|A(B^+\rightarrow \pi^+K^0)|^2-|A(B^-\rightarrow \pi^-\Kbar^0)|^2}{|A(B^+\rightarrow \pi^+K^0)|^2+|A(B^-\rightarrow \pi^-\Kbar^0)|^2} \\
&= \frac{-2\epsilon \rho_{\pi K} \sin\sigma_{\pi K}\sin\gamma}{1+2\epsilon\rho_{\pi K} \cos\sigma_{\pi K}\cos\gamma+\epsilon^2\rho_{\pi K}^2} \ ,
\end{split}
\end{equation}
while the expression for the direct CP asymmetry of $B^+ \rightarrow K^+ \Kbar^0$ can be obtained 
straightforwardly by making the following replacements: 
\begin{equation}
	\epsilon \to -1 \ , \quad \rho_{\pi K} \to \rho_{KK} \ , \quad \sigma_{\pi K} \to \sigma_{KK} \ .
\end{equation}

\noindent The experimental averages for the direct CP asymmetry are given by HFAG \cite{Amh14} as
\begin{align}
\mathcal{A}^{\textrm{dir}}_{\rm CP}(B^\pm\rightarrow \pi^\pm K)&= 0.017 \pm 0.016 \ , \nonumber \\
\mathcal{A}^{\textrm{dir}}_{\rm CP} (B^\pm\rightarrow K^\pm K)&=0.087 \pm 0.100 \ , 
\end{align}
while the branching ratios are listed in Table~\ref{tab:BtohhDecayModes}. We note that both CP asymmetries have switched signs with respect to their values in 2007~\cite{Fle07}.  

\begin{figure}
\centering
\includegraphics[width=0.5\linewidth]{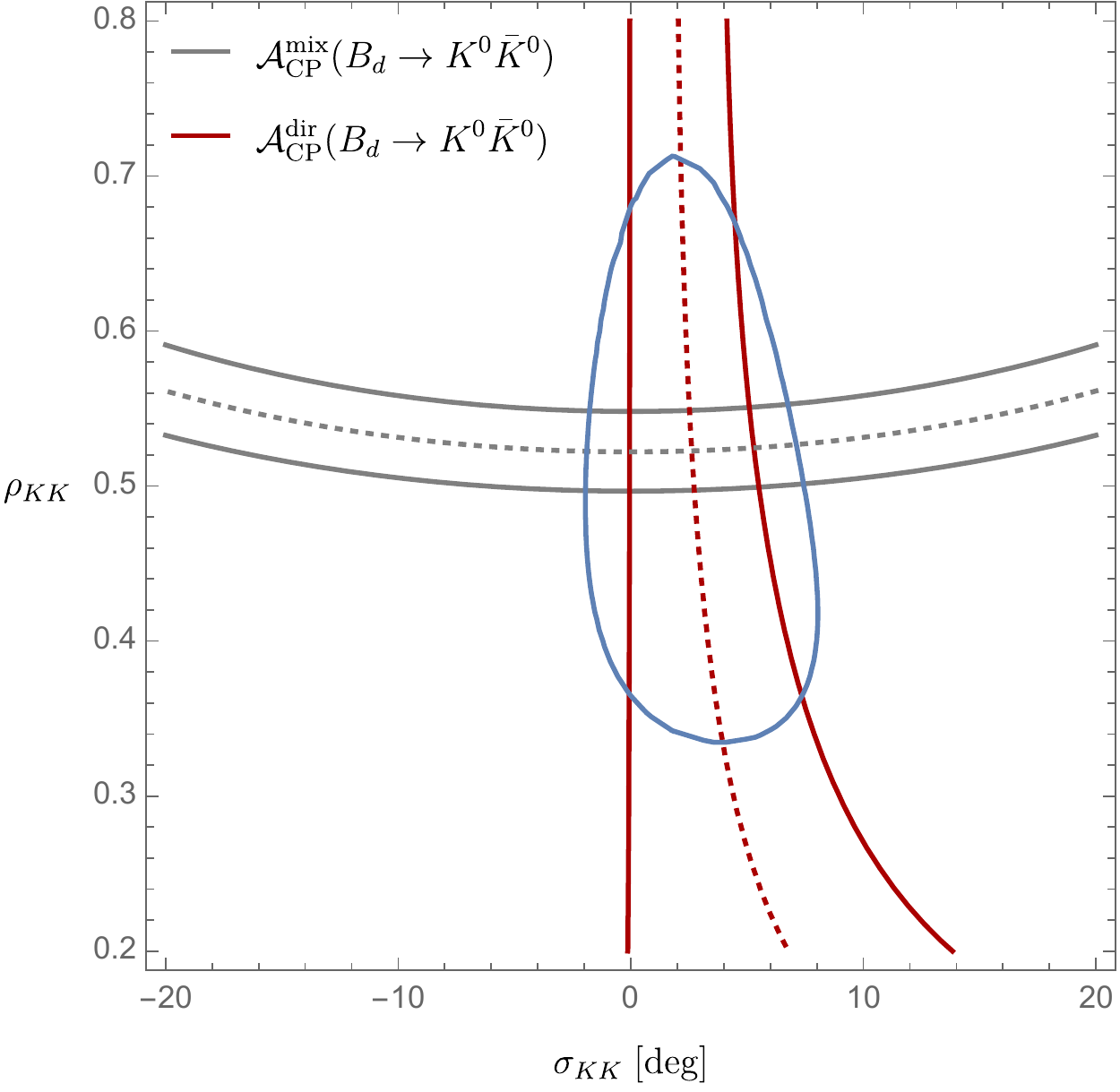}
\caption{Results from a $\chi^2$-minimalization fit to the current data as described in text. The blue contour shows the $1\sigma$ constraint from the fit. The red (gray) contour shows the expected constraint from the direct (mixing-induced) CP asymmetries in $B_d \rightarrow K^0\Kbar^0$ with an anticipated error of $0.05$.}
\label{fig:Hpikk}
\end{figure}

As before, we introduce
\begin{align}
	H_{\pi K}^{KK}  &\equiv  \frac{1}{\epsilon} \left| \frac{\mathcal{P}_{\pi K}}{\mathcal{P}_{KK}} \right|^2 \left[ \frac{\Phi(m_\pi/m_B, m_K/m_B)}{\Phi(m_K/m_B, m_K/m_B)}\right] \frac{\mathcal{B}(B^\pm\rightarrow K^\pm K)}{\mathcal{B}(B^\pm\rightarrow \pi^\pm K)}  \ , \nonumber \\
&=  \frac{1-2\rho_{KK} \cos\sigma_{KK}\cos\gamma+\rho_{KK}^2}{1+2\epsilon\rho_{\pi K} \cos\sigma_{\pi K}\cos{\gamma}+\epsilon^2 \rho^2_{\pi K}} = 0.57 \pm 0.11 \ ,
\end{align}
where we used the following result arising within factorization \cite{Kho04}:
\begin{equation}
 \left|\frac{\mathcal{P}_{KK}}{\mathcal{P}_{\pi K}} \right|_{\rm{fact}} = \left[\frac{m_B^2-m_K^2}{m_B^2-m_\pi^2}\right]\left[\frac{F_0^{BK}(m_K^2)}{F_0^{B\pi}(m_K^2)}\right] = 1.35 \pm 0.11 \ .
 \end{equation}
Combining the CP asymmetries of $B^+ \rightarrow K^+ \Kbar^0$ and $B^+\rightarrow \pi^+ K^0$ with $H_{\pi K}^{KK}$, and assuming the $U$-spin relation
\begin{equation}
 \rho_{KK}=\rho_{\pi K} \quad\quad \sigma_{KK}= \sigma_{\pi K } \ ,
 \end{equation}
 we find the constraints for $\rho_{KK}$ and $\sigma_{KK}$ shown in Fig.~\ref{fig:Hpikk}, which were
obtained through a $\chi^2$-minimalization fit where also $\gamma =(70\pm 7)^\circ$ was added as
a constraint. The best fit result favours interestingly a smaller value of $\gamma=60^\circ$, which is 
caused by the small value of $H_{\pi K}$. This  feature has already been noted in Ref.~\cite{Fle07}. 
Assuming Gaussian distributions, we obtain from the fit
\begin{equation}\label{eq:rhokk}
 \rho_{KK} = 0.52 \pm 0.2 \ , \quad \quad \sigma_{KK} = (2.6\pm 4.6)^\circ \ .
\end{equation} 
These values are in agreement with the estimates in Ref.~\cite{BF-95} and the general hierarchy of
decay topologies discussed in Refs.~\cite{Gro94, Gro95}. We will discuss the 
implications for $|1+r_P|$ and $\Xi_P$ in Subsection~\ref{sec:impl2}.

Using the strong isospin symmetry to relate the up spectator quark in $B^+ \rightarrow K^+ \Kbar^0$ to 
the down spectator quark in \BdtoKantiK  gives the relation 
\begin{equation}
d_{KK} = \rho_{KK}, \quad \quad \theta_{KK} = \sigma_{KK} \ .
\end{equation}
We shall assume these relations, which we expect to hold with excellent precision, for the remainder 
of this section.  Using Eq.~\eqref{eq:rhokk}, we may calculate the CP-violating observables of the 
\BdtoKantiK decay:
\begin{eqnarray} \label{eq:expetedBdtoKantiKCP}
\mathcal{A}^{\textrm{mix}}_{\rm CP}(\uBdtoKantiK) &	= -0.32 \pm 0.39   , \nonumber \\
\mathcal{A}^{\textrm{dir}}_{\rm CP} (\uBdtoKantiK)  &=  0.05 \pm 0.09 ,
\end{eqnarray}
where the errors are dominated by the uncertainty of $\rho_{KK}$. These values are in agreement with the current experimental measurements given in Table~\ref{tab:CPsBdtoKantiK}, although the experimental
uncertainties are unfortunately too large to draw any conclusions.

Improved CP violation measurements in \BdtoKantiK would allow a powerful and theoretically clean 
determination of $\rho_{KK}$ and $\sigma_{KK}$, as illustrated in Fig.~\ref{fig:Hpikk}. Here we have 
added the contours from the expected CP asymmetries in \BdtoKantiK with an assumed 
error of $0.05$ in the era of Belle II and the LHCb upgrade. We observe that in particular the 
mixing-induced CP asymmetry of $\BdtoKantiK$ has the potential to constrain $\rho_{KK}$ much further, thereby reducing the uncertainty for an important parameter of our strategy. A measurement of the mixing-induced CP asymmetry of \BdtoKantiK with a precision of $0.1$ would allow a determination of 
$\rho_{KK}$ with a precision of $0.1$, which would be a significant improvement over the current 
precision in Eq.~\eqref{eq:rhokk}. 

Using in addition a future measurement of the CP asymmetries of the \BstoKantiK channel would 
allow a clean determination of $d_{KK}'$ and $\theta_{KK}'$, thereby offering an interesting test of the
$U$-spin symmetry in these penguin parameters. The observable $H_{KK}$ is not needed for this
analysis, but offers instead further insights into the $U$-spin symmetry for the QCD penguin topologies.

\boldmath
\subsection{$\BdtoKpi$ and $\BstoKpi$} \unboldmath \label{sec:tilde}

The decays $\BdtoKpi$ and $\BstoKpi$ receive only contributions from tree and penguin topologies
and are related to each other through the $U$-spin symmetry \cite{GroRo,Fle07}. We have already
encountered the \BdtoKpi channel in Subsection~\ref{sec:proof}, while the amplitude for \BstoKpi
takes the form
\begin{equation}
A(\BstoKpi) = \euler^{\imaginary\gamma}\tilde{\mathcal{C}}\left[1-\tilde{d}\euler^{\imaginary\tilde{\theta}}\euler^{-\imaginary\gamma}\right] \ ,
\end{equation}
where $\tilde{\mathcal{C}}$ and $\tilde{d}\euler^{\imaginary\tilde{\theta}}$ are defined in analogy to 
Eq.~\eqref{eq:cdtilpdef}.

As the final states are flavour-specific, only direct CP violation can occur. The expressions for the direct CP asymmetry can be obtained by making suitable replacements in Eq.~\eqref{eq:cpasym}. The current direct CP asymmetries as given by the PDG are \cite{PDG}:
\begin{subequations}\label{eq:upmix}
	\begin{align}
	\mathcal{A}_{\textrm{CP}}^{\textrm{dir}}(\BdtoKpi) &= 0.082\pm0.006  \;\;\; (0.082\pm 0.003) \ , \\
	\mathcal{A}_{\textrm{CP}}^{\textrm{dir}}(\BstoKpi) &=-0.26\pm0.035 \;\; (-0.26 \pm 0.006) \ .
	\end{align}
\end{subequations} 
In parentheses, we give a future scenario for the Belle II and LHCb upgrade era \cite{LHCbup, Belle-II}.

\begin{figure}
	\centering
	\includegraphics[width=0.5\linewidth]{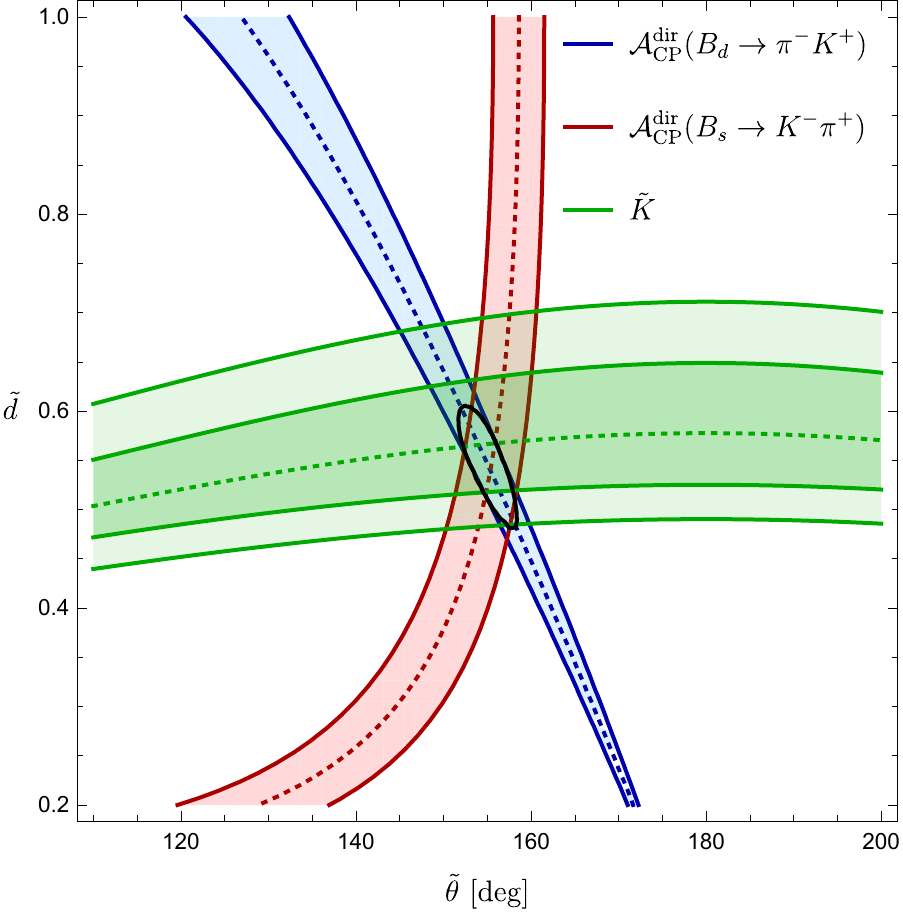}
	\caption{Current constraints on the penguin parameters $\tilde{d}$ and $\tilde{\theta}$ from the $\BdtoKpi$, $\BstoKpi$ CP asymmetries and ratio $\tilde{K}$. The black contour gives the constraints from a $\chi^2$ fit to the CP asymmetries and $\gamma = (70\pm 7)^\circ$. For $\tilde{K}$, we consider $|\tilde{\mathcal{C}}/\tilde{\mathcal{C}}'|$ in factorization (wide band) and $|\tilde{\mathcal{C}}/\tilde{\mathcal{C}}'| = 1$ (small band). }
	\label{fig:Ktilde}
\end{figure}

For the current data, the CP asymmetries combined with the $U$-spin relation 
\begin{equation}
\tilde{d}\euler^{\imaginary\tilde{\theta}} = \tilde{d}'\euler^{\imaginary\tilde{\theta}'}
\end{equation}
give the constraints for $(\tilde{d}, \tilde{\theta})$ shown in Fig.~\ref{fig:Ktilde}. They are obtained using a $\chi^2$ fit with $\gamma = (70\pm 7)^\circ$ added as a constraint. We find
\begin{equation} \label{eq:dthetatildenow}
\tilde{d} =  0.54\pm 0.06 \ , \quad\quad \tilde{\theta} = (155.4\pm 3.3)^\circ \ . 
\end{equation} 
For the upgrade scenario in Eq.~\eqref{eq:upmix} with $\gamma = (70\pm 1)^\circ$, the fit gives
\begin{equation} \label{eq:dthetatilde}
\tilde{d} = 0.54\pm0.02  \ , \quad\quad \tilde{\theta} = (155.4\pm0.6)^\circ  \ . 
\end{equation} 
These determinations agree with the picture arising from CP violation in $\Bdtopipi$ and the values 
in Eq.~\eqref{eq:primed}. Specifically, the parameter
\begin{equation}
\zeta \equiv |\zeta|\euler^{\imaginary\omega}
\end{equation}
relates the hadronic parameters. Using Eq.~\eqref{eq:dval}, we find for the upgrade scenario 
\begin{equation}\label{eq:xis}
|\zeta| \equiv \tilde{d}/d =0.93\pm 0.05 \ , \qquad \omega \equiv \tilde{\theta}-\theta =(4.0 \pm 1.3 )^\circ \ , 
\end{equation}
showing an impressive accuracy for the picture assumed in the era of Belle II and the LHCb upgrade.

Let us now utilize again the information provided by semileptonic decays. In order to complement 
the ratio $\tilde{R}_K$ defined in Eq.~\eqref{eq:rktilde}, we introduce
\begin{equation}
\tilde{R}^\prime_K \equiv \frac{\Gamma(\BstoKpi)}{d\Gamma(B_s^0\rightarrow K^- \ell^+ \nu_\ell)/
dq^2|_{q^2 = m_\pi^2} }\ ,
\end{equation}
which requires the measurement of the semileptonic differential rate of the decay
$B_s^0 \rightarrow K^- \ell^+ \nu_\ell$, which we require also for our key observable $R_K$. 
In analogy to our new strategy, we may determine the parameters $\tilde{d},\tilde{\theta}$ and 
$\tilde{d}', \tilde{\theta}'$, which allow an interesting test of the $U$-spin symmetry in the 
dominant tree and penguin topologies.

Lacking at the moment a measurement of $B_s^0 \rightarrow K^- \ell^+ \nu_\ell$, we might also consider the 
ratio of branching ratios, as we discussed for the \Bdtopipi, \BstoKK system:
\begin{align} \label{eq:kTilde}
\tilde{K} &\equiv \frac{1}{\epsilon}\left|\frac{\tilde{\mathcal{C}}}{\tilde{\mathcal{C}}^\prime}\right|^2\left[\frac{m_{B_d}}{m_{B_s}}\frac{\Phi(m_{K^\pm}/m_{B_s},m_{\pi^\pm}/m_{B_s})}{\Phi(m_{\pi^\pm}/m_{B_d},m_{K^\pm}/m_{B_d})}\frac{\tau_{B_s}}{\tau_{B_d}}\right]\frac{\mathcal{B}(\uBdtoKpi)}{\mathcal{B}(\uBstoKpi)_\text{theo}}, \nonumber \\
&= \frac{1+2(\tilde{d}^\prime/\epsilon)\cos\tilde{\theta}^\prime\cos\gamma+(\tilde{d}^\prime/\epsilon)^2}{1-2\tilde{d}\cos\tilde{\theta}\cos\gamma+\tilde{d}^2} \myexp{exp} 63.6_{-12.3}^{+20.1} \ ,
\end{align}
where we used the factorization approximation to obtain
\begin{equation} \label{eq:ctctp}
\left|\frac{\tilde{\mathcal{C}}}{\tilde{\mathcal{C}}'}\right|_\text{fact} = \frac{f_\pi}{f_K}\left[\frac{m_{B_s}^2-m_K^2}{m_{B_d}^2-m_\pi^2}\right]\left[\frac{F_0^{B_sK}(m_\pi^2)}{F_0^{B_d\pi}(m_K^2)}\right] = 0.99_{-0.08}^{+0.15} \ .
\end{equation}
The ratio of form factors $F_0^{B_sK}(0)/F_0^{B_d\pi}(0)=1.15^{+0.17}_{-0.09}$ follows from an LCSR calculation \cite{Dub08}, and $f_K/f_\pi = 1.1928\pm 0.0026$ \cite{Ros15}. It is interesting to note that the form factors and decay constants enter Eq.~\eqref{eq:ctctp} in such a way that they almost cancel.

The uncertainty of Eq.~\eqref{eq:kTilde} is dominated by the form factors. If we assume a perfect determination of $|\tilde{\mathcal{C}}/\tilde{\mathcal{C}}'|  = 1$, we find $\tilde{K} = 65.1 \pm 7.3$. Combining the ratio $\tilde{K}$ with $\gamma= (70\pm 7)^\circ$ gives an additional constraint on $(\tilde{d}, \tilde{\theta})$, which we have added to Fig.~\ref{fig:Ktilde}. There, the wide band and central value follow from Eq.~\eqref{eq:kTilde}, while the small band corresponds to the situation for $|\tilde{\mathcal{C}}/\tilde{\mathcal{C}}'| = 1$. We find good agreement with the constraints following from the measurements of direct CP violation in the $\BdtoKpi$ and $\BstoKpi$ decays, which we also give in Fig.~\ref{fig:Ktilde}. The latter are not affected by form factor uncertainties. The consistent picture in Fig.~\ref{fig:Ktilde} is remarkable and does not point towards any anomalously 
large $U$-spin-breaking effects.

\section{Insights into Exchange and Penguin Annihilation Dynamics} \label{sec:InformationFromBtohhDecays}

\begin{table}[t]
	\centering
	\begin{tabular}{  l | c | cccc | c}
		Decay & $\mathcal{C}$ & \multicolumn{4}{c|} {Topologies} & Specific use: \\
		& & $T$ & $P$ & $E$ & $PA$ & \\
		\hline\hline
		\Bdtopipi & $\mathcal{C}$ & x & x & x & x & Determine $d$ and $\theta$ ($\gamma$ and $\phi_d$ as input) \\
		\BdtoKpi & $\mathcal{\tilde{C}}'$ & x & x & & & Direct determination of $T+P$ \\
		\BdtoKK & $\mathcal{\hat{C}}$ & &  & x & x & Direct determination of $E$+$PA$ \\
		\hline
		\BstoKK & $\mathcal{C}'$ & x & x & x &  x & Determination of $\phi_s$, $d'$, $\theta'$ \\ 
		\BstoKpi & $\mathcal{\tilde{C}}$ & x & x & & & Non-factorizable effects in $T$ and $P$ \\ 
		\Bstopipi & $\mathcal{\hat{C}}'$ & &  & x &  x & Non-factorizable effects in $E$ and $PA$ \\ 
		\hline
		\BdtoKantiK & $\mathcal{C}_{KK}$ & & x & &  x & Direct determination of penguin ratio $\rho_P$ \\ 
		\BstoKantiK & $\mathcal{C}'_{KK}$ & & x & &  x & Non-factorizable effects in penguin ratio $\rho_P$ \\ 
		$B^+ \rightarrow \pi^+K^0$ & $\mathcal{P}_{\pi K}$ & & x & &  & Alternative determination of $\rho_P$ \\ 
		$B^+ \rightarrow K^+\Kbar^0$ & $\mathcal{P}_{KK}$ & & x & & & Alternative determination of	$\rho_P$ \\ 
	\end{tabular}
	\caption{Compilation of various $B\rightarrow hh$ channels ($h=\pi,K$) with their decay topologies
	and their use in the context of our strategy.}
	\label{tab:decaymodes}
\end{table}

The exchange and penguin annihilation contributions enter our new strategy through the parameter $\xi_{\rm NF}^a$. Consequently, we need information about these topologies to assess the theoretical precision.
Specifically, we study the parameters $x$ (see Eq.~\eqref{eq:x}) and $\zeta$ (see Eq.~\eqref{eq:Repa}) and their $U$-spin partners, which enter $\Xi_x$ and $\Xi_P$, respectively. Fortunately, we may use 
experimental data to determine the size of these contributions and do not have to rely on model-dependent
assumptions. In Table~\ref{tab:decaymodes}, we give an overview of the relevant $B\rightarrow hh$ decays
($h=\pi, K$) and the topologies that are used to obtain insights into the different contributions to our strategy.

The \BdtoKK and \Bstopipi modes emerge only from exchange and penguin-annihilation topologies.
Consequently, this allows us to explore these contributions in a direct way. Unfortunately, the current 
experimental data is not yet sufficient to make full use of the potential of these decays although important
constraints can already be obtained, with excellent future prospects. In view of this situation, we discuss also
alternative indirect determinations of the exchange and penguin-annihilation topologies in 
Subsections~\ref{sec:indDetermOfx} and \ref{sec:rpa}. In Subsection~\ref{sec:impl}, we 
return to the \BdtoKK and \Bstopipi decays, discussing future scenarios for the era of Belle II and
the LHCb upgrade. 


\boldmath
\subsection{Direct Determination from \BdtoKK and \Bstopipi} \label{sec:hat}
\unboldmath

The decays \BdtoKK and \Bstopipi receive only contributions from exchange and penguin annihilation topologies. Their amplitudes are given by
\begin{align} \label{eq:ABdtoKK}
A(\BdtoKK) &= \euler^{\imaginary\gamma}\hat{\mathcal{C}}\left[1-\hat{d}\euler^{\imaginary\hat{\theta}}\euler^{-\imaginary\gamma}\right] \\	
A(\Bstopipi) &= \sqrt{\epsilon}\euler^{\imaginary\gamma}\hat{\mathcal{C}}^\prime\left[1+\frac{1}{\epsilon}\hat{d}^\prime\euler^{\imaginary\hat{\theta}^\prime}\euler^{-\imaginary\gamma}\right] \ ,
\end{align}
with
\begin{equation} \label{eq:CdHat}
\hat{\mathcal{C}} \equiv \lambda^3AR_b\left[\hat{E}+\hat{PA}^{(ut)}\right] \ , \qquad \hat{d}\euler^{\imaginary\hat{\theta}} \equiv \frac{1}{R_b}\left[\frac{\hat{PA}^{(ct)}}{\hat{E}+\hat{PA}^{(ut)}}\right].
\end{equation}
The parameters $\hat{\mathcal{C}}^\prime$ and $\hat{d}^\prime$ are given by analogous expressions. The CP asymmetries can be obtained from Eq.~\eqref{eq:cpasym} by replacing $d (\theta) \rightarrow \hat{d} (\hat{\theta})$ and equivalently $d' (\theta') \rightarrow \hat{d}' (\hat{\theta'})$. Since these CP 
asymmetries have not yet been measured, we explore the currently available information by considering 
\begin{align} \label{eq:Kexp}
\hat{K} &= \frac{1}{\epsilon} \left|\frac{\hat{\mathcal{C}}}{\hat{\mathcal{C}}^\prime}\right|^2\left[\frac{m_{B_s}}{m_{B_d}}\frac{\Phi(m_K/m_{B_d},m_K/m_{B_d})}{\Phi(m_\pi/m_{B_s},m_\pi/m_{B_s})}\frac{\tau_{B_d}}{\tau_{B_s}}\right]\frac{\mathcal{B}(\uBstopipi)_\text{theo}}{\mathcal{B}(\uBdtoKK)}  \nonumber \\
&=\frac{1}{\epsilon^2} \frac{\epsilon^2+2\epsilon\hat{d}^\prime\cos\hat{\theta}^\prime\cos\gamma+\hat{d}^{\prime 2}}{1-2\hat{d}\cos\hat{\theta}\cos\gamma+\hat{d}^2} \myexp{exp} 224.6 \pm 50.2 \ , 
\end{align}
where we have used the scaling factor \cite{Bob14}
\begin{equation} \label{eq:su3fact_Chat_Chatprime}
\frac{\hat{\mathcal{C}}}{\hat{\mathcal{C}}^\prime} \approx \frac{f_{B_d}f_{K^\pm}^2}{f_{B_s}f_{\pi^\pm}^2} 
\end{equation}
with $f_{B_s}/f_{B_d}=1.192\pm 0.006$ \cite{Ros15}. Since there is no effective lifetime measurement for \Bstopipi available, we used the experimental branching ratio for simplicity. A more sophisticated analysis
can be performed by using the expression of ${\cal A}_{\Delta\Gamma}(B_s\to\pi^-\pi^+)$ in terms of the
hadronic parameters to convert the experimental into the theoretical branching ratio, applying the formulae
given in Subsection~\ref{ssec:untagged}. 

 Assuming the $U$-spin relation
 \begin{equation} \label{eq:hatduspin}
 \hat{d}\euler^{\imaginary\hat{\theta}} = \hat{d}^\prime\euler^{\imaginary\hat{\theta}^\prime}
 \end{equation}
 gives
\begin{equation} \label{eq:dHat_vs_thetaHat}
\hat{d} = \frac{\epsilon}{1-\epsilon^2\hat{K}} \left[-\cos\hat{\theta}\cos\gamma\left(1+\epsilon\hat{K}\right)\pm\sqrt{\cos^2\hat{\theta}\cos^2\gamma\left(1+\epsilon\hat{K}\right)^2-\left(1-\epsilon^2\hat{K}\right)\left(1-
\hat{K}\right)}\right]\ .
\end{equation}
In analogy to the $K$ observable for the  \Bdtopipi, \BstoKK system, $\hat{K}$ is not a clean observable because it depends on $|\hat{\mathcal{C}}/\hat{\mathcal{C}}^\prime|$. This ratio is sensitive to both 
factorizable and non-factorizable $U$-spin-breaking corrections.

\begin{figure}[t]
	\centering
	\includegraphics[width=0.6\linewidth]{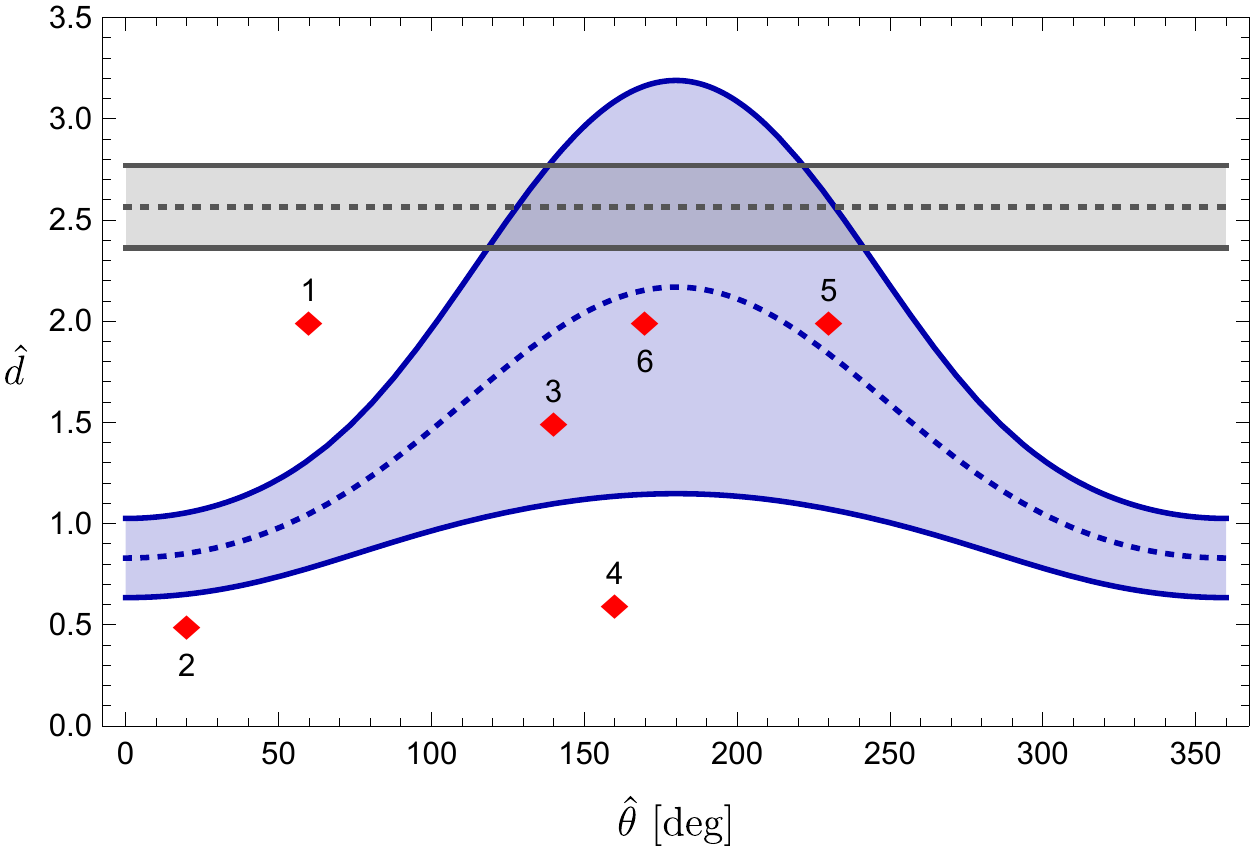}
	\caption{Current constraints on $\hat{d}$ as a function of $\hat{\theta}$. The horizontal band gives the naive constraint on $\hat{d}$ in Eq.~\eqref{eq:constraintdhat}, discussed in the text. The various diamond points represent the different future scenarios discussed in Section~\ref{sec:InformationFromBtohhDecays}. }
	\label{fig:dHat_vs_thetaHat}
\end{figure}

\begin{figure}[t]
	\centering
	\includegraphics[scale=.45]{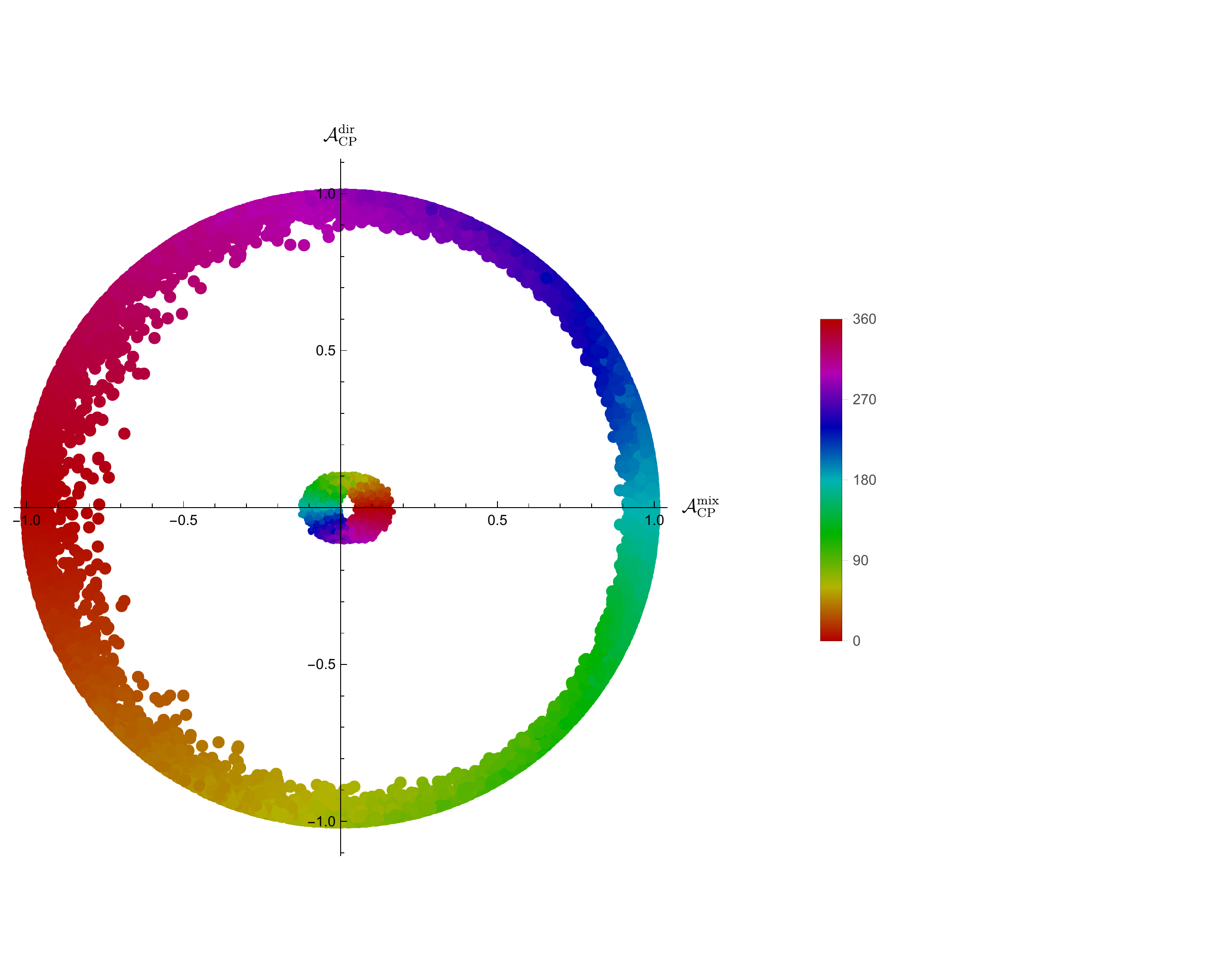}
	\caption{Correlation between the predicted direct and mixing-induced CP asymmetries of \BdtoKK (outer region) and for \Bstopipi (inner region). The colour coding indicates the value of the strong phase $\hat{\theta}$ [deg].}
	\label{fig:correlations_CPasymmetries_BdKK_BsPiPi}
\end{figure}

Fig.~\ref{fig:dHat_vs_thetaHat} shows the relation between $\hat{d}$ and $\hat{\theta}$ with $1\sigma$ error bands for the current data.  As the penguin-annihilation topologies are loop suppressed while the
exchange contributions arise at the tree level, we obtain the following naive -- but plausible -- upper bound:
\begin{equation} \label{eq:constraintdhat}
\hat{d} \lsim \frac{1}{R_b} \approx 2.56 \pm 0.20,
\end{equation}
which we have included as a constraint in Fig.~\ref{fig:dHat_vs_thetaHat}. 
Measurements of the CP-violating observables of these channels will allow a clean determination of 
the hadronic parameters $\hat{d}$ and $\hat{\theta}$. In order to explore their expected ranges, we 
employ the correlation between $\hat{d}$ and $\hat{\theta}$ in Fig.~\ref{fig:dHat_vs_thetaHat} to calculate
a correlation between the direct and mixing-induced CP asymmetries. To this end, we use 
$\gamma=(70\pm7)^\circ, \phi_d = (43.2\pm 1.8)^\circ$ and $\phi_s= -(0.68\pm 2.2)^\circ$ 
as determined from experiment. We obtain a surprisingly constrained situation, as shown in 
Fig.~\ref{fig:correlations_CPasymmetries_BdKK_BsPiPi}. The general relation between the 
CP asymmetries in Eq.~\eqref{eq:sumrule} implies
\begin{equation} \label{eq:CPconstraint}
[\mathcal{A}_\text{CP}^\text{dir}(B_s\to\pi^-\pi^+)]^2 + [\mathcal{A}_\text{CP}^\text{mix}(B_s\to\pi^-\pi^+)]^2 
= 1 - [\mathcal{A}_{\Delta\Gamma}(B_s\to\pi^-\pi^+)]^2 \leq 1.
\end{equation}
Interestingly, we find CP asymmetries of the \BdtoKK channel that are 
scattered pretty close to this relation. 

Future measurements of CP violation in \BdtoKK and \Bstopipi can unambiguously determine 
the parameters $\hat{d}$ and $\hat{\theta}$ and their $U$-spin counterparts $\hat{d}'$ and $\hat{\theta}'$, without making use of $U$-spin assumptions or relying on the $\hat{K}$ observable. Using then 
these parameters in the expression for $\hat{K}$ in Eq.~\eqref{eq:Kexp}, we may extract the amplitude
ratio $|\hat{\mathcal{C}}/\hat{\mathcal{C}}'|$. These studies will allow us to explore $U$-spin-breaking 
effects in exchange and penguin annihilation topologies and will offer valuable further insights into the
dynamics of these contributions.


\boldmath
\subsection{Indirect Determinations of $x$} \unboldmath \label{sec:indDetermOfx}
The direct determination of the exchange and penguin-annihilation topologies from the
decays \BdtoKK and \Bstopipi can be complemented with indirect information from the 
ratios of branching ratios $\Xi_i^{(\prime)}$ listed in Table~\ref{tab:Xi}:
\begin{equation}
\Xi(B_x \to X X', B_y \to YY')  \equiv \left[\frac{m_{B_x}}{m_{B_y}}\frac{\Phi(m_Y/m_{B_y},m_{Y'}/m_{B_y})}{\Phi(m_X/m_{B_x},m_{X'}/m_{B_x})}\frac{\tau_{B_y}}{\tau_{B_x}}\right]\frac{\mathcal{B}(B_x \to X X')}{\mathcal{B}(B_y \to Y Y')} \ ,
\end{equation}
where $\Phi$ is the phase-space function in Eq.~\eqref{eq:phaseSpaceFunction}. Although the
theoretical interpretation of these quantities is affected by $U$-spin-breaking corrections, we 
have plenty of data available, allowing us to constrain the parameter $x$. For this analysis, 
also the penguin parameters $(d, \theta)$ and their counterparts are required. Future data will allow us
to probe $x'$ through the $\Xi_i'$ ratios. In Subsections~\ref{sec:impl} and \ref{sec:impl2}, 
we will discuss the optimal strategy for a future determination of $\Xi_x$ and $\Xi_P$, respectively.

\begin{table}[t]
	\centering
	\begin{tabular}{l|l|c|c|l}
		 & Definition  & \multicolumn{1}{l|}{Input} & \multicolumn{1}{l|}{Factor} & \\
		\hline\hline
		$\Xi_1$ & $\Xi(\uBdtoKK, \uBdtopipi)$ & $(d,\theta), \hat{d}$ & $|\frac{x}{1+x}|$ vs $\hat{\theta}$ & Figs.~\ref{fig:xcons}(a)~and~\ref{fig:xversussigma} \\
		$\Xi_2$ & $\Xi(\uBdtopipi, \uBstoKpi)$ & $(d, \theta), (\tilde{d}, \tilde\theta) $ & $|1+x|$   & Fig.~\ref{fig:xversussigma} \\
		$\Xi_3$ & $\Xi(\uBdtoKK, \uBstoKpi)$ & $(\tilde{d}, \tilde{\theta}), \hat{d}$ & $|x| $ vs $\hat{\theta}$  & Figs.~\ref{fig:xcons}(b)~and~\ref{fig:xversussigma} \\
			
		\hline
		$\Xi_1'$ & $\Xi(\uBstopipi, \uBstoKK)$ & - (*) &$|\frac{r_{PA}'}{1+r_{PA}'}|$ & Figs.~\ref{fig:rPA_ComplexPlane}~and~\ref{fig:asfig} \\	
		$\Xi_2'$ & $\Xi(\uBstoKK, \uBdtoKpi)$ & - (*) & $|1+r_{PA}'|$   &  Fig.~\ref{fig:asfig} \\	
		$\Xi_3'$ & $\Xi(\uBstopipi, \uBdtoKpi)$ & - (*)& $|r_{PA}'|$   &  Fig.~\ref{fig:asfig}
	\end{tabular}
	\caption{Definitions of the ratios of $B\rightarrow hh$ branching ratios and the parameters that they constrain in the current situation. At the moment, the $\Xi_i'$ ratios constrain $r_{PA}'$. In the future, when independent information on the penguin parameters will be available, these ratios can be used to determine $x'$ as well, as indicated by the asterix. }
	\label{tab:Xi}
\end{table}

\begin{figure}
  \centering
 \subfloat[]{ \label{fig:xOverOnePlusX_vs_thetaHat}\includegraphics[scale=0.55]{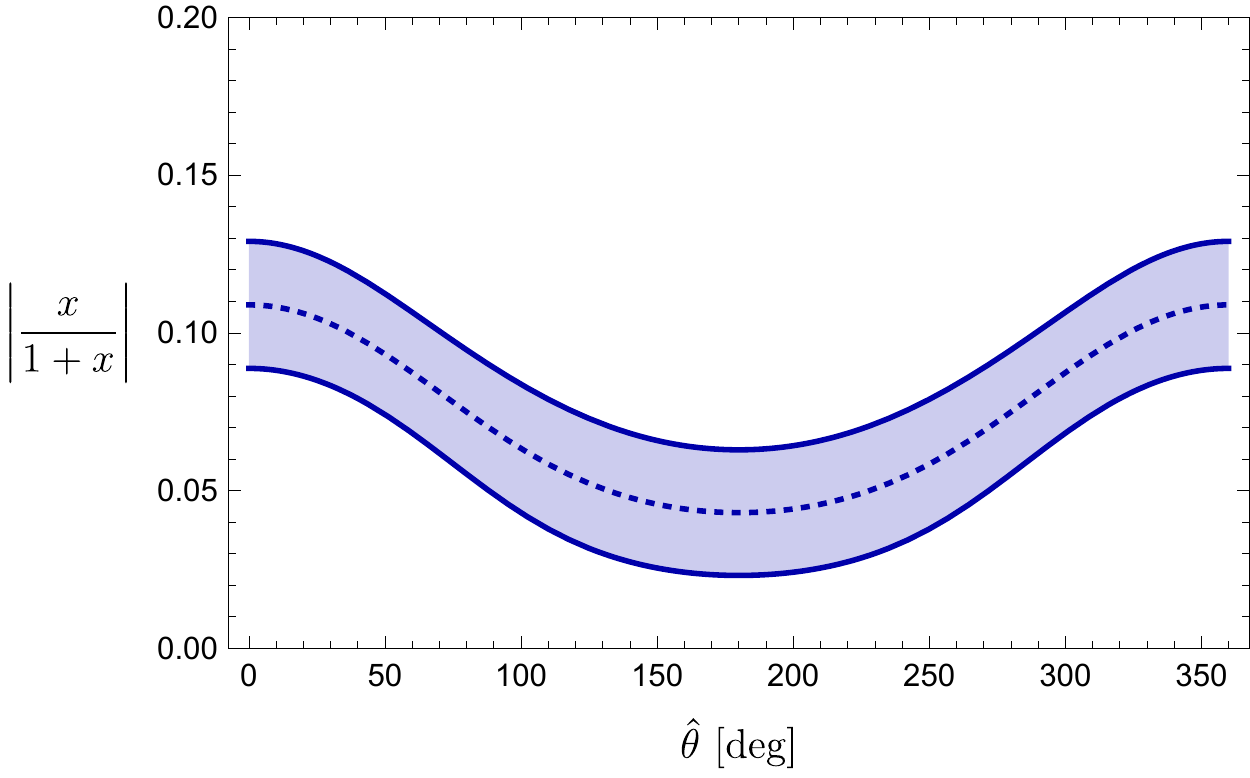}}
\subfloat[]{\label{fig:Xi3}  \includegraphics[scale=0.52]{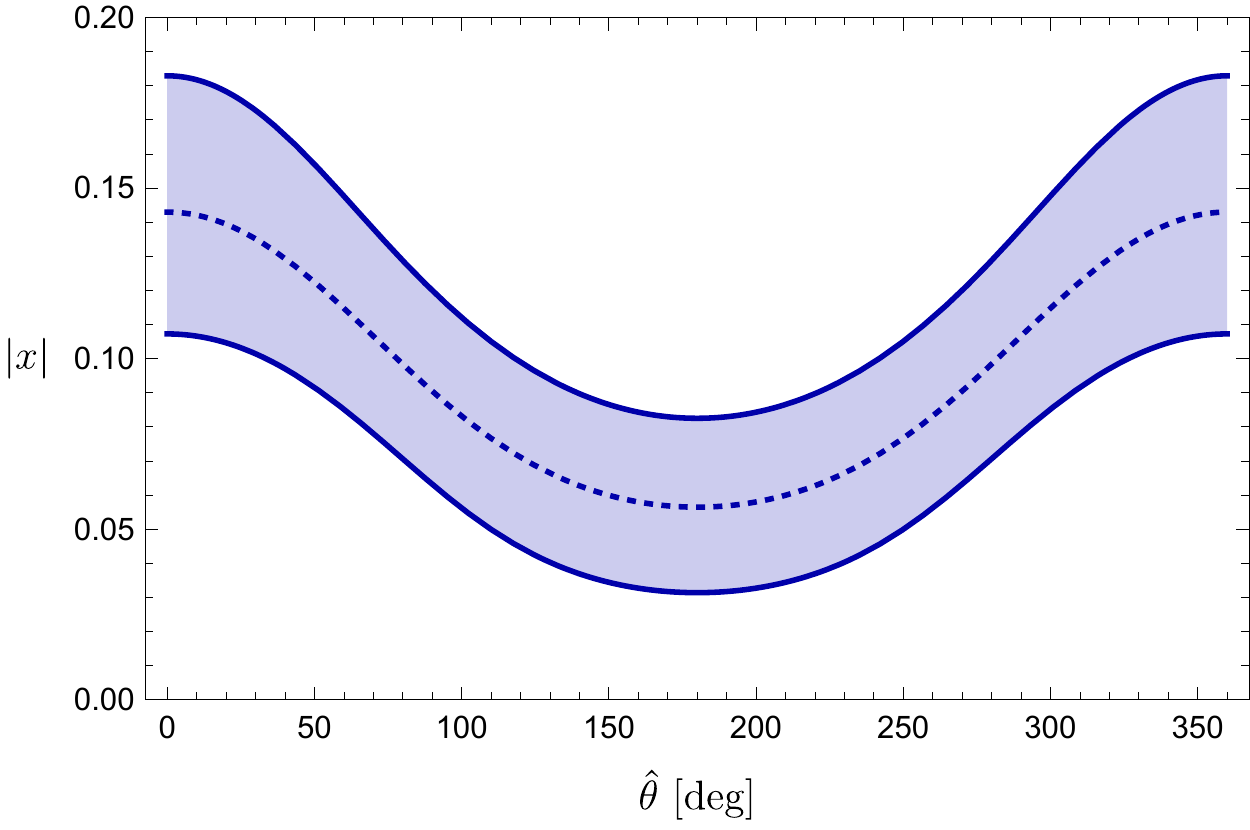}}
	\caption{Constraints on (a) the ratio $|x|/|1+x|$ and (b) $|x|$ as a function of $\hat{\theta}$. }
	\label{fig:xcons}
\end{figure}

Let us first consider the ratio
\begin{equation}
\Xi_1 = \left|\frac{\hat{\mathcal{C}}}{\mathcal{C}}\right|^2\left[\frac{1-2\hat{d}\cos\hat{\theta}\cos\gamma+\hat{d}^2}{1-2d\cos\theta\cos\gamma+d^2}\right] \ .
\end{equation}
Defining 
\begin{equation} \label{eq:etadefval}
\eta \equiv	 \left|\frac{\hat{E}+\hat{PA}^{(ut)}}{E+PA^{(ut)}}\right| \sim \left(\frac{f_K}{f_\pi}\right)^2 = 1.423 \pm 0.006\ ,
\end{equation}
where we have used the decay constants to estimate the non-factorizable topologies \cite{Bob14}, yields 
\begin{equation}
	\left|\frac{\hat{\mathcal{C}}}{\mathcal{C}}\right|^2 =
\left|\frac{x}{1+x}\right|^2\eta^2 \ .
\end{equation}
Consequently, we write 
\begin{equation}
\Xi_1 =	\left|\frac{x}{1+x}\right|^2  \eta^2\left[\frac{1-2\hat{d}\cos\hat{\theta}\cos\gamma+\hat{d}^2}{1-2d\cos\theta\cos\gamma+d^2}\right] \myexp{exp}0.016\pm0.003\ ,
\end{equation}
where the numerical value refers to the experimental branching ratios in Table~\ref{tab:BtohhDecayModes}.
Using $d$ and $\theta$ as determined from the CP-violating observables of the $\Bdtopipi$ channel 
and $\hat{d}$ as a function of $\hat{\theta}$, as described by Eq.~\eqref{eq:dHat_vs_thetaHat} 
and shown in Fig.~\ref{fig:dHat_vs_thetaHat}, we may determine $|x|/|1+x|$ as a function of 
$\hat{\theta}$. The corresponding constraints are shown in Fig.~\ref{fig:xcons}(a). 

Let us next consider the ratio
\begin{equation}
\Xi_2=\left|\frac{\mathcal{C}}{\tilde{\mathcal{C}}}\right|^2\left\langle\left|\frac{1-d\euler^{\imaginary\theta}\euler^{-\imaginary\gamma}}{1-\tilde{d}\euler^{\imaginary\tilde{\theta}}\euler^{-\imaginary\gamma}}\right|^2\right\rangle \ ,
\end{equation}
where
\begin{equation} \label{eq:COverTildeC}
\left|\frac{\mathcal{C}}{\tilde{\mathcal{C}}}\right|^2 = |1+x|^2\rho^2 
\end{equation}
with
\begin{equation} \label{eq:rhodef}
\rho \equiv \left|\frac{T+P^{(ut)}}{\tilde{T}+\tilde{P}^{(ut)}}\right| \ .
\end{equation}
We estimate $\rho$ by considering the ratio of the relevant colour-allowed tree amplitudes 
in factorization, i.e.\
\begin{equation} \label{eq:rhofact}
\rho \sim \left|\frac{T}{\tilde{T}}\right|_{\textrm{fact}} = \left[\frac{m_{B_d}^2-m_\pi^2}{m_{B_s}^2-m_K^2}\right] \left[\frac{F_0^{B_d\pi}(m_\pi^2)}{F_0^{B_sK}(m_\pi^2)}\right] = 0.85_{-0.13}^{+0.07} \ ,
\end{equation}
where we have again used $F_0^{B_sK}(0)/F_0^{B_d\pi}(0)=1.15^{+0.17}_{-0.09}$ from LCSR
calculations \cite{Dub08}, which agrees with the analysis of Ref.~\cite{Kho04}. Finally, we use the unprimed equivalent of Eq.~\eqref{eq:penguinSU3relation}, which leads to
\begin{equation} \label{eq:ddtildeis1}
\frac{1-d\euler^{\imaginary\theta}\euler^{-\imaginary\gamma}}{1-\tilde{d}\euler^{\imaginary\tilde{\theta}}\euler^{-\imaginary\gamma}} = \frac{1-\tilde{d}\euler^{\imaginary\tilde{\theta}}\euler^{-\imaginary\gamma}/\zeta}{1-\tilde{d}\euler^{\imaginary\tilde{\theta}}\euler^{-\imaginary\gamma}} \approx 1
\end{equation}
for $\zeta\sim 1$.

\begin{figure}
	\centering
	\includegraphics[scale=0.8]{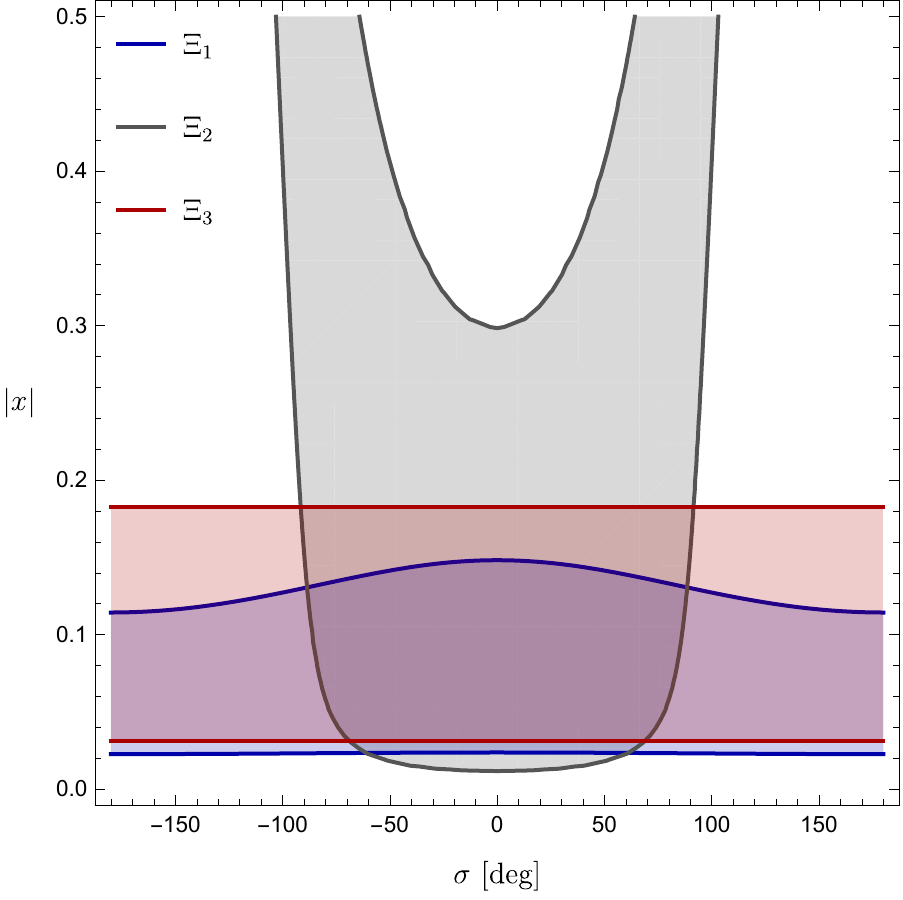}
	\caption{Constraints on $|
x|$ and the phase $\sigma$ from $\Xi_2$. The horizontal lines are conservative bounds from $\Xi_1$ and $\Xi_3$, as explained in the text.  }
	\label{fig:xversussigma}
\end{figure}

From the current experimental data, we extract
\begin{equation}
\Xi_2\approx |1+x|^2\rho^2 \myexp{exp} 0.90 \pm 0.10 \ ,
\end{equation}
which yields
\begin{equation}\label{eq:oneplusx2}
	|1+x|= 1.12 ^{+0.18}_{-0.11} \ .
\end{equation}
The large uncertainty comes from the form factors, and actually makes this ratio less powerful. 
However, we can nevertheless use it to constrain the phase of $x$ introduced in Eq.~(\ref{eq:x}), 
as shown in Fig.~\ref{fig:xversussigma}. 

Finally, we consider
\begin{equation}
\Xi_3 = \left|\frac{\hat{\mathcal{C}}}{\tilde{\mathcal{C}}}\right|^2\left[\frac{1-2\hat{d}\cos\hat{\theta}\cos\gamma +\hat{d}^2 }{1-2\tilde{d}\cos\tilde{\theta}\cos\gamma+\tilde{d}^2}\right] 
\end{equation}
with
\begin{equation}
\left|\frac{\hat{\mathcal{C}}}{\tilde{\mathcal{C}}}\right|^2 = \eta^2\rho^2|x|^2 \ .
\end{equation}
Using the branching ratios in Table~\ref{tab:BtohhDecayModes} gives 
\begin{equation}
\Xi_3 \simeq \eta^2\rho^2|x|^2\left[\frac{1-2\hat{d}\cos\hat{\theta}\cos\gamma +\hat{d}^2 }{1-2\tilde{d}\cos\tilde{\theta}\cos\gamma+\tilde{d}^2}\right] \myexp{exp} 0.014 \pm 0.003 \ .
\end{equation}
If we use $\tilde{d}$ and $\tilde{\theta}$ as determined in Subsection~\ref{sec:tilde} and $\hat{d}$ 
from Eq.~\eqref{eq:dHat_vs_thetaHat}, we may calculate $|x|$ as a function of $\hat{\theta}$, as 
shown in Fig.~\ref{fig:xcons}(b). The bound on $|x|$ varies between $0.03$ and $0.18$, which is 
consistent with the determination shown in Fig.~\ref{fig:xcons}(a).

For obtaining a complete picture, we have added the constraints from Fig.~\ref{fig:xcons} to Fig.~\ref{fig:xversussigma}. Lacking information about the phase $\hat{\theta}$, we have conservatively used the upper bound at $\hat{\theta} = 0^\circ$ and the lower bound at $\hat{\theta} = 180^\circ$ from Fig.~\ref{fig:xcons}, since the values of $|x|/|1+x|$ and $|x|$ are largest and smallest there, respectively.

Unfortunately, the phase $\sigma$ is only poorly constrained. More interesting is the current constraint of 
$|x|<0.2$ from $\Xi_3$. Combining all constraints gives 
\begin{equation}\label{eq:oneplusx}
|1+x| = 1.1 \pm 0.1 \ .
\end{equation} 
We further discuss this parameter and its implications for the ratio $\Xi_x$ in Subsection~\ref{sec:impl}.  


\boldmath
\subsection{Indirect Information on $r_{PA}'$} \label{sec:rpa}
\unboldmath
At the moment, only the ratios $\Xi_i'$ defined in Table~\ref{tab:Xi} can be used to study $r_{PA}'$. 
We may simplify the following discussion by assuming that the quantity $\epsilon$, which enters the $\Xi_i'$, is small in comparison with the penguin parameters. 

\begin{figure}
	\centering
	\includegraphics[width=0.55\textwidth]{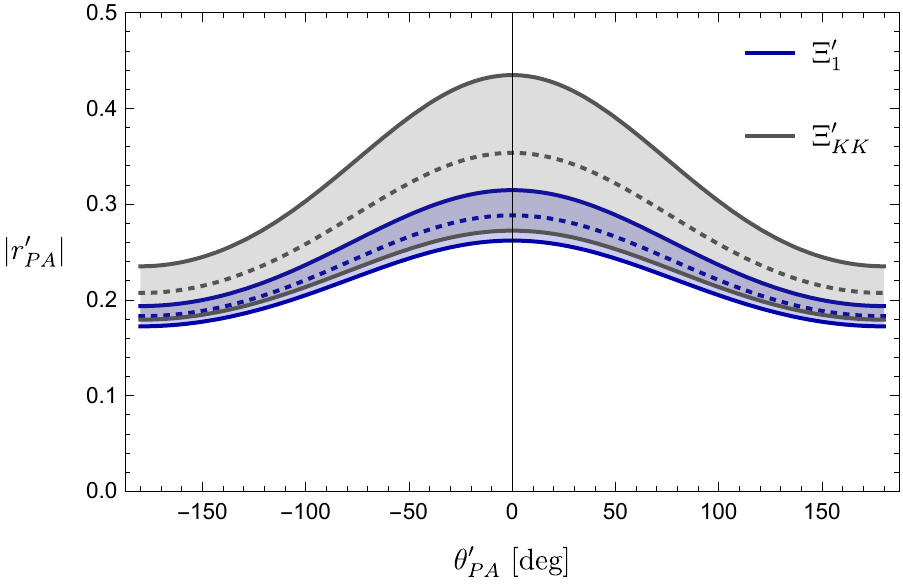}
	\caption{Constraints on $|r_{PA}'|$ and the phase $\theta_{PA}'$ from \Bstopipi, \BstoKK and \Bstopipi, \BstoKantiK with $1\sigma$ error bands.}
	\label{fig:rPA_ComplexPlane}
\end{figure}

Let us first consider
\begin{equation}
\Xi_1'= \left|\frac{\hat{\mathcal{C}}^\prime}{\mathcal{C}^\prime}\right|^2\left[\frac{\epsilon^2+2\epsilon\hat{d}^\prime\cos\hat{\theta}^\prime\cos\gamma+\hat{d}^{\prime 2}}{\epsilon^2+2\epsilon d^\prime\cos\theta^\prime\cos\gamma+d^{\prime 2}}\right] \approx \left|\frac{\hat{\mathcal{C}}^\prime}{\mathcal{C}^\prime}\right|^2\left(\frac{\hat{d}^\prime}{d^\prime}\right)^2 \ ,
\end{equation}
where we have ignored terms of $\mathcal{O}(\epsilon)$. We parametrize the penguin-annihilation amplitudes through
\begin{equation}
\eta' \equiv \left|\frac{\hat{PA}^{(ct) \prime}}{PA^{(ct) \prime}}\right| \sim \left(\frac{f_\pi}{f_K}\right)^2 = 0.703 \pm 0.003 \ ,
\end{equation}
where we have used an approximation similar to Eq.~\eqref{eq:etadefval}. Note that in this approximation $\eta' = 1/\eta$. We find
\begin{equation}
\Xi_1' = \left|\frac{r_\text{PA}^\prime}{1+r_\text{PA}^\prime}\right|^2\eta^{\prime 2} \myexp{exp} 0.025\pm0.004,
\end{equation}
which leads to a contour in the complex plane of
\begin{equation}
r_{PA}' \equiv |r_{PA}'| e^{\imaginary\theta_{PA}'} \ ,
\end{equation} 
as shown in Fig.~\ref{fig:rPA_ComplexPlane}.

In addition, we can consider
\begin{equation}
\Xi_{KK}^\prime = \Xi(\Bstopipi, \BstoKantiK) \sim \left|\frac{\hat{\mathcal{C}}^\prime}{\mathcal{C}_{KK}^\prime}\right|^2 \hat{d}^{'2} \ ,
\end{equation}
where we have neglected the penguin contribution $d_{KK}'$ from $\BstoKantiK$ since it is suppressed by $\epsilon$. Using the experimental branching ratio for \BstoKantiK given in Eq.~\eqref{eq:BRBstoKantiK} yields
\begin{equation}
 \Xi_{KK}^\prime \sim \left(\frac{f_\pi}{f_K}\right)^4\left|\frac{r_\text{PA}^\prime}{1+r_\text{PA}^\prime}\right|^2 \myexp{exp} 0.034\pm 0.011 \ .
\end{equation}
The constraint from this ratio is in perfect agreement with that obtained from $\Xi_1'$, as illustrated 
in Fig.~\ref{fig:rPA_ComplexPlane}. This shows once again the importance of $\BstoKantiK$ and
the potential of future measurements of this decay.
\begin{figure}[t]
	\centering
	\includegraphics[scale=0.8]{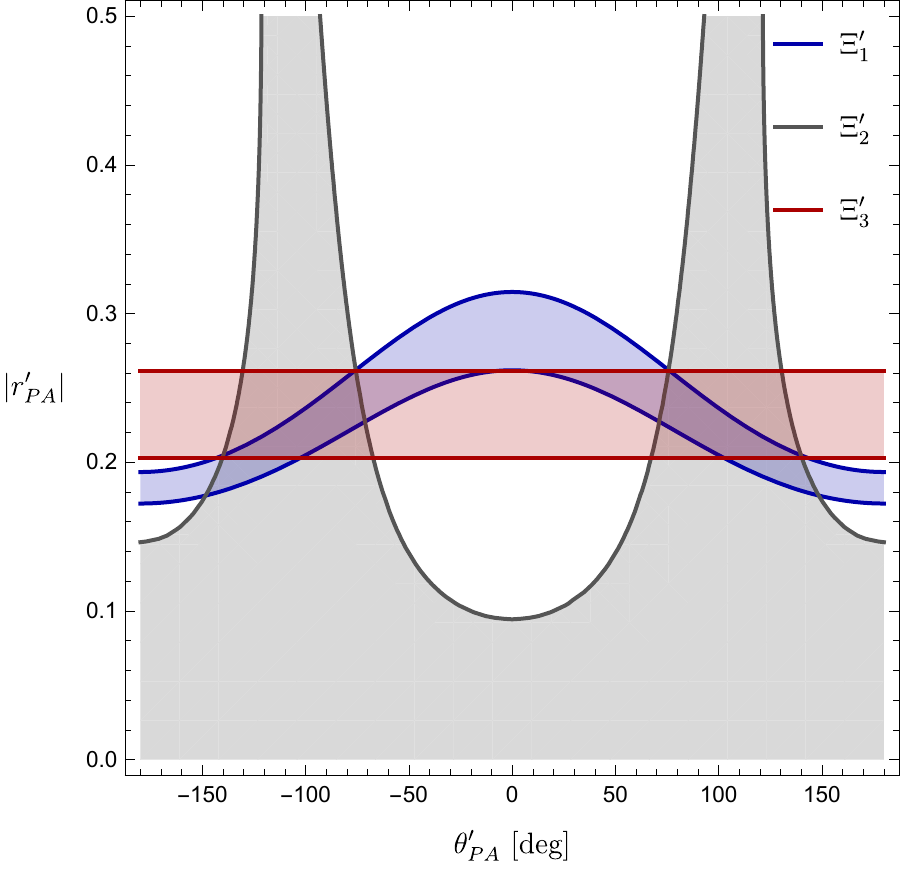}
	\caption{Bounds on $|r_{PA}'|$ and the phase $\theta_{PA}'$ using $\Xi_1', \Xi_2'$ and $\Xi_3'$. }
	\label{fig:asfig}
\end{figure}
Next, we consider the ratio
\begin{equation}
\Xi_2'= \left|\frac{\mathcal{C}'}{\tilde{\mathcal{C}}'}\right|^2\left\langle\left|\frac{1+d'/\epsilon\euler^{\imaginary\theta'}\euler^{-\imaginary\gamma}}{1+\tilde{d}'/\epsilon\euler^{\imaginary\tilde{\theta}'}\euler^{-\imaginary\gamma}}\right|^2\right\rangle
\end{equation}
with
\begin{equation} \label{eq:CPrimeOverTildeCPrime}
\left|\frac{\mathcal{C}^\prime}{\tilde{\mathcal{C}}^\prime}\right|^2 = |1+ x^\prime|^2\rho^{\prime 2} \ ,
\end{equation}
where $\rho'$ is the equivalent of $\rho$ defined in Eq.~\eqref{eq:rhodef}. Making the same approximations for $\rho'$ as for $\rho$, we find 
\begin{equation}
\rho' = 1/\rho = 1.18^{+0.17}_{-0.09}. 
\end{equation}
Neglecting again $\mathcal{O}(\epsilon)$ terms gives
\begin{equation}
\frac{1+\frac{1}{\epsilon}d^\prime\euler^{\imaginary\theta^\prime}\euler^{-\imaginary\gamma}}{1+\frac{1}{\epsilon}\tilde{d}^\prime\euler^{\imaginary\tilde{\theta}^\prime}\euler^{-\imaginary\gamma}} = \frac{\epsilon+\tilde{d}^\prime\euler^{\imaginary\tilde{\theta}^\prime}\euler^{-\imaginary\gamma}(\zeta')^{-1}}{\epsilon+\tilde{d}^\prime\euler^{\imaginary\tilde{\theta}^\prime}\euler^{-\imaginary\gamma}} \approx \frac{1+r_{PA}^\prime}{1+x^\prime} \ .
\end{equation}
Using the experimental branching ratios in Table~\ref{tab:BtohhDecayModes}, we obtain
\begin{equation}
\Xi_2' \approx {\rho'}^{2}|1+r_{PA}^\prime|^2  \myexp{exp} 1.41\pm0.10 \ ,
\end{equation}
which leads to
\begin{equation}\label{eq:oneplusrpa2}
|1+r_{PA}^\prime|= 1.01_{-0.15}^{+0.09} \ .
\end{equation}
We write $r_{PA}' \equiv |r_{PA}'| e^{\theta_{PA}'}$ and give the constraints from $\Xi_2'$ in Fig.~\ref{fig:asfig}. 
In analogy to $\Xi_2$, we observe that the constraint for $|1+r_{PA}^\prime|$ suffers from large uncertainties due to the required form-factor information. Consequently, the ratios $\Xi_2$ and $\Xi_2'$ are at the moment only useful to constrain the phases of $x$ and $r_{PA}'$, respectively. Information on their actual magnitude is more stringently constrained by the ratios $\Xi_1^{(\prime)}$ and $\Xi_3^{(\prime)}$. 

Finally, we have the ratio
\begin{equation}
\Xi_3' = \left|\frac{\hat{\mathcal{C}}'}{\tilde{\mathcal{C}}'}\right|^2\left[\frac{\epsilon^2+2\epsilon\hat{d}^\prime\cos\hat{\theta}^\prime\cos\gamma+\hat{d}^{\prime 2}}{\epsilon^2+2\epsilon\tilde{d}^\prime\cos\tilde{\theta}^\prime\cos\gamma+\tilde{d}^{\prime 2}}\right] \approx \left|\frac{\hat{\mathcal{C}}'}{\tilde{\mathcal{C}}'}\right|^2 \left(\frac{\hat{d}'}{\tilde{d}'}\right)^2 \ ,
\end{equation}
where we neglect once again terms of $\mathcal{O}(\epsilon)$. Defining 
\begin{equation}
\tilde{\rho}' \equiv \left|\frac{P^{(ct) \prime}}{\tilde{P}^{(ct) \prime}}\right|,
\end{equation}
and making the approximation $\tilde{\rho}' \approx \rho'$ gives
\begin{equation}
\Xi_3' \approx \tilde{\rho}^{\prime 2} \eta^{\prime 2} |r_{PA}'|^2 \myexp{exp} 0.035\pm 0.004 \ ,
\end{equation}
yielding
\begin{equation}
|r_{PA}'| = 0.23_{-0.04}^{+0.02} \ .
\end{equation}
In Fig.~\ref{fig:asfig}, we show the contour fixed through this ratio in the complex plane.

We have also added the constraint from $\Xi_1^\prime$ to Fig.~\ref{fig:asfig}, and conclude that 
the current data favour slightly the regions around $\theta_{PA}' = \pm 100^\circ$, while the 
constraint for $|r_{PA}'|$ is governed by the $\Xi_3'$ ratio.


\boldmath
\subsection{Determination of $\Xi_x$} \label{sec:impl}
\unboldmath

\begin{figure}[t]
	\centering
	\includegraphics[scale=1.0]{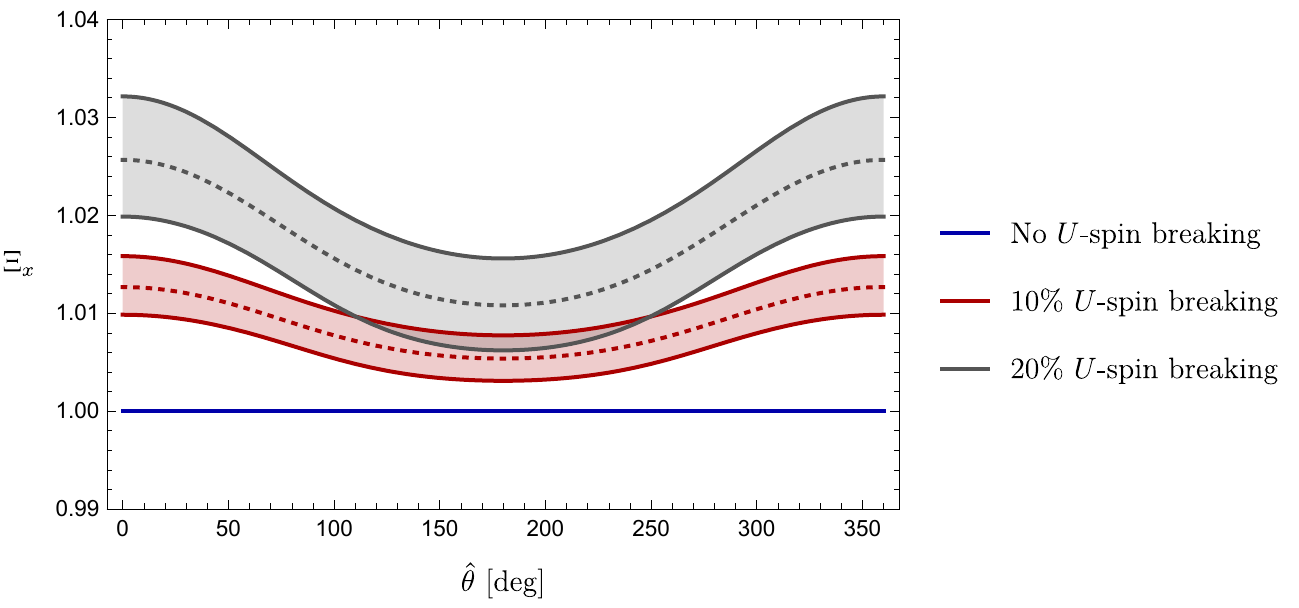}
 	\caption{The ratio $\Xi_x$ as a function of $\hat{\theta}$ for different $U$-spin-breaking effects.}
	\label{fig:xiXthetahatComp}
\end{figure}

\begin{figure}[t]
	\centering
	\includegraphics[width=0.84\textwidth]{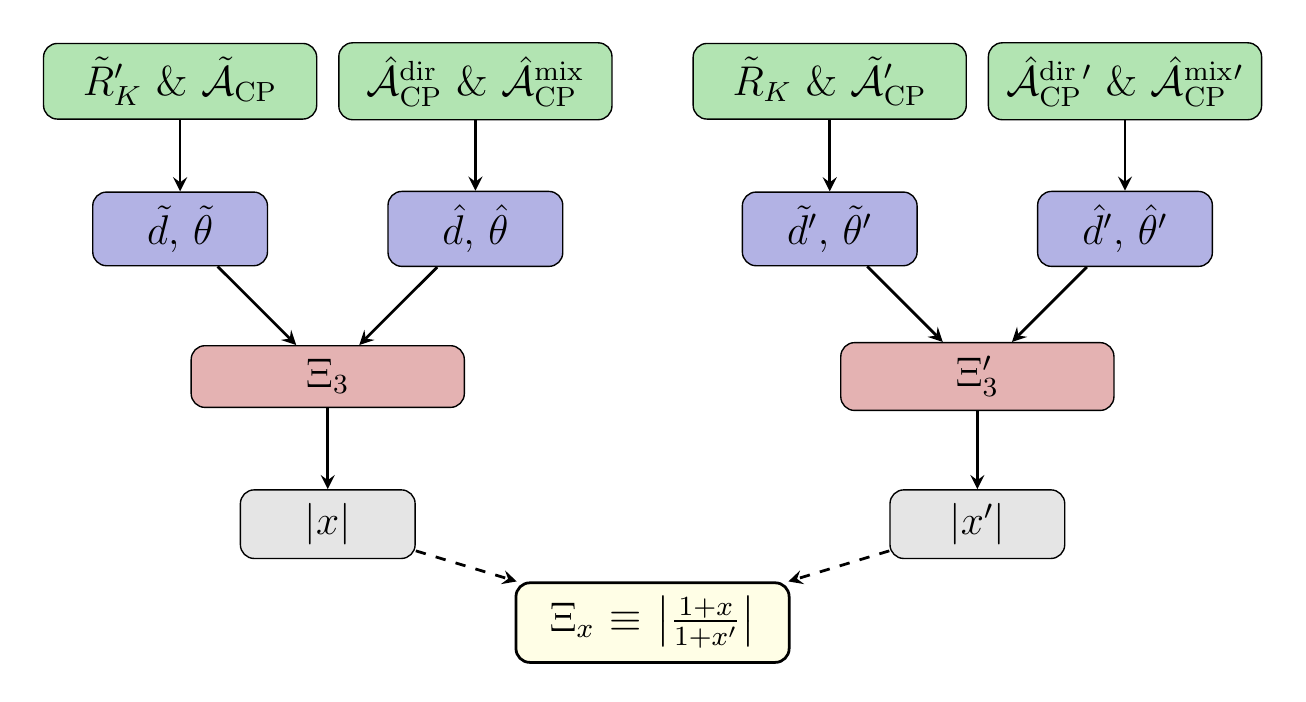}
	\caption{Strategy to determine $\Xi_x$. The $\tilde{\mathcal{A}}_\text{CP}'$, $\tilde{\mathcal{A}}_\text{CP}$ denote the direct CP asymmetries in $\BdtoKpi$ and \BstoKpi, respectively, and $\hat{\mathcal{A}}_\text{CP}^\text{dir}$, $\hat{\mathcal{A}}_\text{CP}^\text{mix}$ and ${\hat{\mathcal{A}}_\text{CP}^{\text{dir}}}{}^\prime$,${\hat{\mathcal{A}}_\text{CP}^{\text{mix}}}{}^\prime$ are the CP asymmetries of \BdtoKK and \Bstopipi, respectively. }
	\label{fig:flowchartXix}
\end{figure}

The previous studies allow us to determine $\Xi_x$ defined in Eq.~\eqref{eq:defrat}
with the help of current data. The ratios $\Xi_1, \Xi_2$ and $\Xi_3$ provide information on $|x|$ and its phase $\sigma$. Independent information on $x'$ is currently not available, but can be obtained from future measurements of CP violation in \Bstopipi. We consider
\begin{equation}
\Xi_x = \left|\frac{1+x}{1+x'}\right| = 1+x\xi_x+\mathcal{O}(x^2),
\end{equation}
where $\xi_x$ is an $SU(3)$-breaking parameter defined through $x' = x(1-\xi_x)$. An important advantage of our strategy is that the exchange and penguin-annihilation topologies only contribute through the ratio $\Xi_x$. Since $x$ is a small quantity, $\Xi_x$ is very robust with respect to $U$-spin-breaking effects. This feature is illustrated in Fig.~\ref{fig:xiXthetahatComp}, which shows the ratio $\Xi_x$ as a function of $\hat{\theta}$ for different $U$-spin-breaking effects. Allowing for $20\%$ $U$-spin-breaking only gives an uncertainty of $\mathcal{O}(4 \%)$ for $\Xi_x$. However, especially around $\hat{\theta}=180^\circ$, which is actually the expected region, the effect can be much smaller. Future determinations of the CP asymmetries in the \BdtoKK, \Bstopipi system can pinpoint these effects further, as illustrated in Fig.~\ref{fig:flowchartXix}. The \BdtoKK, \Bstopipi CP asymmetries allow a determination of $\hat{d}^{(\prime)}$, $\hat{\theta}^{(\prime)}$, while the semileptonic ratios $\tilde{R}_K'$ and $\tilde{R}_K$ would allow an independent determination of $\tilde{d}'$, $\tilde{\theta}'$. Finally, $|x^{(\prime)}|$ can be determined using $\Xi_3^{(\prime)}$ and $\tilde{d}^{(\prime)}$, $\tilde{\theta}^{(\prime)}$. This would give a clean determination of both $|x|$ and $|x'|$ independently, allowing a direct determination of $\Xi_x$, without any $U$-spin assumptions.

\begin{figure}[t]
	\centering
	\subfloat[\BdtoKK]{\label{fig:corrScatterScenBdKK} \includegraphics[width=0.44\textwidth]{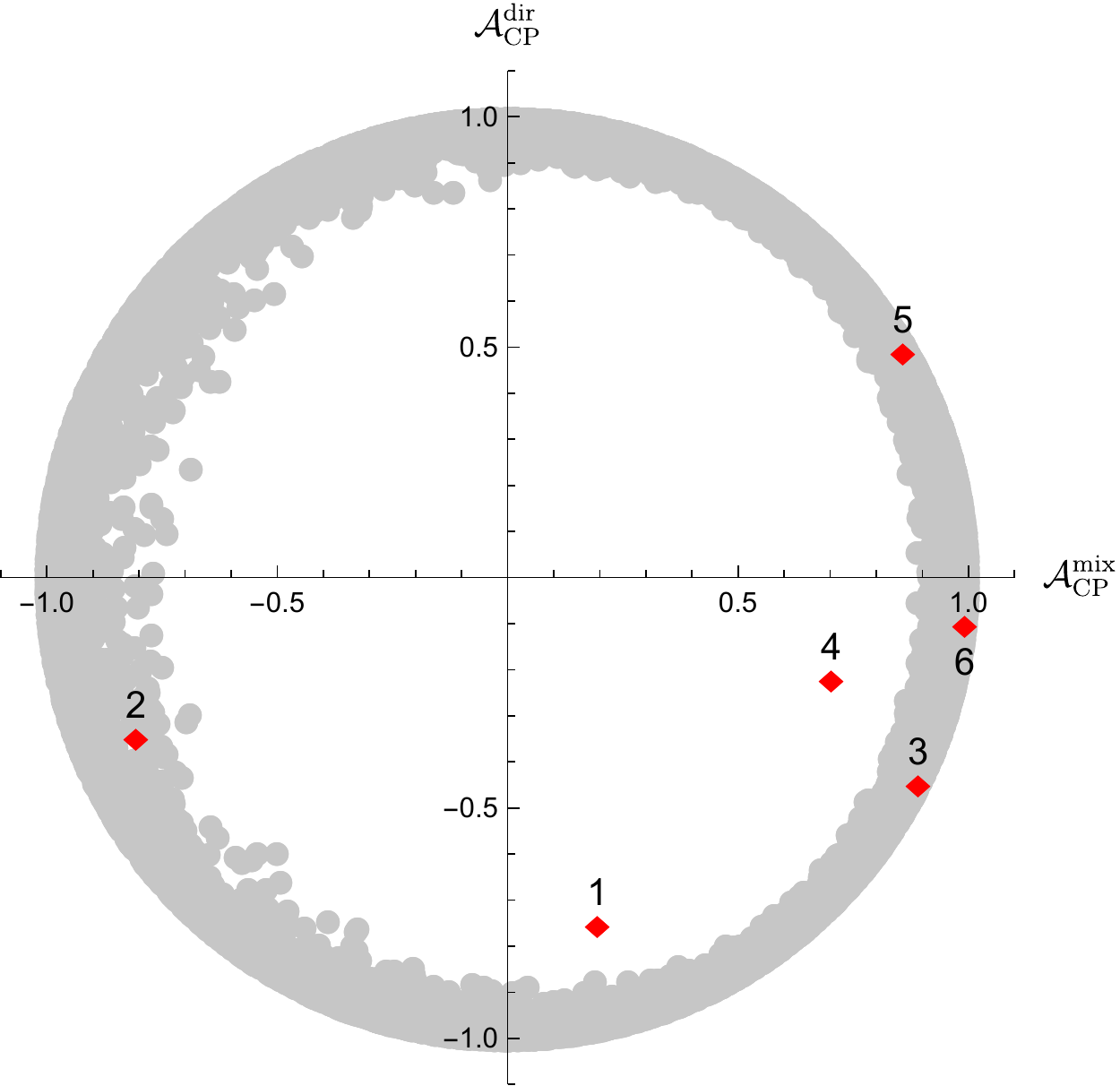}}
	\subfloat[\Bstopipi]{\label{fig:corrScatterScenBspipi} \includegraphics[width=0.44\textwidth]{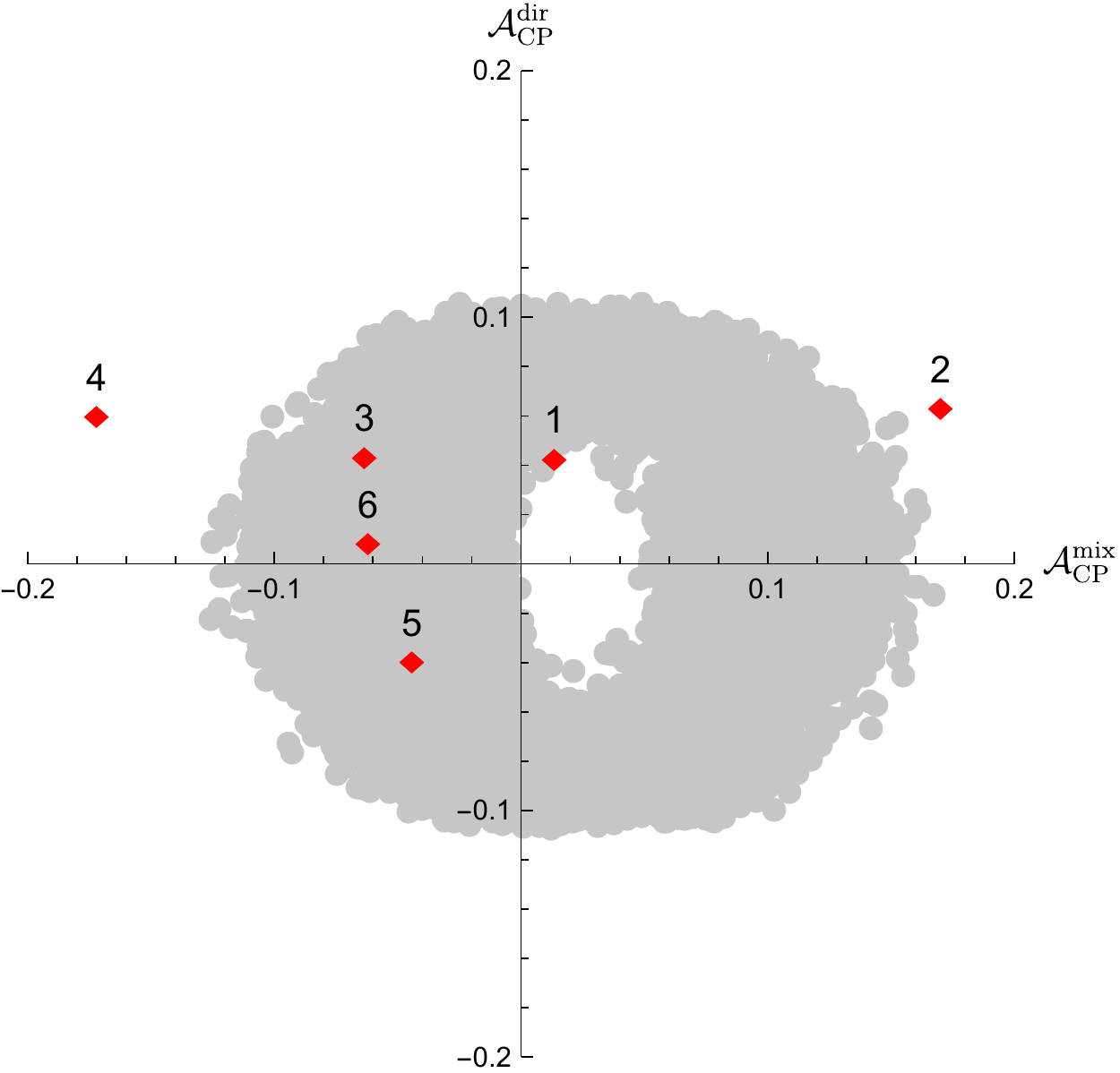}}
	\caption{Correlation between the direct and mixing-induced CP asymmetries of \BdtoKK and of \Bstopipi as in Fig.~\ref{fig:correlations_CPasymmetries_BdKK_BsPiPi}, with the different scenarios indicated by diamonds.}
	\label{fig:cor2}
\end{figure}

We further illustrate the use of the \BdtoKK, \Bstopipi CP asymmetries by discussing six possible future scenarios, given in Table~\ref{tab:BdKKBspipiScenarios}. The specific scenarios are also indicated in Fig.~\ref{fig:cor2}, and we assume the same relative uncertainties as those of the current measurements of the \Bdtopipi, \BstoKK CP asymmetries.

\begin{table}
	\centering
	\begin{tabular}{c|c|c|c|c|c}
		 & $\hat{\mathcal{A}}_\text{CP}^\text{dir}$ & $\hat{\mathcal{A}}_\text{CP}^\text{mix}$ & $\hat{d}$ & $\hat{\theta}$ [deg] & \\
		No. & $\hat{\mathcal{A}}_\text{CP}^{\text{dir} \prime}$ & $\hat{\mathcal{A}}_\text{CP}^{\text{mix} \prime}$ & $\hat{d}'$ & $\hat{\theta}'$ [deg] & $|\hat{\mathcal{C}}/\hat{\mathcal{C}}'|$ \\
		\hline\hline
		 & $-0.75 \pm 0.12$ & $0.20 \pm 0.02$ & $2.0 \pm 0.4$ & $60.0 \pm 7.6$ & \\
		$1$ & $0.043 \pm 0.034$ & $0.014 \pm 0.006$ & $2.0 \pm 1.2$ & $60.0 \pm 22.6$ & $1.44 \pm 0.87$ \\
		\hline
		 & $-0.35 \pm 0.06$ & $-0.81 \pm 0.07$ & $0.50 \pm 0.07$ & $20.0 \pm 3.6$ & \\
		$2$ & $0.064 \pm 0.050$ & $0.17 \pm 0.07$ & $0.50 \pm 0.20$ & $20.0 \pm 16.3$ & $0.80 \pm 0.30$ \\
		\hline
		 & $-0.45 \pm 0.07$ & $0.89 \pm 0.08$ & $[0.9,3.1]$ & $[121,149]$ & \\
		$3$ & $0.044 \pm 0.034$ & $-0.063 \pm 0.027$ & $[1.0,2.8]$ & $[114,170]$ & $[0.41,2.85]$ \\
		\hline
		 & $-0.22 \pm 0.04$ & $0.70 \pm 0.06$ & $0.60 \pm 0.09$ & $160.0 \pm 3.3$ & \\
		$4$ & $0.060 \pm 0.047$ & $-0.17 \pm 0.07$ & $0.60 \pm 0.25$ & $160.0 \pm 16.9$ & $0.66 \pm 0.28$ \\
		\hline
		 & $0.49 \pm 0.08$ & $0.86 \pm 0.08$ & $[0.9,3.1]$ & $[214,244]$ & \\
		$5$ & $-0.039 \pm 0.031$ & $-0.044 \pm 0.019$ & $[1.3,4.2]$ & $[194,255]$ & $[0.54,4.33]$ \\
		\hline
		 & $-0.10 \pm 0.02$ & $0.99 \pm 0.09$ & $[1.0,4.4]$ & $[163,173]$ & \\
		 $6$ & $0.0089 \pm 0.0070$ & $-0.062 \pm 0.027$ & $[1.3,4.3]$ & $[154,178]$ & $[0.39,3.91]$
	\end{tabular}
	\caption{Overview of the different scenarios for the CP-violating observables of the \BdtoKK 
	and \Bstopipi decays.}\label{tab:BdKKBspipiScenarios}
\end{table}

\begin{figure}
	\centering
	\subfloat[Scenario 1]{\label{fig:scenhat1} \includegraphics[width=0.4\textwidth]{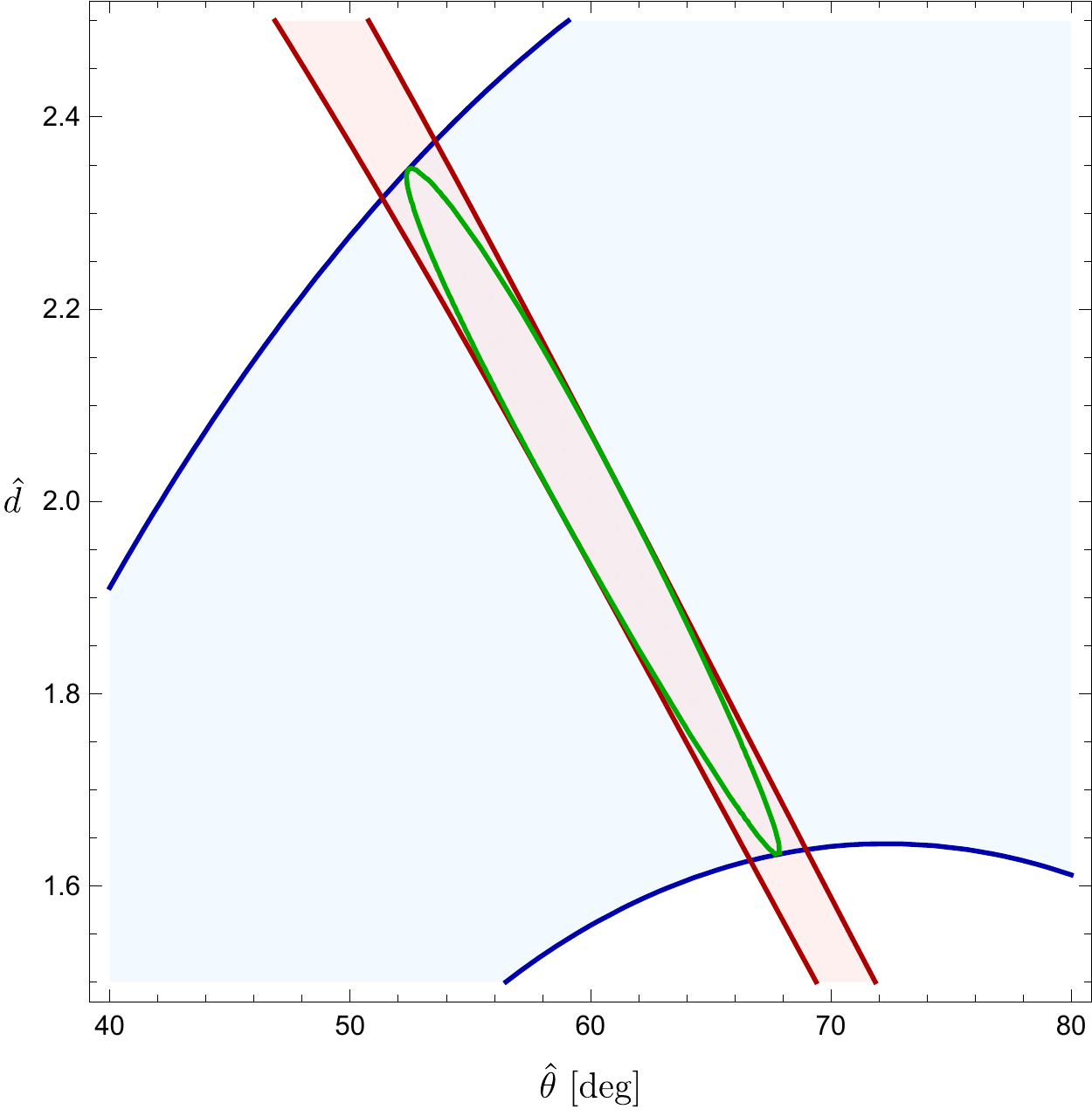}}
	\subfloat[Scenario 2]{\label{fig:scenhat2} \includegraphics[width=0.4\textwidth]{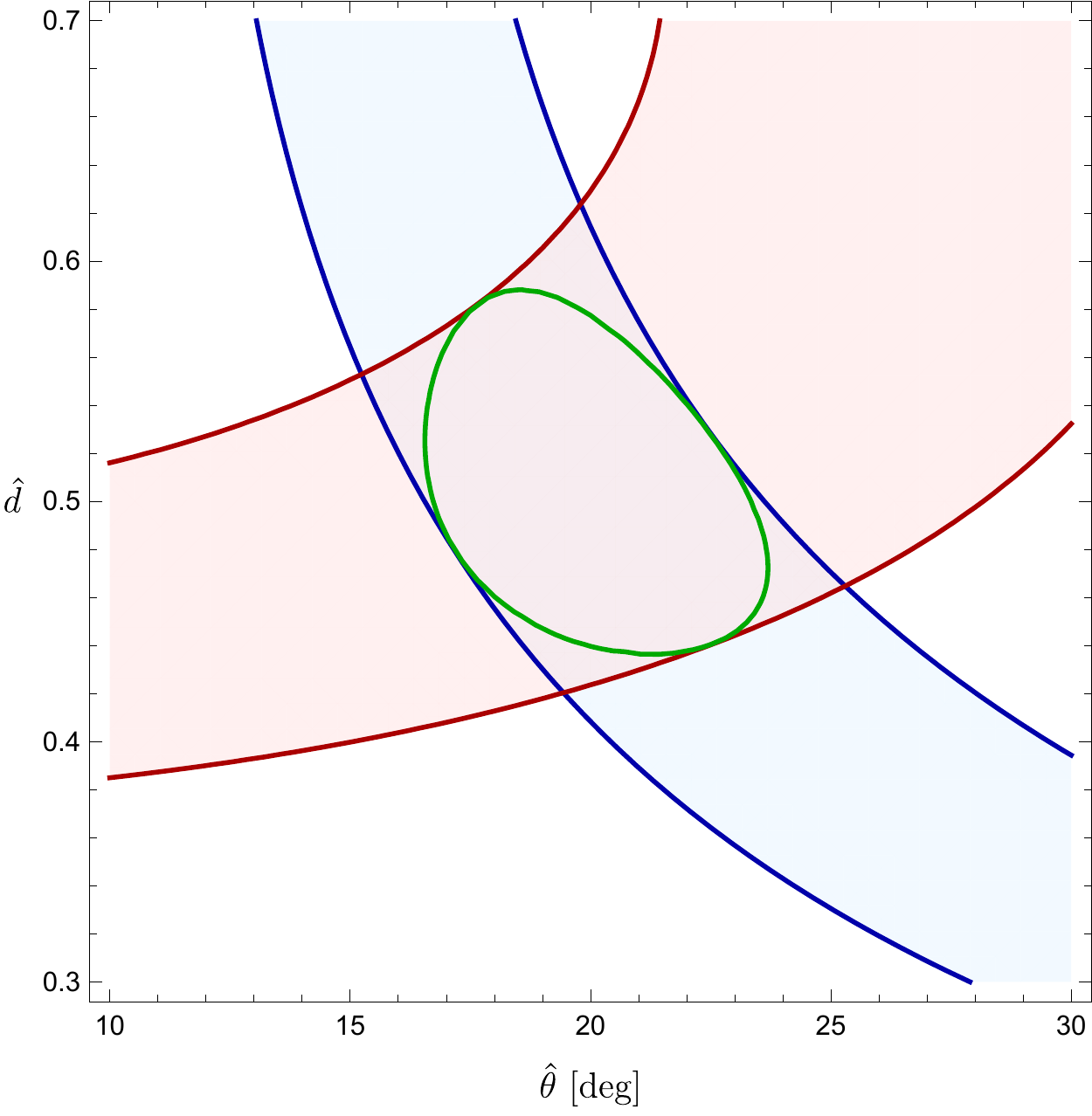}}\\
	\subfloat[Scenario 3]{\label{fig:scenhat3} \includegraphics[width=0.4\textwidth]{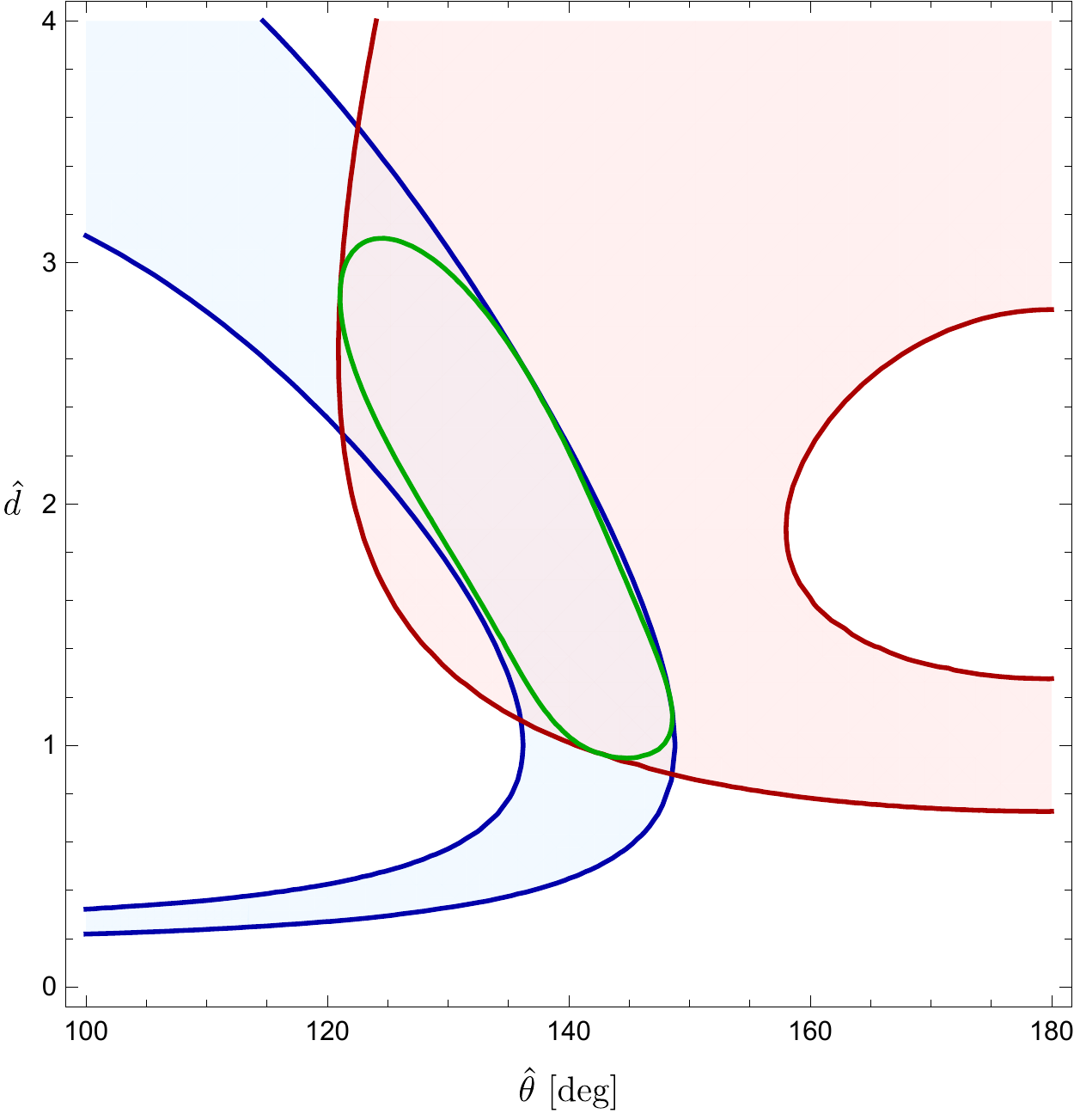}}
	\subfloat[Scenario 4]{\label{fig:scenhat4} \includegraphics[width=0.4\textwidth]{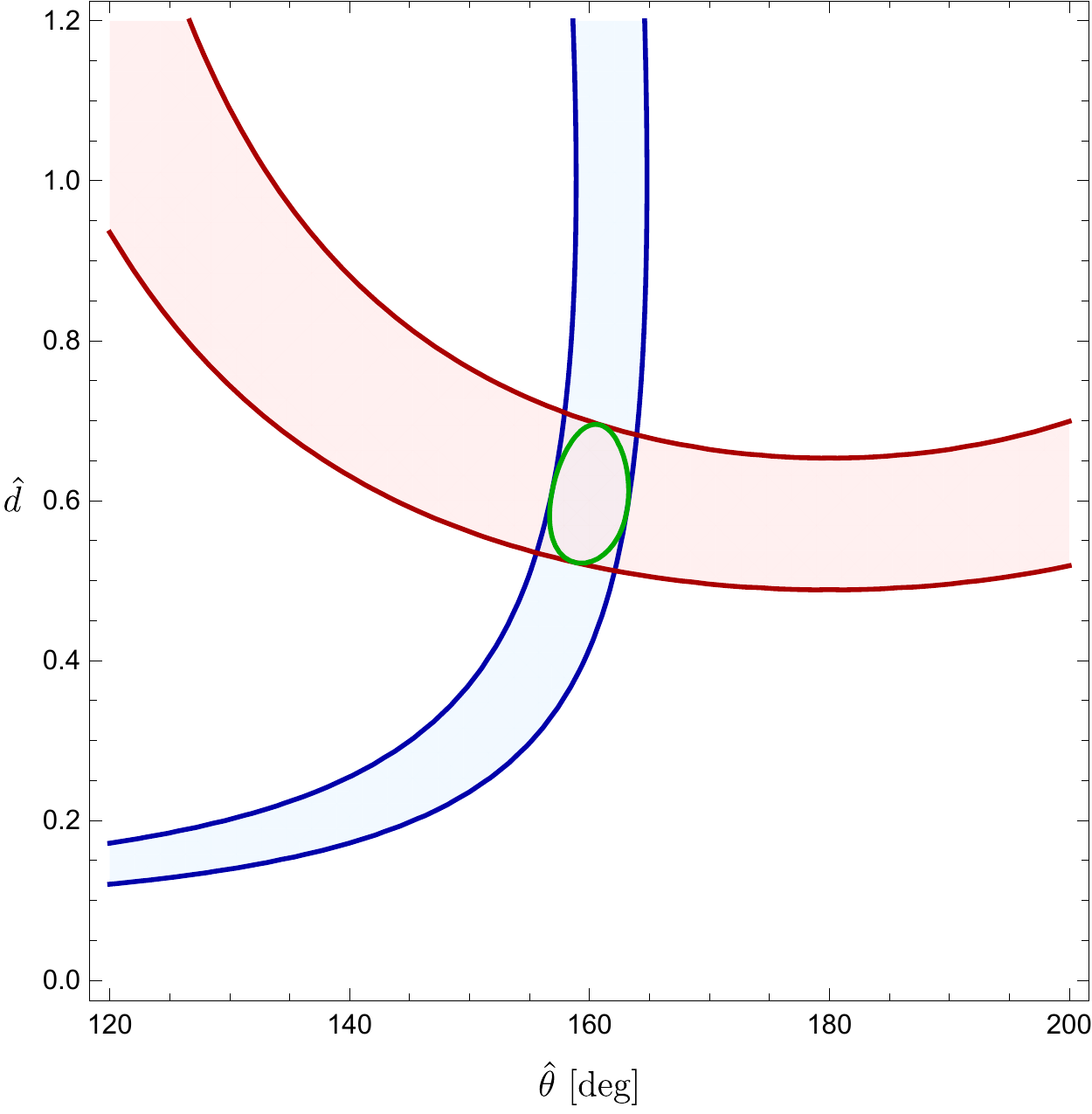}}\\
	\subfloat[Scenario 5]{\label{fig:scenhat5} \includegraphics[width=0.4\textwidth]{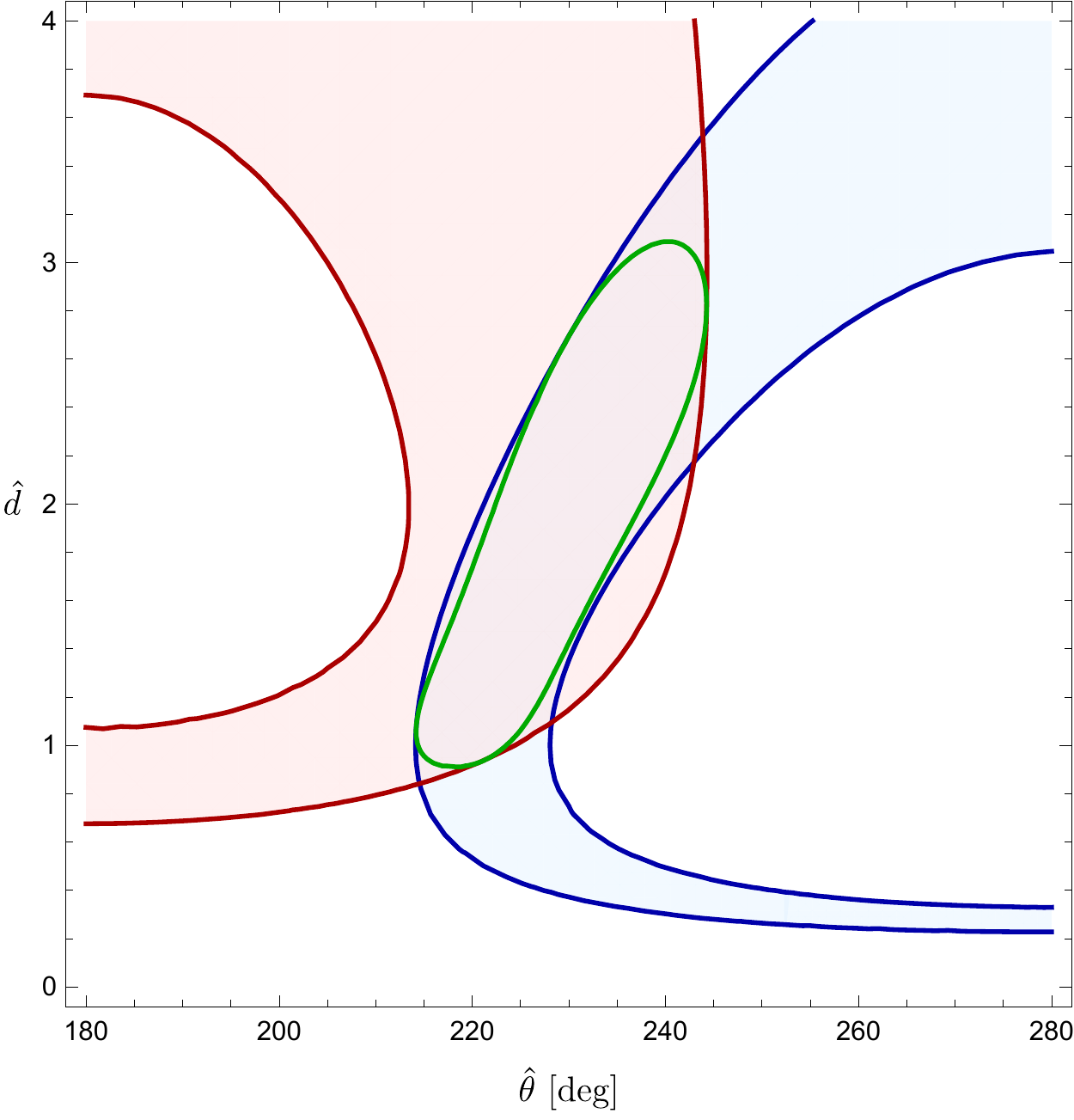}}
	\subfloat[Scenario 6]{\label{fig:scenhat6} \includegraphics[width=0.4\textwidth]{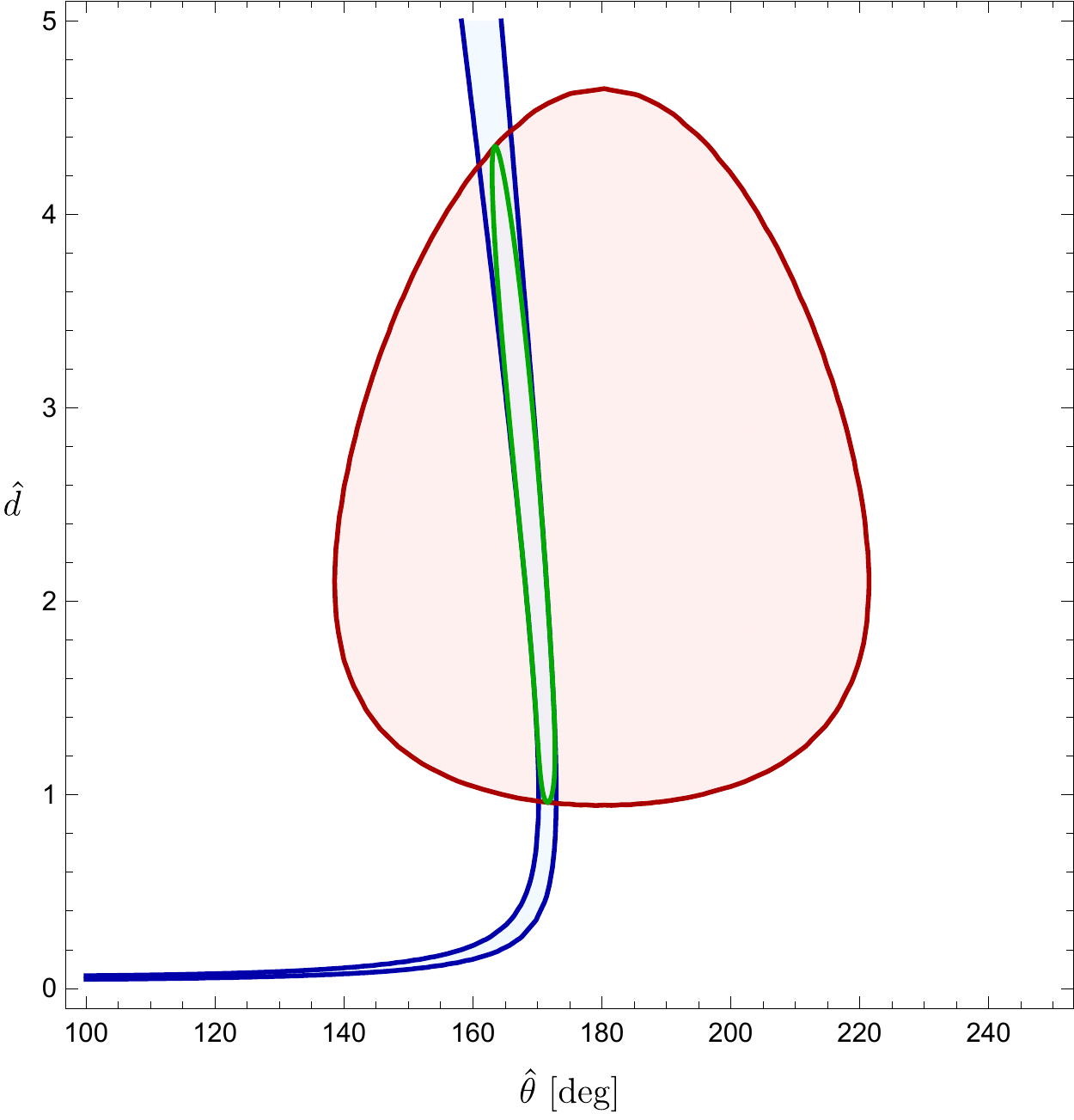}}
\caption{Determination of $\hat{d}$ and $\hat{\theta}$ from the CP-violating observables of \BdtoKK. The blue and red contours follow from the direct and mixing-induced CP asymmetries, respectively. The $1\sigma$ contours resulting from a  $\chi^2$ fit are shown in green.}
\label{fig:scenhat}
\end{figure}

For the different scenarios, $\hat{d}$ and $\hat{\theta}$ are extracted from the \BdtoKK CP asymmetries, using $\gamma = (70 \pm 1)^\circ$ and $\phi_d= (43.2\pm 0.6)^\circ$ as before. This gives two solutions, where
we discard the one which leads to anomalously large $U$-spin-breaking effects. The results are collected in Table~\ref{tab:BdKKBspipiScenarios}. For scenarios 1, 2 and 4, the analytic expression is used to obtain the uncertainty. However, for scenarios $3, 5$ and $6$, the $1\sigma$ ranges are obtained from a $\chi^2$ fit to take into account the correlated erros (see Fig.~\ref{fig:scenhat}). The different values that were obtained are also indicated in Fig.~\ref{fig:dHat_vs_thetaHat}. In addition, the parameters $\hat{d}'$ and $\hat{\theta}'$ 
are determined from the CP asymmetries of the \Bstopipi channel, using the central value of the current 
PDG average $\phi_s=-(0.68 \pm 0.5)^\circ$ with an error expected for the era of Belle II and the LHCb upgrade.

Some of the obtained  parameters $(\hat{d}, \hat{\theta})$ in Table~\ref{tab:BdKKBspipiScenarios} have 
large uncertainties. In particular scenarios 5 and 6 fall into this category as the mixing-induced CP asymmetries
are close to 1. Since the CP asymmetries in \BdtoKK saturate the relation in Eq.~\eqref{eq:CPconstraint}, 
the corresponding direct CP asymmetries are constrained to values around $0$. This feature 
reduces significantly the sensitivity to $(\hat{d}, \hat{\theta})$. The various scenarios are also illustrated in Fig.~\ref{fig:scenhat}, which shows the contours in the $\hat{d}$--$\hat{\theta}$ plane
following from the direct (blue) and mixing-induced (red)  CP asymmetries of the \BdtoKK channel,
along with the $1\sigma$ contour from a $\chi^2$ fit of these two observables.

We notice that the amplitude ratio $|\hat{\mathcal{C}}/\hat{\mathcal{C}}'|$ in Eq.~\eqref{eq:Kexp} can 
unfortunately only be determined with limited precision in our scenarios. The results are 
summarized in Table~\ref{tab:BdKKBspipiScenarios}, where the ranges correspond to the allowed 
regions of the penguin parameters.

Finally, implementing the strategy illustrated in Fig.~\ref{fig:flowchartXix}, we can determine $|x|$ and $|x'|$. 
Based on the definition in Eq.~\eqref{eq:CdHat}, we expect the strong phases $\hat{\theta}^{(\prime)}$ 
to take values around $180^\circ$. Let us therefore consider scenario 6, where in addition $\hat{d}$ is close to the prediction from Eq.~\eqref{eq:dHat_vs_thetaHat}, and scenario 4, where $\hat{d}$ is closer 
to the value of $d$.

With the input from scenario 6 $(S_6)$, we find
\begin{equation} \label{eq:xoneplusxandxScenario}
	\left|\frac{x}{1+x}\right|_{S_6} = [0.024,0.071], \qquad |x|_{S_6}= [0.031,0.093],
\end{equation}
where the range corresponds to the allowed region of $\hat{d}$ and $\hat{\theta}$. If we assume scenario 4 $(S_4)$, we find
\begin{equation} \label{eq:xoneplusxandxScenario4}
	\left|\frac{x}{1+x}\right|_{S_4} = 0.087 \pm 0.009, \qquad |x|_{S_4} = 0.11 \pm 0.02,
\end{equation}
which has remarkably  small uncertainties. Most important, even though the uncertainty for the extracted
value of $\hat{d}$ might be significant, the impact on the determination of $|x|$ is small. 

Interestingly, we can now also determine $|x'|$ with the help of $\Xi_3'$. At the moment, we cannot
determine $\tilde{d}'$ and $\tilde{\theta}'$ in an independent way. However, as 
discussed in Subsection~\ref{sec:tilde} and illustrated in Fig.~\ref{fig:flowchartXix}, 
measurements of the semileptonic decay rates will change this situation. To illustrate this 
future determination, we consider the results in Eq.~\eqref{eq:dthetatilde}, yielding
\begin{equation}
	|x'|_{S_6} = [0.028,0.092], \qquad |x'|_{S_4} = 0.20_{-0.09}^{+0.08} \ .
\end{equation}
These results are in impressive agreement with the constraints for $|x|$ in Eq.~\eqref{eq:xoneplusxandxScenario}, and suggest small $U$-spin-breaking effects.


\boldmath
\subsection{Determination of $\Xi_P$} \label{sec:impl2}
\unboldmath
It is instructive to write the ratio $\Xi_P$ introduced in Eq.~\eqref{eq:defrat} as
\begin{equation}\label{eq:xipuspin}
\Xi_P =	\frac{1+r_P}{1+r_P'} = 1+r_P\xi_r+\mathcal{O}(r_P^2) \ ,
\end{equation}
where $\xi_r$ is an $U$-spin-breaking parameter defined through 
\begin{equation}
r_P' = r_P (1- \xi_r) \ .
\end{equation} 
As in Eq.~\eqref{eq:onePlusrPzetarho}, we may write $r_P$ as a function of $(d,\theta)$ and $\zeta$:
\begin{equation}
1+r_P = \frac{1}{1-\zeta d e^{i\theta} \rho_P} \ ,
\end{equation}
where
\begin{equation}
\zeta \equiv 
|\zeta| e^{i \omega} \equiv \frac{1+x}{1+r_{PA}} \ ;
\end{equation}
an analogous expression holds for $1+r_P'$.

In our new strategy, we eventually determine $d'$ and $\theta'$ from the data, while $d$ and $\theta$ are fixed
through the CP asymmetries of the $B^0_d\to \pi^-\pi^+$ decay. Starting with $\Xi_P =1$, as in 
the strict $U$-spin limit, we may include these effects in an iterative way. 

The parameter $\zeta$ can be determined from our previous analysis. Taking $|1+x|= 1.1 \pm 0.1$ 
from Eq.~\eqref{eq:oneplusx} and $|1+r_{PA}'| = 1.01^{+0.09}_{-0.15}$ as given in 
Eq.~\eqref{eq:oneplusrpa2} yields
\begin{equation}\label{eq:zeta}
|\zeta| = 1.09^{+0.19}_{-0.14} \ .
\end{equation}
Furthermore, $\zeta$ relates the penguin parameters in \Bdtopipi and \BdtoKpi through
\begin{equation}\label{eq:relRdtilde}
\tilde{d}e^{i\tilde{\theta}} = \zeta de^{i \theta} \ ,
\end{equation}
which is only affected by $SU(3)$-breaking effects at the spectator-quark level (see Eq.~\eqref{eq:su3rel}). 

We use now Eq.~\eqref{eq:relRdtilde} to write
\begin{equation}\label{eq:rpdef}
r_P = \frac{\rho_{P}\tilde{d} e^{i(\theta_{P}+\tilde{\theta})} }{1-\rho_{P}\tilde{d} e^{i(\theta_{P}+\tilde{\theta})}}
\end{equation}
and
\begin{equation} \label{eq:rpdif}
|1+r_P| = \left|\frac{1}{1-\rho_{P}\tilde{d} e^{i(\theta_{P}+\tilde{\theta})}} \right| \ . 
\end{equation}
Applying the results for the penguin ratio $\rho_P$ in Eq.~\eqref{eq:rhokk}, and using $(\tilde{d}, \tilde{\theta})$ from Eq.~\eqref{eq:dthetatildenow}, we find
\begin{equation}\label{rP-range}
|r_P| = 0.22 \pm 0.07
\end{equation}
and 
\begin{equation} \label{eq:onePlusrPval}
|1+r_P| = 0.79\pm 0.07\ ,
\end{equation}
where the uncertainties are dominated by those of $\rho_P$ and $\theta_P$. 
Using the numerical range in (\ref{rP-range}) and $\xi_r=0.2$, i.e. assuming $U$-spin-breaking effects
of 20\%, the favourable structure of the $\Xi_P$ ratio in Eq.~\eqref{eq:xipuspin} reduces these uncertainties to the 
$5 \%$ level.

\begin{figure}
	\centering
	\includegraphics[scale=0.75]{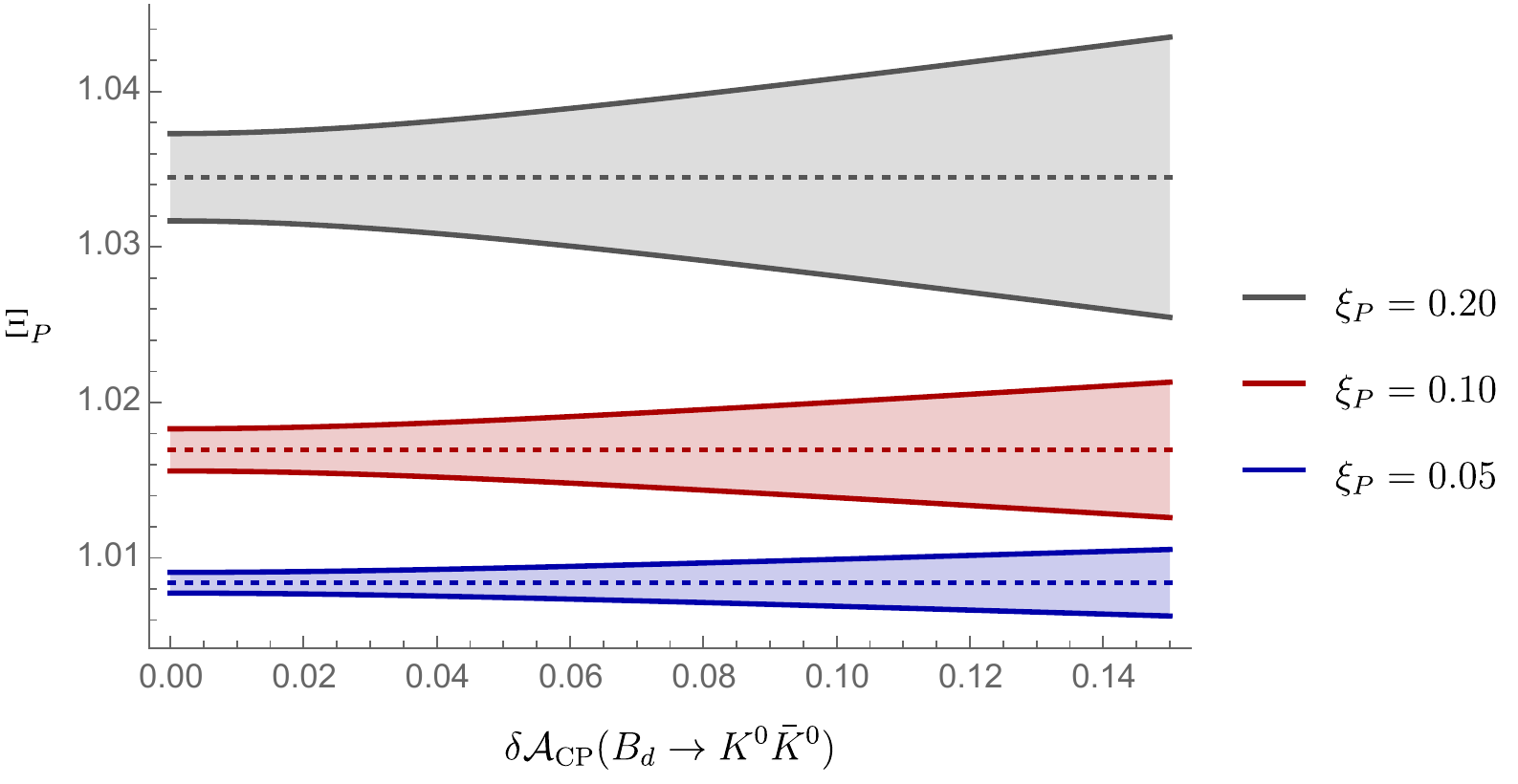}
	\caption{The ratio $\Xi_P$ as a function of the precision of the CP asymmetries of the decay \BdtoKantiK for different $U$-spin-breaking effects.}
	\label{fig:xiPUBPlot}
\end{figure}

Let us now explore how we may reduce the uncertainty of $\Xi_P$ further through sophisticated
analyses provided by future experimental data. We aim at an independent precise determination of $r_P$ and its primed counterpart $r_P'$, whose uncertainties are dominated by $(\rho_P, \theta_P)$. In Subection~\ref{sec:tilde}, we discussed the achievable precision for the \BdtoKantiK penguin parameters $(d_{KK}, \theta_{KK})$, which -- using the $SU(3)$ flavour symmetry -- are equivalent to $(\rho_{P}, \theta_{P})$. A determination of the CP asymmetries at the 0.05 level would lead to a determination of 
$\rho_P$ with $0.03$ uncertainty, giving in turn $r_P = 0.22 \pm 0.02$. The CP asymmetries for 
\BstoKantiK, which have not yet been measured, would allow a determination of $\rho_P'$, thereby providing full information on $U$-spin-breaking effects in these penguin topologies. However, also improved information from the CP asymmetries in \BdtoKantiK alone would already significantly reduce the uncertainty for $\Xi_P$, as determined from Eq.~\eqref{eq:xipuspin} and shown in Fig.~\ref{fig:xiPUBPlot}. There the relation between $\Xi_P$ and the uncertainty of the \BdtoKantiK asymmetries is shown for different $U$-spin-breaking effects between $r_P$ and $r_P'$ defined by $\xi_P$. Consequently, the CP asymmetries in \BdtoKantiK have the potential to reduce the uncertainty for $\Xi_P$ significantly below the $4\%$ level. 

\begin{figure}[t]
	\centering
	\includegraphics[width=0.84\textwidth]{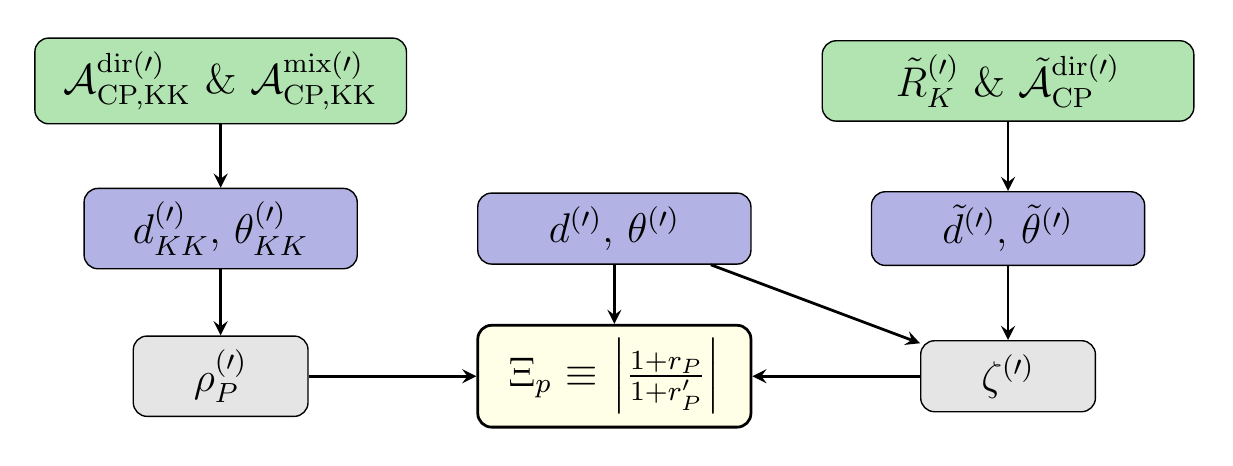}
	\caption{Strategy to determine $\Xi_P$. On the left-hand side, the strategy to determine $\rho_P$
		and  $\rho_P'$ is illustrated, while we show on the right hand side the strategy to improve $\zeta$ and $\zeta'$ using the semileptonic decay ratios. }
	\label{fig:flowchartXiP}
\end{figure}

In addition, the input $(\tilde{d}, \tilde{\theta})$ and their primed analogues can be independently determined via the semileptonic ratios $\tilde{R}_K$ and $\tilde{R}_K^\prime$ through the strategy illustrated in Fig.~\ref{fig:flowchartXiP}. Using Eq.~\eqref{eq:relRdtilde}, we may 
determine $\zeta$ and $\zeta'$, providing additional information into $U$-spin-breaking effects in exchange and penguin-annihilation topologies. In order to illustrate the future 
precision of this method, we consider $(d, \theta)$ and $(\tilde{d}, \tilde{\theta})$ for the upgrade 
scenario as given in Eqs.~\eqref{eq:dval} and \eqref{eq:dthetatilde}, respectively, which leads to 
an impressive precision of $|\zeta|= 0.93\pm 0.05$ and $\omega  = (4.0 \pm 1.3)^\circ$ as given 
in Eq.~\eqref{eq:xis}.

\section{Prospects of the New Strategy} \label{sec:Discussion}
The precision for $\phi_s$ achievable with the new strategy depends on experimental and theoretical uncertainties. Experimentally, the precision with which the semileptonic ratios $R_K$ and $R_\pi$ can be determined dominate the uncertainty. In Subsection~\ref{sec:DeterminationOfDeltaPhiKK}, we showed that a relative precision for $R_K$ and $R_\pi$ at the $5\%$ level allows an impressive $0.5^\circ$ uncertainty for $\Delta\phi_{KK}$. With the information obtained in the previous sections, we can now quantify the theoretical error for $\xi_\text{NF}^a$. This uncertainty arises from $U$-spin-breaking effects in the ratios $\Xi_x$ and $\Xi_P$. Fortunately, these ratios are very robust with respect to these effects and can be obtained from experimental data.

For the current data, we obtain an uncertainty of $5 \%$ for $\Xi_P$, which can be further reduced with more precise data for the $\BdtoKantiK$ and $\BstoKantiK$ CP asymmetries (see \ref{sec:rhokk}~and~\ref{sec:impl2}). In addition, we find an uncertainty of $4 \%$ for $\Xi_x$, which can also be further improved using the CP asymmetries of \Bdtopipi, \BstoKK (see \ref{sec:impl}). The last source of uncertainty is related to the non-factorizable $U$-spin-breaking effects in the ratio of the colour-allowed tree topologies, which are theoretically well-behaved and give an error at the $1 \%$ level (see \ref{sec:DeterminationOfDeltaPhiKK}). Finally, adding up the individual errors in quadrature we find a precision of about $7 \%$ for $\xi_\text{NF}^a$. 

\begin{figure}[t]
	\centering
	\includegraphics[scale=0.8]{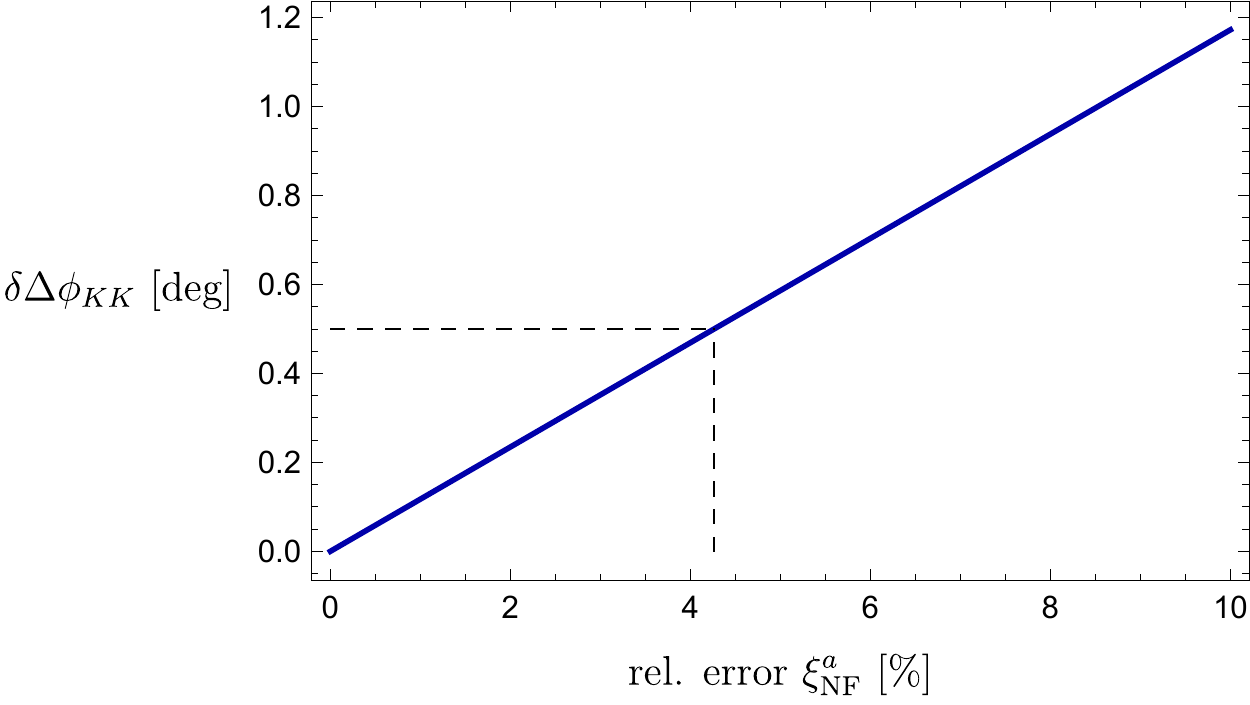}
	\caption{The uncertainty for $\Delta\phi_{KK}$ as a function of the relative error of $\xi_\text{NF}^a$, assuming a perfect experimental situation.}
	\label{fig:phis_2}
\end{figure}

Fig.~\ref{fig:phis_2} gives the precision of $\Delta\phi_{KK}$ as a function of the relative error of $\xi_\text{NF}^a$, assuming a perfect experimental situation. We observe that a $7\%$ precision for $\xi_\text{NF}^a$ gives a theoretical uncertainty at the $0.8^\circ$ level for $\Delta\phi_{KK}$. Recalling that $\phi_s = \phi_s^{\rm eff} - \Delta\phi_{KK}$ and that a precision of $0.5^\circ$ for $\phi_s^{\rm eff}$ can be reached in the upgrade era (Eq.~\eqref{eq:phiseff}), we aim for a similar theoretical precision for $\Delta\phi_{KK}$, which is indicated by the dashed line in Fig.~\ref{fig:phis_2}. Such a precision requires an $\mathcal{O}(4 \%)$ determination of $\xi_{\rm NF}^a$, which is within reach in the upgrade era. 

\begin{figure}[t]
\centering
	\includegraphics[scale=0.85]{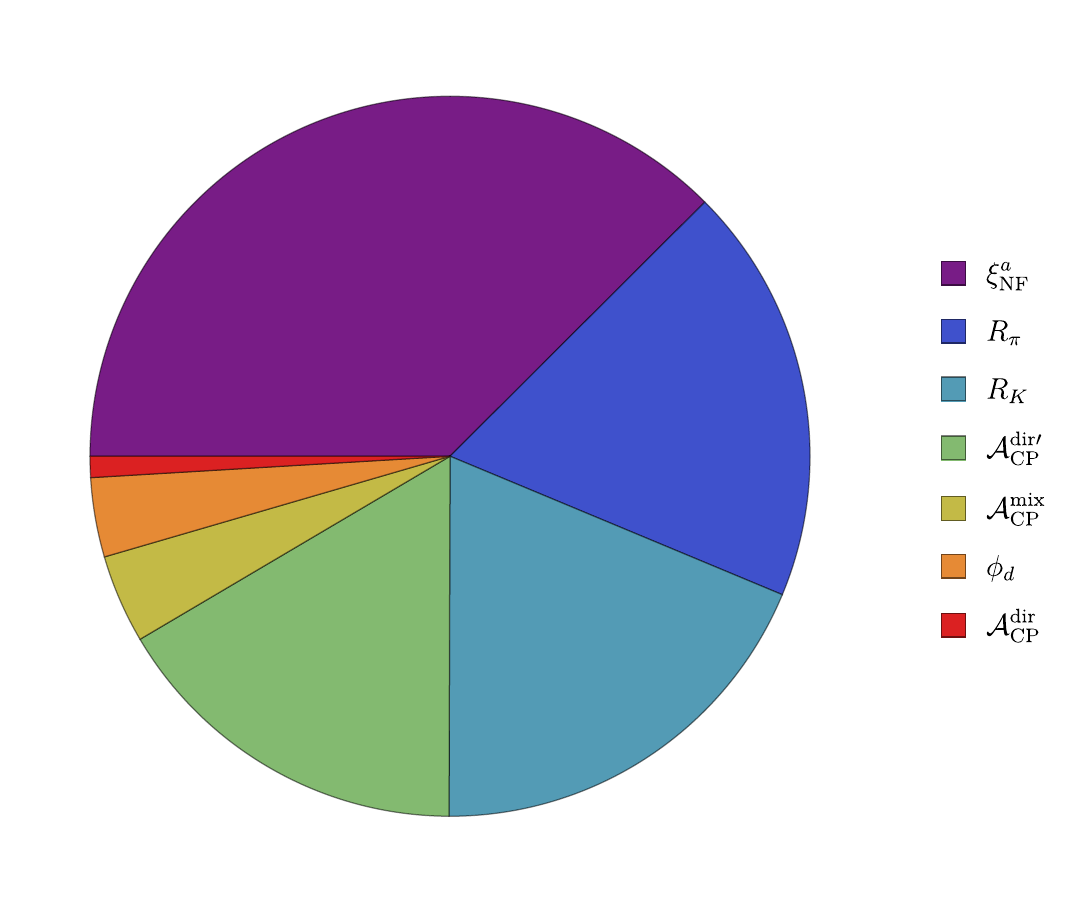}
	\caption{Error budget of the hadronic phase shift $\Delta\phi_{KK}$.}
	\label{fig:phis_withXi}
\end{figure}

Combining now the experimental and theoretical uncertanties, assuming a relative precision of $5 \%$ for the relevant parameters $R_\pi$, $R_K$ and $\xi_\text{NF}^a$, results in an impressive uncertainty of $0.8^\circ$ for $\Delta\phi_{KK}$. The error budget of $\Delta\phi_{KK}$ in this scenario is given in Fig.~\ref{fig:phis_withXi}. This allows a determination of $\phi_s$ with a similar precision, which is a major improvement with respect to the current situation in Eq.~\eqref{eq:Uspinnow}.

Interestingly, our new method allows also the determination of the hadronic parameters $d'$ and $\theta'$. Assuming that $R_K, R_\pi$ and $\xi_{\rm{NF}}^a$ can be determined with $5\%$ uncertainty, we find
\begin{equation} \label{eq:dpthetapup}
d' = 0.58 \pm 0.04 \ , \qquad \theta' = (151.4 \pm 3.5)^\circ \ ,
\end{equation} 
showing a very impressive precision and providing valuable insights into the $U$-spin symmetry. In particular, we may now determine the $U$-spin-breaking parameters $\xi$ and $\Delta$ in Eq.~\eqref{eq:uspinbreaking}. For the upgrade scenario, $\xi$ can be extracted with an uncertainty at the $0.07$ level.

In addition, our method offers a test of QCD factorization in the \Bdtopipi and \BstoKK decays through the information for $r_P$ and $x$. We have given the current experimental value for $|a_{\rm NF}|$ 
of the $\Bdtopipi$ decay in Eq.~\eqref{eq:anf}:
\begin{equation}
|a_{\rm NF}| =  |1+r_P||1+x||a_\text{NF}^T| = 0.73 \pm 0.06\ .
\end{equation} 
Using $|1+x| = 1.1\pm 0.1$ from Eq.~\eqref{eq:oneplusx} and $|1+r_P|=0.8\pm 0.08$ from Eq.~\eqref{eq:onePlusrPval} yields 
\begin{equation}
|a_{\rm{NF}}^T| = 0.82 \pm 0.13 \,,
\end{equation}
which agrees with the QCD factorization calculation in Eq.~\eqref{QCDF-calc} at the 
$1\,\sigma$ level.

A key element in the new strategy are the semileptonic differential rates and the corresponding $R_\pi$ and $R_K$ ratios. Since the $B^0_s\to K^-\ell^+\nu_\ell$ decay has not yet been measured, it is interesting to come back to the ratio $K$ and the use of form-factor calculations as input. In this case, the ratios $R_\pi$ and $R_K$ are no longer required and we can write
\begin{equation}
    r_K = r_\pi K,
\end{equation}
where $K$ is given in Eq.~\eqref{eq:kObservable}. The only difference with respect to our new strategy is that we have now to rely on theoretical input for the form-factor ratio $F^{B_sK}(m_K^2)/F^{B_d\pi}(m_\pi^2)$, replacing the ratio $R_\pi/R_K$ which can be determined by means of experimental data. The non-factorizable 
$U$-spin-breaking effects are again described by the parameter $\xi_{\rm NF}^a$. The current 
determination of the form-factor ratio from LCSR, $F_0^{B_sK}(0)/F_0^{B_d\pi}(0)=1.15^{+0.17}_{-0.09}$
\cite{Dub08}, has still a significant uncertainty. However, dedicated efforts using lattice QCD and
progress with LCSR analyses may lead to a sharper picture of $F^{B_sK}(m_K^2)/F^{B_d\pi}(m_\pi^2)$
in the future. 

\begin{figure}
    \centering
        \includegraphics[width=.85\linewidth]{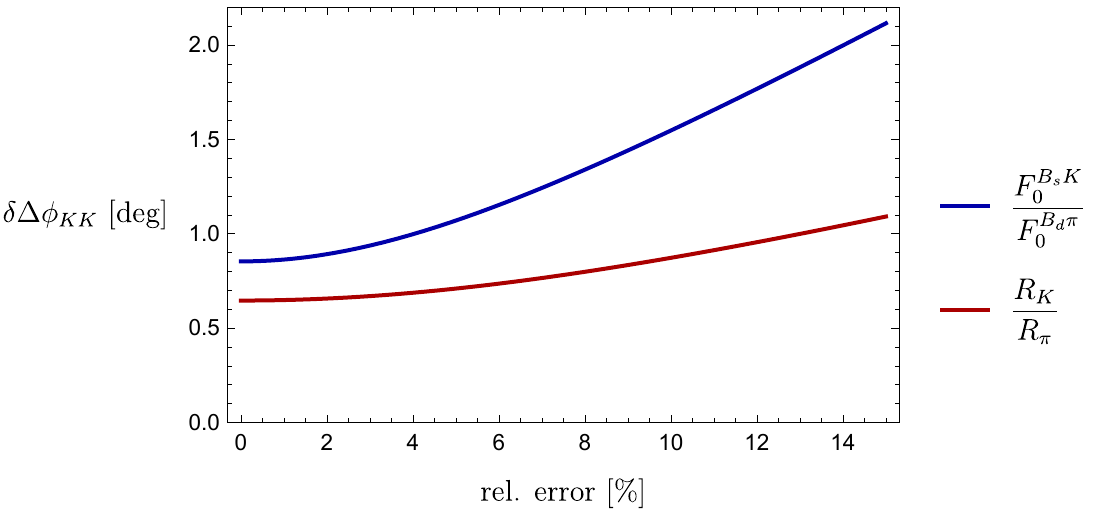}
        \caption{The dependence of the uncertainty of $\Delta\phi_{KK}$ on the relative error of the ratio of form factors using the ratio $K$, and $R_K/R_\pi$ using the semileptonic decays.}
    \label{fig:stratWithKerrxiNFa}
\end{figure}

Let us consider the LHCb upgrade scenario, assuming $\xi_\text{NF}^a = 1.00 \pm 0.05$. In Fig.~\ref{fig:stratWithKerrxiNFa}, we show the precision of $\Delta\phi_{KK}$ as a function of the relative uncertainty of the form-factor ratio in comparison with relative precision of $R_K/R_\pi$ 
using the new strategy. We observe that a good precision can be reached using the ratio $K$, provided 
it is possible to calculate form-factor ratio with a precision at the $5\%$ level. However, it will be 
challenging to go beyond the precision of our new strategy, even if the experimental ratio $R_K/R_\pi$ 
would only be known with $15\%$ precision. Consequently, the new strategy, which does not rely on 
non-perturbative input for the form factors, is most powerful for extracting $\phi_s$. 

We may actually use our new strategy to determine $F^{B_sK}(m_K^2)/F^{B_d\pi}(m_\pi^2)$. 
Using the values of $d$, $\theta$ and $d'$, $\theta'$, we may calculate $K$ with the help of
Eq.~\eqref{eq:kObservable}, which allows us to extract $|\mathcal{C}/\mathcal{C}^\prime|$ from the ratio of the \BstoKK, \Bdtopipi branching ratios,
and write
\begin{equation}
    \left|\frac{\mathcal{C}}{\mathcal{C}^\prime}\right| \equiv  \left|\frac{T}{T^\prime}\right|_\text{fact} \xi_{\rm{NF}}^{a} = \frac{f_\pi}{f_K}\left[\frac{m_{B_d}^2-m_\pi^2}{m_{B_s}^2-m_K^2}\right]\left[\frac{F_0^{B_d\pi}(m_\pi^2)}{F_0^{B_sK}(m_K^2)}\right]\xi_{\rm{NF}}^{a} \ .
\end{equation}
For the current data, using Eqs.~\eqref{eq:dvalnow} and \eqref{eq:dthetatildenow} and 
$\gamma=(70\pm 7)^\circ$, we find
\begin{equation}
K = 57.2 \pm 14.4 \ ,
\end{equation}
where we assumed $d' = \tilde{d}$ and $\theta'= \tilde{\theta}$, neglecting tiny exchange and 
penguin-annihilation topologies. Using $\xi_\text{NF}^a = 1.00 \pm  0.07$ gives 
\begin{equation}
\frac{F_0^{B_sK}(m_K^2)}{F_0^{B_d\pi}(m_\pi^2)} = 1.09 \pm 0.16 ,
\end{equation}
which is in interesting agreement with the LCSR calculation. For the LHCb upgrade scenario, we expect that the precision for the ratio of form factors can be reduced to the $0.06$ level.

\section{Conclusions} \label{sec:Conclusion}
The $U$-spin relation between the \Bdtopipi and \BstoKK decays has originally been proposed to 
extract the UT angle $\gamma$ and the mixing phase $\phi_s$ \cite{Fle99, Fle07, Fle10}. The current
experimental picture is already impressive, in agreement with the SM and uncertainties at the $7^\circ$
level. The theoretical precision is limited by $U$-spin-breaking corrections, which do not allow us -- unless 
there is significant progress to calculate them -- to take full advantage of the data to be collected in the 
era of Belle II and the LHCb upgrade. 

In view of this situation, we proposed a new strategy to fully exploit the physics potential of the non-leptonic \BstoKK and \Bdtopipi decays to extract $\phi_s$ \cite{SHORT}. The strategy utilizes the $U$-spin relation between these two decays for theoretically well behaved quantities, thereby resulting in a very robust
situation with respect to $U$-spin-breaking effects. The new key elements are the differential rates of the
semileptonic decays $B_d^0\rightarrow \pi^-\ell^+\nu_\ell$ and $B_s^0\rightarrow K^-\ell^+\nu_\ell$, 
which enter ratios with the \Bdtopipi and \BstoKK decay rates $R_\pi$ and $R_K$, respectively. In fact, only the double ratio $R_\pi$/$R_K$ enters our strategy, which is an advantage from the experimental point
of view as uncertainties cancel. A theoretical advantage is that the form factors now enter only in a double ratio, which is equal to 1 with excellent precision. In our new strategy, non-factorizable $U$-spin-breaking
corrections to the notoriously difficult to calculate penguin, exchange and penguin-annihilation topologies 
only contribute through the ratios $\Xi_P$ and $\Xi_x$. As we have shown, these quantities are very 
robust with respect to $U$-spin-breaking effects. On the other hand, the original strategy is limited by 
leading non-factorizable $U$-spin-breaking effects that emerge from penguin topologies.

The CP-violating observables of \BstoKK allow us to determine the ``effective" \Bs--\Bsb 
mixing phase $\phi_s^\text{eff}$, which is a pure experimental quantity. In order to extract the 
mixing phase $\phi_s$ from $\phi_s^\text{eff}$, we have to  subtract the hadronic phase shift 
$\Delta\phi_{KK}$, which depends on non-perturbative quantities. In the upgrade era, $\phi_s^\text{eff}$ 
can be measured with an uncertainty at the $0.5^\circ$ level. Consequently, our goal is to match
this very impressive experimental precision by theory, determining $\Delta\phi_{KK}$ with similar 
uncertainty.

Unfortunately, the $B_s^0\rightarrow K^-\ell^+\nu_\ell$ decay, a key input for our new strategy, has 
not yet been measured. We strongly advocate analyses of this channel at Belle (II) and LHCb, preferably 
extracting $R_K$ or the ratio $R_\pi/R_K$ directly from the experimental data. In order to illustrate the 
strength of our new method, we use data for \BdtoKpi. This decay is related to \BstoKK by a 
$U$-spin relation at the spectator quark level if the small contributions from exchange and penguin annihilation topologies are neglected. We find a precision for $\Delta\phi_{KK}$ of $0.6^\circ$, which 
shows impressively the power of our strategy. Moreover, we obtain excellent agreement with the picture of the 
$U$-spin symmetry, excluding anomalously large corrections. 

The determination of $\Delta\phi_{KK}$ is affected by experimental and theoretical uncertainties. 
For a perfect theoretical situation, measurements of $R_\pi$ and $R_K$ with $5\%$ precision 
are required to obtain a $0.5^\circ$ precision for $\Delta\phi_{KK}$. The theoretical precision is 
limited by $U$-spin-breaking corrections to quantities which have very favourable structures. 
In order to fully exploit the precision of our strategy, we need information both for the 
penguin ratio $r_P$ and for the exchange and penguin-annihilation parameter $x$. 

The penguin parameter $r_P$ can be studied with the help of the pure penguin decays \BdtoKantiK 
and \BstoKantiK, which offer an interesting laboratory for the upgrade era. Since the 
current data for these modes are limited, we have also used the charged decays  
$B^+ \rightarrow K^+ \Kbar^0$ and $B^+\rightarrow \pi^+ K^0$ to constrain the size of $r_P$. 
In summary, using these decays, we find an uncertainty for the relevant ratio $\Xi_p$ at the $5 \%$ 
level. We have presented a strategy to further reduce this uncertainty, as illustrated 
in Fig.~\ref{fig:flowchartXiP}. 

Future measurements of the CP asymmetries of \BdtoKK and \Bstopipi allow us to determine the 
exchange and penguin-annihilation contributions with high precision. We have discussed the correlation between these CP asymmetries following from the current data, resulting in an interesting picture for
the future data taking, and presented scenarios of future measurements and their use to pin down the exchange and penguin-annihilation contributions even further. For the current data, we use ratios of 
different $B \to hh$ ($h = \pi, K$) decays and find a contribution of $x\sim0.1$, which results in a 
theoretical uncertainty of $\mathcal{O}(4 \%)$ for the exchange and penguin-annihilation ratio $\Xi_x$. 

Combining the different sources of theoretical uncertainty, we find a theoretical precision of 
$\Delta\phi_{KK}$ at the $0.8^\circ$ level. We have discussed different strategies to reduce 
this uncertainty further with future experimental data, and have illustrated them with various scenarios, showing that a future ultimate precision at the $0.5^\circ$ level is within reach. Consequently, the
new strategy has the potential to extract $\phi_s$ from CP violation in  \BstoKK 
with a theoretical precision matching experiment. The key question is whether the corresponding
value will eventually show a discrepancy with respect to the clean SM prediction $\phi_s^{\rm SM}$
and determinations from other decays, in particular $\B_s^0 \to J/\psi \phi$. Since \BstoKK is 
dominated by QCD penguin topologies, which are sensitive to possible new heavy particles, we may 
actually find a surprise, fully exploiting the excellent experimental precision attainable at Belle II and the LHCb upgrade.


\section*{Acknowledgements}
We would like to thank Kristof De Bruyn for very useful discussions and Vincenzo Vagnoni for 
correspondence. This work is supported by the Foundation for Fundamental Research on Matter 
(FOM) and by the Deutsche Forschungsgemeinschaft (DFG) within research unit FOR 1873 (QFET).


\end{document}